\documentstyle[12pt]{article}

\catcode`@=11
\def\un#1{\relax\ifmmode\@@underline#1\else
        $\@@underline{\hbox{#1}}$\relax\fi}
\catcode`@=12

\let\du=\d                      
\let\um=\H                      

\def\a{\alpha}
\def\b{\beta}

\def\d{\delta}
\def\e{\epsilon}
\def\f{\phi}
\def\g{\gamma}
\def\h{\eta}

\def\j{\psi}
\def\k{\kappa}
\def\l{\lambda}

\def\r{\rho}
\def\s{\sigma}
\def\t{\tau}

\def\x{\xi}
\def\z{\zeta}

\def\F{\Phi}
\def\G{\Gamma}
\def\J{\Psi}
\def\L{\Lambda}
\def\O{\Omega}

\def\S{\Sigma}

\def\X{\Xi}

\def\ca{{\cal A}}
\def\cb{{\cal B}}

\def\cv{{\cal V}}

\font\ooo=lcircle10                      
\font\ro=manfnt                         
\def\kcl{{\hbox{\ro 6}}}                
\def\kcr{{\hbox{\ro 7}}}                
\def\ktl{{\hbox{\ro \char'134}}}        
\def\ktr{{\hbox{\ro \char'135}}}        
\def\kbl{{\hbox{\ro \char'136}}}        
\def\kbr{{\hbox{\ro \char'137}}}        

\def\ip{{=\!\!\! \mid}}                                    
\def\bo{{\raise.15ex\hbox{\large$\Box$}}}               
\def\pa{\partial}                                       
\def\pr{\prod}                                          
\def\TH{{\raise.2ex\hbox{$\displaystyle \bigodot$}\mskip-4.7mu \llap H \;}}
\def\face{{\raise.2ex\hbox{$\displaystyle \bigodot$}\mskip-2.2mu \llap {$\ddot
        \smile$}}}                                      
\def\dg{\sp\dagger}                                     

\def\sp#1{{}^{#1}}                              
\def\Tilde#1{{\widetilde{#1}}\hskip 0.03in}                     
\def\Hat#1{\widehat{#1}}                        
\def\Bar#1{\overline{#1}}                       
\def\leftrightarrowfill{$\mathsurround=0pt \mathord\leftarrow \mkern-6mu
        \cleaders\hbox{$\mkern-2mu \mathord- \mkern-2mu$}\hfill
        \mkern-6mu \mathord\rightarrow$}
\def\dvec#1{\vbox{\ialign{##\crcr
        \leftrightarrowfill\crcr\noalign{\kern-1pt\nointerlineskip}
        $\hfil\displaystyle{#1}\hfil$\crcr}}}           
\def\dt#1{{\buildrel {\hbox{\LARGE .}} \over {#1}}}     

\def\frac#1#2{{\textstyle{#1\over\vphantom2\smash{\raise.20ex
        \hbox{$\scriptstyle{#2}$}}}}}                   
\def\ha{\frac12}                                        
\def\sfrac#1#2{{\vphantom1\smash{\lower.5ex\hbox{\small$#1$}}\over
        \vphantom1\smash{\raise.4ex\hbox{\small$#2$}}}} 
\def\bfrac#1#2{{\vphantom1\smash{\lower.5ex\hbox{$#1$}}\over
        \vphantom1\smash{\raise.3ex\hbox{$#2$}}}}       
\def\afrac#1#2{{\vphantom1\smash{\lower.5ex\hbox{$#1$}}\over#2}}    

\newskip\humongous \humongous=0pt plus 1000pt minus 1000pt
\def\caja{\mathsurround=0pt}
\def\eqalign#1{\,\vcenter{\openup2\jot \caja
        \ialign{\strut \hfil$\displaystyle{##}$&$
        \displaystyle{{}##}$\hfil\crcr#1\crcr}}\,}
\newif\ifdtup
\def\panorama{\global\dtuptrue \openup2\jot \caja
        \everycr{\noalign{\ifdtup \global\dtupfalse
        \vskip-\lineskiplimit \vskip\normallineskiplimit
        \else \penalty\interdisplaylinepenalty \fi}}}
\def\li#1{\panorama \tabskip=\humongous                         
        \halign to\displaywidth{\hfil$\displaystyle{##}$
        \tabskip=0pt&$\displaystyle{{}##}$\hfil
        \tabskip=\humongous&\llap{$##$}\tabskip=0pt
        \crcr#1\crcr}}

\def\ref#1{$\sp{#1)}$}

\parskip=\medskipamount                 
\thicklines                         

\def\oldheadpic{                                
        \setlength{\unitlength}{.4mm}
        \thinlines
        \par
        \begin{picture}(349,16)
        \put(325,16){\line(1,0){4}}
        \put(330,16){\line(1,0){4}}
        \put(340,16){\line(1,0){4}}
        \put(335,0){\line(1,0){4}}
        \put(340,0){\line(1,0){4}}
        \put(345,0){\line(1,0){4}}
        \put(329,0){\line(0,1){16}}
        \put(330,0){\line(0,1){16}}
        \put(339,0){\line(0,1){16}}
        \put(340,0){\line(0,1){16}}
        \put(344,0){\line(0,1){16}}
        \put(345,0){\line(0,1){16}}
        \put(329,16){\oval(8,32)[bl]}
        \put(330,16){\oval(8,32)[br]}
        \put(339,0){\oval(8,32)[tl]}
        \put(345,0){\oval(8,32)[tr]}
        \end{picture}
        \par
        \thicklines
        \vskip.2in}
\def\oldtitle#1#2#3#4{\oldheadpic\begin{center}\vglue.5in{\large\bf #1}\\[.6in]
        {#2}\\[.1in] {\it Department of Physics and Astronomy}\\
        {\it University of Maryland, College Park, MD 20742}\\[.6in]
        Physics Publication \#{#3}\\ {#4}\\[1.5in] {\bf ABSTRACT}\\[.1in]
        \end{center} \begin{quotation}}                 
\def\oldTitle#1#2#3#4#5#6#7{\oldheadpic\begin{center} \vglue .4in
        {\large\bf #1}\\[.4in]
        {#2}\\[.1in] {\it Department of Physics and Astronomy}\\
        {\it University of Maryland, College Park, MD 20742}\\[.1in]
        {#3}\\[.1in] {\it {#4}}\\ {\it {#5}}\\[.4in]
        Physics Publication \#{#6}\\ {#7}\\[.5in] {\bf ABSTRACT}\\[.1in]
        \end{center} \begin{quotation}}                 
\def\border{                                            
        \setlength{\unitlength}{1mm}
        \newcount\xco
        \newcount\yco
        \xco=-24
        \yco=12
        \begin{picture}(140,0)
        \put(\xco,\yco){$\ktl$}
        \advance\yco by-1
        {\loop
        \put(\xco,\yco){$\kcl$}
        \advance\yco by-2
        \ifnum\yco>-240
        \repeat
        \put(\xco,\yco){$\kbl$}}
        \xco=158
        \yco=12
        \put(\xco,\yco){$\ktr$}
        \advance\yco by-1
        {\loop
        \put(\xco,\yco){$\kcr$}
        \advance\yco by-2
        \ifnum\yco>-240
        \repeat
        \put(\xco,\yco){$\kbr$}}
        \put(-20,11){\tiny University of Maryland Elementary Particle
Physics University of Maryland Elementary Particle Physics University of
Maryland Elementary Particle Physics}
        \put(-20,-241.5){\tiny University of Maryland Elementary
Particle Physics University of Maryland Elementary Particle Physics
University of Maryland Elementary Particle Physics}
        \end{picture}
        \par\vskip-8mm}
\def\bordero{                                           
        \setlength{\unitlength}{1mm}
        \newcount\xco
        \newcount\yco
        \xco=-24
        \yco=12
        \begin{picture}(140,0)
        \put(\xco,\yco){$\ktl$}
        \advance\yco by-1
        {\loop
        \put(\xco,\yco){$\kcl$}
        \advance\yco by-2
        \ifnum\yco>-240
        \repeat
        \put(\xco,\yco){$\kbl$}}
        \xco=158
        \yco=12
        \put(\xco,\yco){$\ktr$}
        \advance\yco by-1
        {\loop
        \put(\xco,\yco){$\kcr$}
        \advance\yco by-2
        \ifnum\yco>-240
        \repeat
        \put(\xco,\yco){$\kbr$}}
        \put(-20,12){\ooo
bacdefghidfghghdhededbihdgdfdfhhdheidhdhebaaahjhhdahba
hgdedge
   hgfdiehhgdigicba}
        \put(-20,-241.5){\ooo
ababaighefdbfghgeahgdfgafagihdidihiidhiagfedhadbfd
ecdcdfa
   gdcbhaddhbgfchbgfdacfediacbabab}
        \end{picture}
        \par\vskip-8mm}
\def\headpic{                                           
        \indent
        \setlength{\unitlength}{.4mm}
        \thinlines
        \par
        \begin{picture}(29,16)
        \put(165,16){\line(1,0){4}}
        \put(170,16){\line(1,0){4}}
        \put(180,16){\line(1,0){4}}
        \put(175,0){\line(1,0){4}}
        \put(180,0){\line(1,0){4}}
        \put(185,0){\line(1,0){4}}
        \put(169,0){\line(0,1){16}}
        \put(170,0){\line(0,1){16}}
        \put(179,0){\line(0,1){16}}
        \put(180,0){\line(0,1){16}}
        \put(184,0){\line(0,1){16}}
        \put(185,0){\line(0,1){16}}
        \put(169,16){\oval(8,32)[bl]}
        \put(170,16){\oval(8,32)[br]}
        \put(179,0){\oval(8,32)[tl]}
        \put(185,0){\oval(8,32)[tr]}
        \end{picture}
        \par\vskip-6.5mm
        \thicklines}
\def\title#1#2#3#4{\border\headpic {\hbox to\hsize{#4 \hfill UMDEPP #3}}\par
        \begin{center} \vglue .5in {\large\bf #1}\\[.6in]
        {#2}\\[.1in] {\it Department of Physics and Astronomy}\\
        {\it University of Maryland, College Park, MD 20742}\\[1.5in]
        {\bf ABSTRACT}\\[.1in] \end{center} \begin{quotation}}  
\def\Title#1#2#3#4#5#6#7{\border\headpic
        {\hbox to\hsize{#7 \hfill UMDEPP #6}}\par
        \begin{center} \vglue .4in {\large\bf #1}\\[.4in]
        {#2}\\[.1in] {\it Department of Physics and Astronomy}\\
        {\it University of Maryland, College Park, MD 20742}\\[.1in]
        {#3}\\[.1in] {\it {#4}}\\ {\it {#5}}\\[.5in] {\bf ABSTRACT}\\[.1in]
        \end{center} \begin{quotation}}                 
\def\endtitle{\end{quotation}\newpage}                  

\def\sect#1{\bigskip\medskip \goodbreak \noindent{\bf {#1}} \nobreak \medskip}
\def\refs{\sect{REFERENCES} \footnotesize \frenchspacing \parskip=0pt}
\def\Item{\par\hang\textindent}

\begin{document}

\def\hati{\hat i} \def\hatj{\hat j} \def\hatk{\hat k}

\def\vq{\vartheta}
\def\gg{{\hbox{\sc g}}}
\def\nt{$~N=2$~}
\def\gg{{\hbox{\sc g}}}
\def\nt{$~N=2$~}
\def\tr{{\rm tr}}
\def\Tr{{\rm Tr}}
\def\mpl#1#2#3{Mod.~Phys.~Lett.~{\bf A{#1}} (19{#2}) #3}

\def\scst{\scriptstyle}
\def\itrema{$\ddot{\scriptstyle 1}$}
\def\st{\Tilde\s}
\def\ve{\varepsilon}
\def\ab{\Bar a}
\def\bb{\Bar b}
\def\cb{\Bar c}
\def\db{\Bar d}
\def\3b{\Bar 3}

\def\Bo{\bo{\hskip 0.03in}}
\def\lrad#1{ \left( A {\buildrel\leftrightarrow\over D}_{#1} B\right) }

\def\kd#1#2{\d{\du{#1}{#2}}}

\def\ula{{\underline a}} \def\ulb{{\underline b}} \def\ulc{{\underline c}}
\def\uld{{\underline d}} \def\ule{{\underline e}} \def\ulf{{\underline f}}
\def\ulg{{\underline g}} \def\ulh{{\underline h}}
\def\ulm{{\underline m}} \def\uln{{\underline n}}
\def\ulp{{\underline p}} \def\ulq{{\underline q}} \def\ulr{{\underline r}}

\def\plpl{{+\!\!\!\!\!{\hskip 0.009in}{\raise -1.0pt\hbox{$_+$}}
{\hskip 0.0008in}}}

\def\mimi{{-\!\!\!\!\!{\hskip 0.009in}{\raise -1.0pt\hbox{$_-$}}
{\hskip 0.0008in}}}

\def\items#1{\\ \item{[#1]}}
\def\ul{\underline}
\def\un{\underline}
\def\-{{\hskip 1.5pt}\hbox{-}}

\def\fracmm#1#2{{{#1}\over{#2}}}
\def\footnotew#1{\footnote{\hsize=6.5in {#1}}}
\def\low#1{{\raise -3pt\hbox{${\hskip 0.9pt}\!_{#1}$}}}

\def\ip{{=\!\!\! \mid}}
\def\ze{\zeta^{+}}
\def\zeb{{\bar \zeta}^{+}}
\def\umb{{\underline {\bar m}}}
\def\unb{{\underline {\bar n}}}
\def\upb{{\underline {\bar p}}}
\def\um{{\underline m}}
\def\up{{\underline p}}
\def\Phib{{\Bar \Phi}}
\def\Phit{{\tilde \Phi}}
\def\Phibt{{\tilde {\Bar \Phi}}}
\def\Db{{\Bar D}_{+}}
\def\gg{{\hbox{\sc g}}}
\def\nt{$~N=2$~}

\border\headpic {\hbox to\hsize{June 1992 \hfill UMDEPP 92--211}}\par
\begin{center}
\vglue .02in

{\large\bf Self--Dual Supersymmetry and Supergravity}\\
{\large\bf in Atiyah--Ward Space--Time}\footnote{Research supported by NSF
grant \# PHY--91--19746}\\[.28in]

Sergei V.~KETOV,\footnote{Address after the October 1st
1992: Institute of Theoretical Physics, University of Hannover, Appelstrasse 2,
D-3000, Hannover 1, Germany}$^,$\footnote{On leave of absence from: High
Current Electronics Institute of the Russian Academy of Sciences, Siberian
Branch, Akademichesky~4, Tomsk 634055, Russia}
{}~Hitoshi NISHINO~ and ~S. James GATES, Jr.\\[.25in]
\baselineskip 10pt
{\it Department of Physics} \\ [0.015in]
{\it University of Maryland at College Park} \\ [0.015in]
{\it College Park, MD 20742-4111, USA}\\ [.08in]
and  \\[.08in]
{\it Department of Physics and Astronomy} \\ [0.015in]
{\it Howard University} \\ [0.015in]
{\it Washington, D.C. 20059, USA} \\ [0.6in]

{\bf ABSTRACT} \\[.025in]

\end{center}
\begin{quotation}

\oddsidemargin=0.03in
\evensidemargin=0.01in
\hsize=5.8in
\textwidth=5.8in

We study supersymmetry and self-duality in a four-dimensional space-time with
the signature $\,(2,2)$, that we call the {\it Atiyah-Ward} space-time. Dirac
matrices and spinors, in particular {\it Majorana-Weyl} spinors, are
investigated in detail. We formulate $\, N\ge 1\,$ supersymmetric
{\it self-dual} Yang-Mills theories and {\it self-dual} supergravities.
An $\, N=1\,$ ``self-dual'' {\it tensor} multiplet is constructed and a
possible ten-dimensional theory that gives rise to the four-dimensional
self-dual supersymmetric theories is found.  Instanton solutions are given
as the zero modes in the $~N=2$~ self-dual Yang-Mills theory.  The $~N=2$~
superstrings are conjectured to have no possible counter-terms at
quantum level to all orders. These self-dual supersymmetric
theories are to generate exactly soluble supersymmetric systems in
lower dimensions.

\baselineskip 10pt
\endtitle
\def\doit#1#2{\ifcase#1\or#2\fi}
\def\[{\lfloor{\hskip 0.35pt}\!\!\!\lceil}
\def\]{\rfloor{\hskip 0.35pt}\!\!\!\rceil}
\def\delsl{{{\partial\!\!\! /}}}
\def\caldsl{{\calD\!\!\! /}}
\def\calO{{\cal O}}
\def\asym{({\scriptstyle 1\leftrightarrow \scriptstyle 2})}
\def\Lag{{\cal L}}
\def\du#1#2{_{#1}{}^{#2}}
\def\ud#1#2{^{#1}{}_{#2}}
\def\dud#1#2#3{_{#1}{}^{#2}{}_{#3}}
\def\udu#1#2#3{^{#1}{}_{#2}{}^{#3}}
\def\calD{{\cal D}}
\def\calM{{\cal M}}
\def\tildef{{\tilde f}}
\def\calDsl{{\calD\!\!\!\! /}}

\def\Hat#1{{#1}{\large\raise-0.02pt\hbox{$\!\hskip0.038in\!\!\!\hat{~}$}}}
\def\hati{{\hat{I}}}
\def\dt{$~D=10$~}
\def\alp{\alpha{\hskip 0.007in}'}
\def\oalp#1{\alp^{\hskip 0.007in {#1}}}
\def\naive{{{na${\scriptstyle 1}\!{\dot{}}\!{\dot{}}\,\,$ve}}}
\def\items#1{\vskip 0.05in\Item{[{#1}]}}
\def\item#1{\Item{#1}}

\def\pl#1#2#3{Phys.~Lett.~{\bf {#1}B} (19{#2}) #3}
\def\np#1#2#3{Nucl.~Phys.~{\bf B{#1}} (19{#2}) #3}
\def\prl#1#2#3{Phys.~Rev.~Lett.~{\bf #1} (19{#2}) #3}
\def\pr#1#2#3{Phys.~Rev.~{\bf D{#1}} (19{#2}) #3}
\def\cqg#1#2#3{Class.~and Quant.~Gr.~{\bf {#1}} (19{#2}) #3}
\def\cmp#1#2#3{Comm.~Math.~Phys.~{\bf {#1}} (19{#2}) #3}
\def\jmp#1#2#3{Jour.~Math.~Phys.~{\bf {#1}} (19{#2}) #3}
\def\ap#1#2#3{Ann.~of Phys.~{\bf {#1}} (19{#2}) #3}
\def\prep#1#2#3{Phys.~Rep.~{\bf {#1}C} (19{#2}) #3}
\def\ptp#1#2#3{Prog.~Theor.~Phys.~{\bf {#1}} (19{#2}) #3}
\def\ijmp#1#2#3{Int.~Jour.~Mod.~Phys.~{\bf {#1}} (19{#2}) #3}
\def\nc#1#2#3{Nuovo Cim.~{\bf {#1}} (19{#2}) #3}
\def\ibid#1#2#3{{\it ibid.}~{\bf {#1}} (19{#2}) #3}

\def\szet{{${\scriptstyle \b}$}}
\def\ula{{\un a}}
\def\ulb{{\un b}}
\def\ulc{{\un c}}
\def\uld{{\un d}}
\def\ulA{{\un A}}
\def\ulM{{\underline M}}
\def\cdm{{\Sc D}_{--}}
\def\cdp{{\Sc D}_{++}}
\def\vTheta{\check\Theta}
\def\Pisl{{\Pi\!\!\!\! /}}

\def\fracmm#1#2{{{#1}\over{#2}}}
\def\gg{{\hbox{\sc g}}}
\def\half{{\fracm12}}
\def\ha{\half}

\def\fracm#1#2{\hbox{\large{${\frac{{#1}}{{#2}}}$}}}
\def\Dot#1{{\hskip 0.3pt}{\raise 7pt\hbox{\large .}}
\!\!\!\!{\hskip 0.3pt}{#1}}

\def\Dot#1{\buildrel{_{_\bullet}}\over{#1}}
\def\dt#1{\Dot{#1}}
\def\Hat#1{\widehat{#1}}

\oddsidemargin=0.03in
\evensidemargin=0.01in
\hsize=6.5in
\textwidth=6.5in

\newpage

\noindent{\bf 1.~~Introduction}
\medskip\medskip

The relevance of the {\it self-dual Yang-Mills} (SDYM) theory in four
dimensions [1] appears to be related with the fascinating conjecture
(presumably, first raised by Atiyah and Ward [2]) that this
theory is likely to be considered as the generating theory for all integrable
or exactly soluble models in two and three dimensions, after some suitable
compactifications or dimensional reductions.  It has recently been extended to
make a connection with string theory {\it via} the observation made by
Ooguri and
Vafa on the basis of string amplitude calculations [3] that the consistent
backgrounds for ~$N=2~$ string propagation correspond to {\it self-dual}
gravity configurations in the case of {\it closed} ~$N=2~$ strings, SDYM
configurations in the case of open strings, and
SDYM configurations coupled to gravity
in the case of ~$N=2~$ {\it heterotic}
strings, in four or lower dimensions.\footnotew{This has been also
reconfirmed by ~$\beta$-function calculation in the last paper in
Ref.~[3].}~~It has been known for a long time that
the $~N=2~$ strings, i.e. strings with ~$N=2~$ extended {\it local}
superconformal
symmetry on a world-sheet, live in 2 space-time dimensions [4]. Later on, it
was emphasized [5] that these are in fact 2 {\it complex} dimensions, and the
correct signature in {\it real} coordinates is ~$(2,2)$. We call
the four-dimensional space-time with this
signature the {\it Atiyah-Ward} (AW) space-time in this
paper.\footnotew{Sometimes we use also the term ~$(2,2)~$ or
{}~$2+2~$ space-time with the same
meaning as $~D=(2,2)$.  The expression $~D=4~$ is also used, when the
signature is not very important.}~~As will be shown below, the
signature $~(+,+,-,-)~$ is crucial for introducing minimal space-time self-dual
supersymmetry. The {\it self-duality} (SD) condition on real gauge fields
makes sense in a space-time with the ~$(2,2)$~ or ~$(4,0)~$ signature, but
not with the ~$(3,1)~$ or ~$(1,3)~$ one.  The previous trials [6] based on
the Minkowski or Euclidean signature are {\it not} appropriate for our
purposes.

The aim of this paper is to extend the SDYM model
and the {\it self-dual gravity} (SDG) to {\it supersymmetric} SDYM and {\it
self-dual supergravity} (SDSG) models in the AW space-time. To formulate these
theories, we need to supersymmetrize the SD condition on the YM and/or gravity
field strengths, and the most appropriate way to do this is to use superspace.
In this paper we formulate both SDYM and SDSG models with simple ~$(N=1)~$ and
extended ($N=2~$ and ~$N=4$) supersymmetry by using superfield formulations of
{\it supersymmetric Yang-Mills} (SYM) theories and {\it supergravities} (SG)
in appropriate superspaces, revised to ~$2 + 2~$ space-time dimensions.
Our approach is based on reviewing the superspace constraints and the Bianchi
identities for those theories in a search for a superfield SD condition. Since
 the SD condition puts the theory on-shell, we can avoid any complications
related to the auxiliary field structure of supersymmetric YM and SG theories
and restrict ourselves to their {\it on-shell} superspace formulations.
Proceeding this, we also construct the ~$N=1~$ supersymmetric {\it self-dual}
scalar and
tensor multiplets, and notice a ``no-go'' barrier preventing an existence of
the self-dual ~$N=2~$ hypermultiplet and the ~$N=4~$ SDYM.

This paper is organized as follows. In sect.~2 we start with the analysis of
basic {\it spinors} and (extended) supersymmetry algebras existing in the
AW space-time.  Sect.~3 is devoted to formulations of the ~$N=1~$
supersymmetric
SDYM model and the simple SDSG. Here various ~$N=1~$ supersymmetric self-dual
multiplets are introduced. In sect.~4 we give formulations
 of the ~$N=2~$ supersymmetric SDYM and ~$N=2~$ SDSG theories in the ~$N=2~$
extended superspace. The construction of the ~$N=4~$ SDSG  in the extended
{}~$N=4~$
superspace is presented in sect.~5. The ``no-go'' barrier is discussed in
sect.~6.  In sect.~7, we give the way to bypass the ``no-go'' barrier,
introducing a propagating multiplier multiplet.  In sect.~8,
we give the instanton
solutions for the SDYM models we have presented.
We give our concluding remarks in sect.~9. Appendix A comprises our
notation and conventions. Details about the Dirac gamma matrices in
$~D=(2,2)$~ are collected in Appendix B.
We briefly discuss the
finiteness of the supersymmetric SDYM theories, SDSG and $~N=2~$
superstrings in Appendix C.
The alternative construction
of SYM and SG theories in the AW space-time {\it via} dimensional reduction
from higher dimensions is outlined in Appendix D.
\vglue.2in

\noindent{\bf 2.~Spinors and Supersymmetry in  ~$2 + 2~$ Dimensions}
\medskip\medskip

To see what spinors can be introduced in $~D=(2,2)$, it is a good idea
to start from the Dirac equation in the AW space-time,\footnotew{We
essentially follow here the construction in Ref.~[7].}
$$\left( i\G^{\ula} \pa_{\ula} + m \right) \J = 0~,
\eqno(2.1) $$
where the $~4\times 4$~ Dirac gamma matrices  ~$\G^{\ula} ~$ satisfy the
Clifford algebra
$$ \left\{ \G^{\ula},\G^{\ulb}\right\} = 2\h\low{(R)}^{\ula\ulb}
\equiv 2\, {\rm diag}~(+,+,-,-)~.
\eqno(2.2)$$
See the Appendix A for our notation and the Appendix B for details about
the Dirac
gamma matrices. One of the key issues in this matter is the {\it charge
conjugation matrix} we are going to discuss in this section (see also
Ref.~[8]
as for a general situation in  ~$s + t ~$ dimensions).

First, one notices that the  ~$\G\-$matrices can be chosen in a way such that
$$(\G^1)^{\dg} = +\G^1,~(\G^2)^{\dg} = +\G^2,~
(\G^3)^{\dg} =-\G^3, ~(\G^4)^{\dg} =-\G^4~.
\eqno(2.3) $$
In particular, the two explicit representations given in the Appendix B do
satisfy eq.~(2.3). It follows that
$$ (\G^{\ula})^{\dg} = -\G_0\G^{\ula}\G_0^{-1}~,
\eqno(2.4) $$
where the ~$(2,2)~$ analogue of the ~$(3 + 1)\-$dimensional
{}~$\G_0\-$matrix has been defined as
$$ \G_0\equiv \G^1\G^2,~\G_0^{-1}=\G_0^{\dg}=-\G_0,~\G^2_0=-1~.
\eqno(2.5)$$

To introduce the charge conjugation matrix ~$C~$ in the usual way, one notices
that both ~$\pm(\G^{\ula})^*~$ and ~$\pm(\G^{\ula})^T~$ separately
form equivalent
representations of the Clifford algebra (2.2). Therefore, there exist
invertible matrices ~$B~$ and ~$C~$ for which
$$ (\G^{\ula})^* = \h B\G^{\ula} B^{-1}~,~~(\G^{\ula})^T = -\h C \G^{\ula}
C^{-1}~,
\eqno(2.6) $$
where the sign ~$\h,~\h=\pm 1$, has been introduced. The correlation of the
overall signs in these two equations appears to be needed for their
consistency with eq.~(2.4). It also yields
$$ C = B^T \G_0~,
\eqno(2.7) $$
modulo a sign factor which is irrelevant here.
It is not difficult to show that the matrix ~$B~$ is unitary and satisfies
[8]
$$ B^* B = 1~.
\eqno(2.8) $$
The charge conjugation matrix ~$C~$ has the properties
$$ C^T = -C~,~~C^{\dg}C=1~,
\eqno(2.9) $$
which follow just from the definitions above. It is optional to take
{}~$C^* =-C$.

In the Majorana representation (B.8) of the ~$\G\-$matrices, the matrix
 ~$B~$ can be chosen to be ~$1~$ while ~$\h=-1$. In that representation the
``Lorentz'' generators ~$\S^{\ula\ulb}=(1/2)\left( \G^\ula\G^\ulb
- \G^\ulb\G^\ula \right)~$
are real. Dividing them into chiral pieces ~$\S_{\pm}^{\ula\ulb}=(1/2)\left( 1
\pm \G_5\right)\S^{\ula\ulb}~$ yields two commuting ~$sl(2,{\bf R})~$ algebras
generated by ~$\S_+^{\ula\ulb}~$ and ~$\S_-^{\ula\ulb}~$ respectively. This
nicely illustrates the well-known isomorphism [9]
$$\Bar{SO(2,2)}\cong SL(2,{\bf R})\otimes SL(2,{\bf R})~,
\eqno(2.10)$$
where ~$\Bar{SO(2,2)}~$ is the covering group of ~$SO(2,2)$.

In the explicit representation (B.10) we have
$$(\G^1)^*=-\G^1, ~(\G^2)^* = +\G^2,~(\G^3)^* = + \G^3, ~(\G^4)^*
=-\G^4~~.
\eqno(2.11) $$
It follows that
$$ {\bf {\rm For}~\h=-1:}~~~~B=\G^3\G^2,~C=\G^1\G^3
=\left(\begin{array}{cc} +\t_2 & 0\\ 0 & +\t_2\end{array}
\right)~~,
\eqno(2.12) $$
and
$$ {\bf {\rm For}~\h=+1:}~~~~B=\G^1\G^4,~C=\G^2\G^4
=\left(\begin{array}{cc} +\t_2 & 0 \\ 0
& -\t_2 \end{array}\right)~~.
\eqno(2.13) $$
In particular, ~$C^* =-C~$ and ~$\st^{\ula}=\t_2\left( \s^{\ula}\right)^T
\t_2$~ (at ~$\h=-1$) in this representation. The ~$\st^{\ula}~$ and
{}~$\s^{\ula}$ are in fact related by the charge conjugation matrix in any
representation of the gamma matrices.

We are now in a position to discuss the simplest spinor representations of the
lowest dimension in the AW space-time. The Dirac equation (2.1) defines
the {\it Dirac} spinor ~$\J~$ whose transformation properties follow from
requiring the covariance of the Dirac equation. The Dirac spinor has 4 complex
components and this representation is clearly reducible because  of the
existence of chiral projection operators. The chiral projectors ~$P_{\pm}=
(1/2)\left(1\pm\G_5\right)~$ can be used to define the {\it chiral} or
{\it Weyl} spinors, ~$\j_{\pm}=\pm\G_5\j_{\pm}$.\footnotew{See eq.~(B.11) for
the definition of $~\G_5$.}~~In the convenient
representation (B.10), we have
$$ \J_{\ul{\a}} =\left( \begin{array}{c} \j_{\a} \\
{\Tilde\j}_{\dt{\a}}
\end{array}\right)~,
\eqno(2.14) $$
where each Weyl spinor, ~$\j~$ or ~$\Tilde{\j}$, has 2 complex components,
$~{\scst \a~=~1,~2;~~\dt{\a}~=~\dt{1},~\dt{2}~;~~{\ul{\a}}~=~(\a,\dt{\a})~}$.
As in $~D=(1,3)$, the antisymmetric
tensors, ~$C^{\a\b},~C_{\a\b}~$ and ~$C^{\dt{\a}\dt{\b}},~C_{\dt{\a}\dt{\b}}$,
defined by chiral pieces of the charge conjugation
matrix in eq.~(2.13), can be used to raise and lower the chiral spinor indices
{}~${\scst \a~}$ and ~${\scst\dt\a}$. It is the specialty of the AW space-time
that the projections ~$P_{\pm}\G^{\ula}~$ {\it separately} transform under the
{}~$B~$ conjugation of eq.~(2.6).

The complex conjugation of the Dirac equation (2.1) yields
$$\left[ i\G^{\ula} \pa_{\ula} +(-\h)m\right]\left( B^{-1}\J^* \right)=0~.
\eqno(2.15) $$
Given ~$(-\h)m= m$, this is the Dirac equation for ~$\J^c\equiv B^{-1}\J^*$.
Since ~$B^* B=1$~, it is consistent to equate
$$ \J^* =B\J ~~{\rm or}~~\J^c = \J~.
\eqno(2.16) $$
The ~$\J^c~$ is known as the {\it Majorana-conjugated} spinor, while eq.~(2.16)
is known as the {\it Majorana} condition. In the massive case, this is only
possible if ~$\h=-1$. In the massless case there is another option, ~$\h=1$,
which allows to define the {\it pseudo-Majorana} spinors as those satisfying
eq.~(2.16). Introducing the ~$(2,2)~$ analogue of the Dirac conjugation as
$$ \Bar{\J} = \J^{\dagger} \G_0 ~~,
\eqno(2.17) $$
we can rewrite the Majorana condition (2.16) to the form which is familiar from
$~D=(1,3)$:
$$\J = \J^c = C\Bar{\J}^T~,
\eqno(2.18) $$
provided the representation of ~$\G\-$matrices, in which ~$C^* =-C$, is used.
In the representation (B.10) the latter takes place, while the Majorana
condition for the spinor (2.14) yields ~$\t_1\j=\j^*~$ and ~$\t_1\Tilde{\j}=
\Tilde{\j}^*$. Clearly, this distinguishes the ~$(2,2)~$ case from its
$(1,3)$
counterpart where we had ~$\j^* =\Tilde{\j}~$ instead.

It is now obvious that we can introduce {\it Majorana-Weyl} (MW) spinors in
$~D=(2,2)$~ without adding additional isospin indices. The chiral parts
 of the Majorana spinor just represent the MW spinors. In other words, the two
constraints provided by the chirality and Majorana conditions can be
simultaneously and consistently imposed on a spinor in the AW space-time!
In the Majorana representation (B.8) with pure imaginary ~$\G^{\ula}$, there
exist {\it real} (Majorana) spinors, which satisfy the real Dirac equation.
Since in the Majorana representation the ~$\G_5~$ is off-diagonal but
{\it real}, the chirality condition makes perfect sense and
leads to a 2-component {\it
real chiral} spinor which is nothing but the MW spinor. In the representation
(B.10), a four-component MW spinor ~$\J~$ takes the form
$$\J_{\ul{\a}} = \left(\begin{array}{c} \j_{\a} \\ 0\end{array}\right),~~~
\j_{\a}=\left(\begin{array}{c} \j \\ \j^*\end{array}\right)~,
\eqno(2.19)$$
where the ~$\j~$ has just one complex component. Finally, from the viewpoint of
the isomorphism (2.10), the MW spinor just realizes the fundamental
(two-dimensional and real) representation of one of the ~$SL\left(2,{\bf R}
\right)~$ factors on the r.h.s. of eq.~(2.10).

It is worthy to mention here that the {\it little} group\footnotew{We
define the little group in the AW space-time as the subgroup of the
``Poincar\' e'' group ~$\Pi$, $~\Pi\equiv SO(2,2)\, \subset\!\!\!\!\!\!\times
\,T^{2,2}$, which leaves invariant a non-zero ``momentum'' vector.
Here the symbol ~$\subset\!\!\!\!\!\!\times~$ means a semi-direct product,
whereas the ~$T^{2,2}~$ denotes the
translation group.} for the AW space-time is just ~$GL(1)$, but {\it not}
{}~$U(1)$.  The ~$GL(1)~$ is an abelian group whose
irreducible representations (irreps) are one-dimensional. To have a non-trivial
spin under the little group, the ~$\Pi\-$irrep should be {\it
massive}. The {\it massless} ~$\Pi\-$irreps consist only of scalars. The
meaning of the massless scalars in reference to the massless states with
non-trivial spins had already been explained by Ooguri and Vafa [3]. Those
scalars appear to be the potentials for the {\it self-dual} fields which may
have non-trivial spin but only one physical degree of freedom [3].

The {\it supersymmetry} can be introduced in the usual way [10] as the
``square root'' of the AW space-time. The difference, however, is in the
existence of {\it two} different square roots in the AW space-time, related
to either SD or {\it anti-self-duality} (ASD). As explained in the Appendix B,
 the SD is associated with
the choice {\bf I} of the ~$\g\-$matrices while the ASD picks up another choice
{\bf II}. Therefore, one can introduce two sets of the (extended) supersymmetry
charges ~$Q~$ and ~$\check{Q}~$ which are linked to the choice {\bf I} and
{\bf II} respectively, and they are completely independent on each other.
The basic
{}~$N\-$extended supersymmetry algebra is described by
$$\eqalign{&\left\{ Q^i_{\a},Q^j_{\b}\right\}=\S_{\a\b}Z^{(ij)}+C_{\a\b}
Z^{\[ij\]}~, \cr
&\left\{ \Tilde{Q}_{\dt{\a}i},\Tilde{Q}_{\dt{\b}j}\right\} = 0~, \cr
&\left\{ Q^i_{\a},\Tilde{Q}_{\dt{\b}j}\right\} =i\d^i{}_j\left(
\s^{\ula}_{{\bf I}}\right)_{\a\dt{\b}}\pa_{\ula}~; \cr
&\left\{ \check{Q}_{\a}^{\bar{i}},\check{Q}_{\b}^{\bar{j}}\right\}
=0~,\cr
&\left\{ \check{\Tilde{Q}}_{\dt{\a}\bar{i}},\check{\Tilde{Q}}_{\dt{\b}\bar{j}}
\right\} = \check{\S}_{\dt{\a}\dt{\b}}\check{Z}_{(\bar{i}\bar{j})}+
C_{\dt{\a}\dt{\b}}\check{Z}_{\[\bar{i}\bar{j}\]}~, \cr
&\left\{ \check{Q}_{\a}^{\bar{i}},\check{\Tilde{Q}}_{\dt{\b}\bar{j}}\right\}
=i\d^{\bar{i}}{}_{\bar{j}}\left( \s^{\ula}_{{\bf II}}\right)_{\a\dt{\b}}
\pa_{\ula}~, \cr }
\eqno(2.20)$$
where $~{\scst i,~j~=~1,~2,~\ldots,~p;~\,\bar{i},~\bar{j}~=~1,~2,~\ldots
,~q}~,~N=p + q$. The appearance of the two independent
sets of supersymmetry generators resembles
the two-dimensional case, where the ~$(p,q)~$ supersymmetry algebras comprise
$~p~$ chiral and ~$q~$ anti-chiral Majorana spinor charges simultaneously
[11].
In eq.~(2.20) we have explicitly introduced the {\it central charges}
{}~$Z^{(i j)},~Z^{\[i j\]},~\check{Z}_{(\bar{i}\bar{j})}~$ and
{}~$\check{Z}_{\[\bar{i} \bar{j}\]}~$ for
more generality; the ~$\check{\S}~$ represents the {\bf II}-analogue of the
symmetric matrix ~$\S~$ (see the Appendix B). In this paper we will restrict
ourselves to the case ~$q=0~$ with {\it all} the central charges vanishing.
However, in some cases, the central charges may really be important. In
particular, it happens in the three-dimensional supersymmetric Chern-Simons
theory where the central charges are active [12].

\bigskip\bigskip

\noindent{\bf 3.~$N=1$ Supersymmetry and Self-Duality}
\medskip\medskip

In this section we consider $~N=1~$ supersymmetric {\it scalar}, {\it tensor},
{\it vector} and {\it supergravity} multiplets, and formulate {\it
supersymmetric} SD conditions for them in the AW space-time. The ~$N=1~$
supersymmetric SDYM theory, which is apparently relevant to integrable models
and comprises  all of the remarkable features mentioned above about
spinors in $~D=(2,2)$, will be in the heart of our discussion.
\vglue.2in

\noindent{\bf 3.1.~Self-Dual Scalar Multiplet}

As an immediate corollary of the existence of MW spinors in $~D=(2,2)$,
we notice the existence of a {\it real scalar} ~$N=1$~ multiplet (SM) in the
AW space-time. The real scalar (or chiral) ~$N=1~$ multiplet consists of a
{\it real} scalar ~$A$, a {\it Majorana-Weyl} spinor ~$\Tilde{\j}~$ and a {\it
real} auxiliary scalar ~$F~$ (in total, ~$2 + 2~$ components off-shell and
{}~$1 + 1~$ components on-shell). The ~$N=1~$ supersymmetry transformation
laws (with the MW parameters ~$\e_{\a}~$ and ~$\Tilde\e_{\dt{\a}}$~ read:
$$\eqalign{
& \d A =  \Tilde\e^{\,\Dot\a} \Tilde\psi_{\Dot\a} ~,\cr
& \d\Tilde{\j}_{\dt{\a}}=
i(\st^{\ula})_{\dt{\a}}{}^{\b}\pa_{\ula}A\e{\low\b} +
F\Tilde{\e}_{\dt{\a}}~,\cr
& \d F =
-i \e^\a (\s^\ula)\du \a{\Dot\b} \partial_\ula \Tilde \psi _{\Dot\b} ~.}
\eqno(3.1)$$
Hence, this $~D=4$~ multiplet is very much like a $~D=2$~ real
SM!  Given the Majorana representation for the
{}~$\g\-$matrices (see
the Appendix B), all the spinors in eq.~(3.1) would be just real. The real
(chiral) SM arises as the result of imposing the reality
condition on a (chiral) SM, which is consistent in the AW
space-time. An existence of the more fundamental real SM, compared
to its $~D=(1,3)$~ complex scalar counterpart [13], is clearly a
remarkable feature of supersymmetry in $~D=(2,2)$. The real SM
does {\it not} allow an action since there is no (undotted) spinor
partner for the (dotted) MW spinor ~$\Tilde{\j}~$ in order to form a
fermionic kinetic term. Therefore, there seems to be a good reason to call
the {\it real chiral} SM {\it ``self-dual scalar multiplet''} (SDSM).  We will
find further evidence to support such identification, when considering the
{}~$N=2~$ supersymmetric SDYM multiplet from the viewpoint of the ~$N=1~$
multiplets it contains (see subsect.~4.2).

        Another point of interest is the existence of $~D=(2,2)$~
models that mimic the structure of the chiral rings seen in conformal field
theory [14].  To show the existence of such models, we first need to
introduce a second ``SDSM''.  From the discussion
above we have learned that $~(A, \Tilde\psi_{\Dot\a}, F)$~ constitute a
representation consistent with SD and supersymmetry.  Now consider the
multiplet $~(B, \Omega_\a, U)$, where
$$\eqalign{&B \equiv F~~, ~~~~ \Omega_\a \equiv -i (\s^\ula)\du
\a{\Dot\g} (\partial _\ula \Tilde\psi_{\Dot\g} ) ~~, \cr
& U \equiv \Bo A~~, \cr }
\eqno(3.2) $$
and it is easy to see that the transformation laws of these are given by
$$\eqalign{&\d B = \e^\a \Omega_\a ~~, \cr
& \d \Omega_\a = - i (\s^\ula) \du \a {\Dot \b} \Tilde\e_{\Dot\b}
(\partial_\ula B) + U \e_\a ~~, \cr
& \d U = i \Tilde\e^{\,\Dot\a} (\Tilde\s^\ula )\du{\Dot\a} \b
(\partial_\ula\Omega_\b ) ~~. \cr }
\eqno(3.3) $$

        Notice now a feature that appears to be unique to the SM.
Even in the presence of the duality constraint, there still
remain {\it auxiliary} fields.  We can eliminate them by making specific
choices.  For example, we now regard $~B,~\Omega_\a$~ and $~U$~ as
independent functions with the transformation laws given by eq.~(3.3).
Introducing independent ``superpotentials'' $~W'(B)$~ and ~$\Tilde
W'(A)$~ in addition,\footnotew{In our notation, a prime as a superscript
always means a differentiation of a function with respect to a given
argument.}~~we impose the constraints
$$F\equiv W'(B)~~, ~~~~ U\equiv \Tilde W'(A) ~~.
\eqno(3.4) $$
We can easily demonstrate an action for these ideas.  As one might
guess from $~D=(1,3)$~ space, an appropriate lagrangian is given by
$$\eqalign{\Lag = \, & 2 A\Bo B + 2i \Omega^\a (\s^\ula)\du \a {\Dot\b}
\partial_\ula \Tilde\psi_{\Dot\b} +  F U \cr
& - F \Tilde W'(A) + \half \Tilde\psi^{\Dot\a} \Tilde \psi_{\Dot\a}
\Tilde W''(A) \cr
& - U  W'(B) + \half \Omega^\a \Omega_\a W '' (B)~~,  \cr }
\eqno(3.5) $$
which clearly follows from the superfield expression
$$ \Lag = \int d^2\theta d^2\Tilde\theta \, \Phi \Psi - \int d^2\Tilde\theta
\, \Tilde W(\Phi) - \int d^2 \theta \, W (\Psi) ~~.
\eqno(3.6) $$
In this expression $~\Phi$~ is a real anti-chiral superfield $~(D_\a
\Phi = 0 )$, while $~\Psi$~ is an independent real chiral superfield $~(
\Tilde D_{\Dot \a} \Psi = 0 )$.  The $~(A,\Tilde \psi_{\Dot\a}, F)$~
components are contained in $~\Phi$~ and the $~(B,\Omega_\a, U)$~
components are contained in $~\Psi$.  From the form of eq.~(3.6) we can
clearly see that as a final generalization a K{\" a}hler potential can
be used
$$\int d^2 \theta d^2\Tilde\theta \, \Phi\Psi ~~ \rightarrow ~~
\int d^2\theta d^2\Tilde\theta \, K(\Phi, \Psi) ~~.
\eqno(3.7) $$

\vglue.2in

\noindent{\bf 3.2~~$N=1$~ Supersymmetric Self-Dual Yang-Mills Theory}

We begin our construction of the ~$N=1~$ supersymmetric SDYM theory with
component considerations, and then transfer the results to superspace.
The superspace approach will be particularly useful when
considering the extended supersymmetry and supergravity in the next sections.

The ~$N=1~$ SYM theory in four space-time dimensions is described off-shell in
terms of the ~$N=1~$ non-Abelian real vector multiplet which
contains (in the Wess-Zumino gauge) a gauge vector
field ~$A_{\ula}\equiv A_{\ula}{}^I t^I$, one Majorana spinor ~$\L~$
and an auxiliary
scalar ~$D$, all in the adjoint representation of the gauge group [15]. The
(anti-hermitian) generating matrices ~$t^I~$ of the gauge group act on the
implicit gauge group indices of the above-mentioned fields. We
omit the gauge indices, unless their absence may cause confusion.

The supersymmetry transformation laws of the ~$N=1~$ SYM multiplet take the
usual form in the four-component notation for spinors:
$$\eqalign{
&\d A_{\ula} = -i \Bar{\e}\G_{\ula}\L ~,\cr
&\d \L =  \S^{\ula\ulb}F_{\ula\ulb}\e + D\e ~,\cr
&\d D = -{\fracm 12}i\bar{\e}\G_5\G^{\ula}\nabla_{\ula}\L ~,}
\eqno(3.8)$$
where the gauge-covariant YM derivatives ~$\nabla_{\ula}=\pa_{\ula} +
A_{\ula}~$ and the non-abelian YM field strength
{}~$F_{\ula\ulb}~$ have been introduced.

The celebrated SD condition [1] on the YM field strength ~$F~$
in four space-time dimensions with the ~$(2,2)~$ signature is given by
$$ F_{\ula\ulb} = \fracm 1 2\e_{\ula\ulb}^{~~\ulc\uld}F_{\ulc\uld}
\eqno(3.9)$$
in the {\it real} notation, where the totally antisymmetric Levi-Civita
symbol ~$\e^{\ula\ulb\ulc\uld}~$ with unit weight has been introduced.
Given the {\it complex} notation (see the Appendix A for our
notation and conventions), eq.~(3.9) is equivalent to the {\it two}
conditions [16,17]
$$ F_{a b} = F_{\ab\bb}=0~,
\eqno(3.10a) $$
$$ \h^{a\bb}F_{a\bb}=0~.
\eqno(3.10b) $$
The first SD condition in eq.~(3.10a) can be
interpreted as an {\it integrability condition} for the existence of the {\it
holomorphic} and {\it anti-holomorphic} spinors satisfying the equations
{}~$\nabla_a \X = 0~$ and ~$\nabla_{\ab}\O  = 0$,
respectively. The YM fields satisfying eq.~(3.10a) can be referred to as the
``{\it hermitian}'' gauge configurations. Eq.~(3.10a) can easily be solved in
terms of the prepotential {\it scalar} fields ~$J~$ and ~$\Bar{J}~$ as [18]
$$ A_a =J^{-1}\pa_{a}J~,~~A_{\ab}=\Bar{J}^{-1}\pa_{\ab}\Bar{J}~.
\eqno(3.11)$$

The second SD condition (3.10b) puts the YM theory {\it on-shell} since it
implies the YM field equations of motion to be satisfied. Therefore, when
discussing SD, it is always enough to consider an on-shell theory.
As for the ~$N=1~$ SYM theory, its equations of motion are given by
$$\li{ &\nabla^{\ulb}F_{\ula\ulb}{}^I=-if^{IJK}\h_{\ula\ulc}\left[
\l_{\b}{}^J C^{\b\a}
(\s^{\ulc})_{\a}{}^{\dt{\a}}\Tilde{\l}_{\dt{\a}}{}^K\right]~,
&(3.12a) \cr
& i(\s^{\ula})_{\a}{}^{\dt{\a}}\nabla_{\ula}\Tilde{\l}_{\dt{\a}}=i(\st^{\ula}
)_{\dt{\a}}{}^{\a}\nabla_{\ula}\l_{\a}=0~,
&(3.12b) \cr
&D = 0~,
&(3.12c) \cr } $$
where the gauge group structure constants ~$f^{I J K}~$ have been introduced.

In the complex notation, the second line of eq.~(3.8) can be rewritten on-shell
as
$$\d\L=\left( 2i\g_3\e^{ab}F_{ab} +
2i\g_{\3b}\e^{\ab\bb}F_{\ab\bb}+\fracm 12
\S\h^{a\bb}F_{a\bb} + \Hat{\S}^{a\bb}F_{a\bb} + D \right)\e ~.
\eqno(3.13) $$
Now we can separate the vanishing SDYM field strengths from the
non-vanishing ones simply by taking the chiral projections of eq.~(3.13),
$$\li{ &\d\l_{\a}= \left( 2i\g_3\e^{ab}F_{ab} +
2i\g_{\3b}\e^{\ab\bb}F_{\ab\bb}+
\fracm 12\S\h^{a\bb}F_{a\bb} + D \right)\du\a\b\e_\b ~~,
&(3.14a) \cr
& \d\Tilde{\l}_{\dt{\a}}=F_{a\bb}(\Hat{\S}^{a\bb})_{\dt{\a}}
{}^{\dt{\b}}\Tilde\e_{\dt{\b}} ~~,
&(3.14b) \cr} $$
because of eqs.~(B.27) and (B.28). This observation immediately gives rise
to the ~$N=1~$ supersymmetric SDYM constraint to the  Majorana spinor
superpartner ~$\L~$ of the
YM field  ~$F~$ in the simple form of the {\it Majorana-Weyl} condition:
$${\fracm 12}\left( 1 + \g_5\right)\L_{\rm SD} =0~,
\eqno(3.15a)$$
while we still have
$$ D = 0 ~.
\eqno(3.15b)$$
Therefore, the ~$N=1$~ supersymmetric SDYM multiplet consists of the SDYM
field ~$F_{\dt{\a}\dt{\b}}~$ and the MW spinor ~$\Tilde{\l}_{\dt{\a}}~$
($1 + 1~$ components on-shell per one gauge index value).

In order to check the component super-SD conditions (3.10) and (3.15), first
we notice the consistency of the SD conditions (3.10) on the YM fields with the
SYM equations of motion (3.12), just because the gaugino source term on the
r.h.s. of the YM field equations is now {\it vanishing}, subject to the
constraint (3.15a). Second, we can also confirm that the gaugino field
equations
(3.12b) come out of the spinorial derivatives ~$\nabla_{\ul{\a}}F_{ab}{}^I
=0~$ and
{}~$\nabla_{\ul{\a}}\left(\h^{a\bb}F_{a\bb}{}^I\right)=0\,$.

Given the second choice $({\bf II})$ in eq.~(B.14) to represent the
{}~$\g\-$matrices, the similar analysis would give rise to the
{\it anti-self-dual} $~N=1~$ SYM theory, characterized by the minus sign
in front of the
r.h.s. of eq.~(3.9) and the constraint:
$${\fracm 12}\left( 1 - \g_5\right)\L_{\rm ASD}=0~.
\eqno(3.16)$$
We thus find the very interesting observation that the two different
ways of defining a spin-structure in the AW space-time are co-related with the
definition of self-dual or anti-self-dual supersymmetry.

In the ~$N=1~$ {\it superspace} associated with the AW space-time, the
superfield formulation of the ({\it non}-self-dual) ~$N=1~$ SYM theory  can be
developed along the lines of the conventional ~$(3 + 1)\-$dimensional case
[19,20]. The SYM superfield potentials ~$\ca_{A}~$ are used to define the
gauge-covariant superspace derivatives ~$\nabla_A\equiv D_A + \ca_A$, where
$~{\scst A=
(\ula,\ul{\a})},~D_A=(\pa_{\ula},D_{\a},\Tilde{D}_{\dt{\a}})$. These
gauge-covariant derivatives define superfield strengths by
$$\left[ \nabla_A,\nabla_B\right\}= F_{A B} + T_{A B}{}^{C}\nabla_C~~,
\eqno(3.17)$$
where the torsion ~$T_{A B}{}^C~$ is the same one as in flat superspace,
and some of the field strengths vanish [19,20]:
$$F_{\a\b}=F_{\dt{\a}\dt{\b}}=F_{\a\dt{\b}}=0~.
\eqno(3.18)$$
Eq.~(3.18) represents the whole set of the ~$N=1~$ SYM superspace
constraints, whose only role is to remove as much independent components as
possible for matching with the component SYM formulation, without getting a
flat theory [20]. Given the constraints on some of the field strengths
{}~$F_{A B}$, the superspace {\it Bianchi identities} (BIds)
$$\nabla_{\[ A}F_{BC)} - T_{\[ A B|}{}^D F_{D|C)} =0~
\eqno(3.19)$$
are used to determine the implications of the constraints to all the other
SYM field strengths. We refer the reader to Refs.~[19,21] for the details
related to the analysis of the SYM Bianchi ``identities'' subject to the
conventional SYM constraints for all ~$N\leq 4$. That analysis is still valid
in $~D=(2,2)$, before imposing any reality conditions. The outcome of
such analysis is usually represented by some kind of the BIds ``solution''
that expresses all of the SYM field strengths in terms of a smaller number of
the independent superfields, still satisfying certain superspace constraints.
In particular, the off-shell ~$N=1~$ SYM theory is well-known to be described
by
a single Majorana spinor superfield ~$W_\a~$ which comprises all of the SYM
field components  and has the ~$\L~$ as the leading one [20]. The MW chiral
constituents ~$w_{\a}~$ and ~$\Tilde{w}_{\dt{\a}}~$ of the ~$W_{\ul{\a}}~$
are chiral and anti-chiral in the gauge-covariant superspace sense,
respectively
$$\Tilde{\nabla}^{\dt{\a}}w_{\b}=0~,~~\nabla^{\a}\Tilde{w}_{\dt{\b}}=0~,
\eqno(3.20)$$
and, in addition, satisfy the constraint [20,21]
$$\nabla^{\a}w_{\a}+\Tilde{\nabla}^{\dt{\a}}\Tilde{w}_{\dt{\a}}=0~.
\eqno(3.21)$$
The non-vanishing SYM field strengths take the form [20,21]
$$\li{&F^{\ula\a}= -i(\s^{\ula})^{\a\dt{\b}}\Tilde{w}_{\dt{\b}}~,~~
F^{\ula\dt{\a}}= -i(\st^{\ula})^{\dt{\a}\b}w_{\b}~,
&(3.22a) \cr
&F_{\ula\ulb}=\fracm 12i\left[ \, \nabla^{\a}(\s_{\ula\ulb})_{\a}{}^{\b}w_{\b}
+ \Tilde{\nabla}^{\dt{\a}}(\st_{\ula\ulb})_{\dt{\a}}{}^{\dt{\b}}
\Tilde{w}_{\dt{\b}}\,\right]~.
&(3.22b) \cr} $$

We are now in a position to find out the SYM superfield SD condition easily. We
simply notice that since the ~$w_{\a}~$ superfield has ~$\l_{\a}~$ as its
$~\theta=0\-$component, which is vanishing in the SD case, the whole superfield
{}~$w_{\a}~$ should vanish, too.  This gauge-chiral superfield ~$w_{\a}~$ just
has the ~$\l_{\a}$, ~$D$~ and the {\it anti-self-dual} part ~$F_{\a\b}~$
of the YM field
 strength as its independent components. The constraint (3.21), whose meaning
was just the reality of the ~$D\-$component, is now trivially satisfied because
of eq.~(3.15b), while the self-dual part ~$F_{\dt{\a}\dt{\b}}~$ of the YM field
strength remains in the decomposition (3.22b). The superspace equations of
motion take formally the same form as in eq.~(3.12b) after the substitution
{}~$\L\rightarrow W$, and they are obviously consistent with the SD condition.
We
have repeated the analysis of the BIds (3.19) subject to the conventional SYM
constraints (3.18) and the superspace SD condition in the most straightforward
 form ~$F_{\a\b}=0$, and reached the conclusion that all those
constraints taken together, in order to be consistent, do imply that the
Weyl part ~$w~$ of the Majorana spinor ~$W~$ should be simultaneously
holomorphic
and anti-holomorphic, ~$\nabla_a w=\nabla_{\ab}w=0$, which means ~$w=0$.
In summary,
we have found that the MW condition on the SYM spinor superfield strength ~$W$,
$${\fracm 12}\left( 1 + \g_5\right)W_{\rm SD}=0~,
\eqno(3.23)$$
is the true ~$N=1~$ SYM SD constraint in superspace. It is usually the case in
superspace that the origin of constraints on bosonic fields can be tracked
back to constraints on some fermionic fields in lower-dimensional sector. We
see that it indeed happens in our case when the first-order differential SD
constraint on the YM field is replaced by the {\it algebraic} constraint on
the spinor SYM superfield. The existence of the MW spinors was crucial in
this respect.

\bigskip\bigskip

\noindent{\bf 3.3 ~$~N=1$~ Self-Dual Supergravity}

The ~$N=1~$ (simple) SDSG  follows the pattern provided by the ~$N=1~$
supersymmetric SDYM theory above.

In the curved ~$N=1~$ superspace associated with the AW space-time, the
covariant derivatives ~$\nabla_A = E_A + \O_A~$ are defined in terms of the
(super)vielbein ~$E_A=E_A{}^N D_N~$ and the (super)connection ~$\O_A=
(\O_A)_{\a}{}^{\b}{\cal M}_{\b}{}^{\a} + (\O_A)_{\dt{\a}}{}^{\dt{\b}}
\Tilde{\cal M}_{\dt{\b}}{}^{\dt{\a}}\,$, where the ~$M$'s are ``Lorentz''
rotation
generators. From the covariant derivatives one defines torsions and curvatures,
$$\[\nabla_A,\nabla_B\left.\right\}= T_{A B}{}^C\nabla_C + R_{A B}
\equiv T\du{A B} C \nabla_C +\half R\du{A B c} d {\cal M}\du d c~~,
\eqno(3.24)$$
which satisfy the well-known ~$N=1~$ SG constraints [20]. By setting the
auxiliary superfields equal to zero, the solution of the SG BIds subject to
the SG constraints [20] assumes the form
$$\eqalign{& \left\{ \nabla_{\a},\nabla_{\b}\right\} =
\left\{\Tilde{\nabla}_{\dt{\a}},
\Tilde{\nabla}_{\dt{\b}}\right\}=0~,~~\left\{ \nabla_{\a},\Tilde{\nabla}_{
\dt{\b}}\right\}=i\nabla_{\a\dt{\b}}~, \cr
&\left[\Tilde{\nabla}_{\dt{\a}},i\nabla_{\b\dt{\b}}\right]
=C_{\dt{\b}\dt{\a}}W_{\b\g} {}^{\d}{\cal M}_{\d}{}^{\g}
{}~,~~\left[\nabla_{\a},i\nabla_{\b\dt{\b}}\right]=C_{\b\a}
\Tilde{W}_{\dt{\b}\dt{\g}}{}^{\dt{\d}}\Tilde{{\cal M}}_{\dt{\d}}
{}^{\dt{\g}}~, \cr
& \left[i\nabla_{\a\dt{\a}},i\nabla_{\b\dt{\b}}\right]=C_{\dt{\a}\dt{\b}}\left[
\fracm 1{4!}\nabla_{(\a}W_{\b\g\d)}{\cal M}^{\g\d}-W_{\a\b}{}^{\g}\nabla_{\g}
\right] + C_{\b\a}\left[ \fracm 1 {4!}\Tilde{\nabla}_{(\dt{\a}}
\Tilde{W}_{\dt{\b}\dt{\g} \dt{\d})}\Tilde{{\cal M}}^{\dt{\g}\dt{\d}}-
\Tilde{W}_{\dt{\a}\dt{\b}}{}^{\dt{\g}} \Tilde{\nabla}_{\dt{\g}}\right]~, \cr }
\eqno(3.25)$$
where all of the torsion and curvature tensors have been expressed in terms of
only two irreducible conformal superfield strengths ~$W_{\a\b\g}~$ and
{}~$\Tilde{W}_{\dt{\a}\dt{\b}\dt{\g}}~$ which are totally symmetric on
their spinor indices. These tensors originate from the identification [20]
$$W_{\a\b\g}=\fracm{1}{12}iR^{\dt{\a}}{}_{(\a\dt{\a},\b\g)}=-\fracm{1}{12}
T_{(\a}{}^{\dt{\a}}{}_{\b\dt{\a},\g)}~,
\eqno(3.26)$$
and similarly for ~$\Tilde{W}_{\dt{\a}\dt{\b}\dt{\g}}\,$, and satisfy the
relations [20]
$$\eqalign{
\Tilde{\nabla}_{\dt{\a}}W_{\a\b\g} & = \nabla^{\a}W_{\a\b\g}  = 0~,\cr
\nabla_{\a}\Tilde{W}_{\dt{\a}\dt{\b}\dt{\g}} & = \Tilde{\nabla}^{\dt{\a}}
\Tilde{W}_{\dt{\a}\dt{\b}\dt{\g}}  = 0~.}
\eqno(3.27)$$
The leading components of the ~$W_{\a\b\g}~$ and ~$\Tilde{W}_{\dt{\a}\dt{\b}
\dt{\g}}~$ just represent the chiral parts of the on-shell {\it
Rarita-Schwinger} field strength, while their symmetrized spinor derivatives
{}~$W_{\a\b\g\d}\equiv (1/4!)\nabla_{(\a}W_{\b\g\d)}~$ and
{}~$\Tilde{W}_{\dt{\a}\dt{\b}\dt{\g}\dt{\d}}\equiv (1/4!)\Tilde{\nabla}_{
(\dt{\a}}\Tilde{W}_{\dt{\b}\dt{\g}\dt{\d})}~$ are the anti-self-dual and
self-dual constituents of the {\it Weyl} curvature, respectively.

The SD condition is imposed on the Riemann curvature tensor, and in the {\it
real} notation it takes the form
$$R_{\ula\ulb}{}^{\ulc\uld}=\fracm{1}{2}\e_{\ula\ulb}{}^{\ulf\ulh}R_{\ulf\ulh}
{}^{\ulc\uld}~.
\eqno(3.28) $$
Given the {\it complex} notation, eq.~(3.28) is rewritten by
$$R_{a b}{}^{\ulc\uld}=R_{\ab\bb}{}^{\ulc\uld}=0~,
\eqno(3.29a)$$
$$\h^{a\bb}R_{a\bb}{}^{\ulc\uld}=0~,
\eqno(3.29b)$$
which is quite similar to its SDYM counterpart in eq.~(3.10). Eq.~(3.29a) is
known as the ``K\"ahlerness'' condition [22] which means that the base manifold
(curved space-time) should be a hermitian manifold with the connection to be
consistent with the complex structure. Eq.~(3.29b) puts the theory on-shell and
it is equivalent to the Ricci-flatness condition. As is well known in
$~D=4$~ [23], a K\"ahlerian and Ricci-flat manifold has a self-dual
curvature tensor and {\it vice versa}.

It now becomes clear that the ~$N=1~$ SDSG constraint in superspace is given by
$$W_{\a\b\g}=0~.
\eqno(3.30)$$
Eq.~(3.30) means that the gravitino field becomes a Majorana-Weyl spinor (only
{}~$\Tilde{W}_{\dt{\a}\dt{\b}\dt{\g}}~$ survives) while the anti-self-dual
constituents of both Weyl and Riemann curvature tensors vanish. As a
consistency check, we notice that both the SG contorsion tensor and the
gravitino stress-energy tensor vanish as a consequence of eq.~(3.30). The SD
condition on the curvature remains unchanged in the ~$N=1~$ SDSG.
Simultaneously, eq.~(3.30) is consistent with the SG BIds and does not render
the theory flat, just self-dual. The second-order differential constraint on
the metric in eq.~(3.28) is implied by the superspace constraint (3.30) which
is the first-order differential equation on the gravitino field.

\bigskip\bigskip

\noindent {\bf 3.4 Self-Dual Tensor Multiplet}

        As a remarkable aspect of our AW space-time, we give here a
completely new multiplet, different from any of the above mentioned
multiplets, which satisfies a SD condition in  a generalized sense.
This multiplet is based on what is called the {\it tensor multiplet} (TM) or
the linear multiplet, with the field content $~(B_{\ulm\uln},\Phi,
\chi_\a,\Tilde\chi_{\Dot{\a}})$.  The superspace BIds are
$$\fracm 16\nabla_{\[ A} G\du{B C D)} \, - \, \fracm 14 T\du{\[ A B|} E
G_{E|C D)}  \equiv 0 ~~,
\eqno(3.31) $$
where $~G_{A B C}$~ is the superfield strength of $~B_{A B}$.  The
constraints and constituency relations that
solve these BIds are
$$\eqalign{&G_{\a\Dot\b \ulc} = -i (\s_\ulc)_{\a\Dot\b} \Phi ~~, \cr
&G_{\a\ulb\ulc} = \fracm 1{\sqrt 2}
(\s_{\ulb\ulc})\du\a \b \chi{\low\b} ~~, ~~~~
G_{\Dot\a \ulb \ulc} = \fracm 1{\sqrt 2}
(\Tilde\s_{\ulb\ulc})\du{\Dot\a}{\Dot\b} \Tilde \chi_{\Dot
\b} ~~, \cr
&\nabla_\a \Phi = -\fracm 1{\sqrt2} \chi_\a ~~, ~~~~
\Tilde\nabla_{\Dot\a} \Phi= -\fracm 1{\sqrt2} \Tilde\chi_{\Dot\a} ~~, \cr
&\nabla_\a \Tilde\chi_{\Dot\b} = -i \fracm 1{\sqrt2}
(\s^\ulc)_{\a\Dot\b} \nabla_\ulc \Phi + i\fracm 1{6{\sqrt2}}
\e^{\ulc\uld\ule\ulf}(\s_\ulf)_{\a\Dot\b} G_{\ulc\uld\ule} ~~, \cr
&\Tilde\nabla_{\Dot\a} \chi{\low\b} = -i \fracm 1{\sqrt 2}
(\s^\ulc)_{\b\Dot\a} \nabla_\ulc \Phi - i\fracm 1{6{\sqrt2}}
\e^{\ulc\uld\ule\ulf}(\s_\ulf)_{\b\Dot\a} G_{\ulc\uld\ule} ~~, \cr
&\nabla_\a\chi{\low\b} = 0~~, ~~, ~~~~
\Tilde\nabla_{\Dot\a} \Tilde \chi_{\Dot\b} = 0 ~~, \cr
&T\du{\ul\a\ulb}\ulc = 0~~, ~~~~ T\du{\ula\ulb} \ulc = 0 ~~, ~~~~
T\du{\ul\a\ul\b}{\ul\g}= 0~~, ~~~~T\du{\ul\a\ulb}{\ul\g}= 0 ~~.  \cr }
\eqno(3.32) $$

The invariant lagrangian of this multiplet is
$$\Lag_{\rm TM}^{N=1}
=  \fracm 1{12} G_{\ulm\uln\ulr} ^2 +
\half \eta^{\ulm\uln} (\partial_\ulm \Phi) (\partial_\uln \Phi)
+i \chi^\a (\s^\ulc)\du\a{\Dot\b} \partial_\ulc\Tilde
\chi_{\Dot\b} ~~.
\eqno(3.33) $$

We turn to the question of SD in this system of TM.
Based on our experience in other multiplets, it is natural to assume that
such a SD is related to the {\it Majorana-Weyl} condition on the field
$~\chi$, namely the condition of vanishing of either $~\chi_\a$~ or
$~\Tilde\chi_{\Dot\a}$~ would generate the SD.  In our case, we find it
convenient to choose
$$ \Tilde\chi _{\Dot\a} = 0 ~~,
\eqno(3.34) $$
in order to have consistent couplings with other self-dual multiplets.
Under eq.~(3.34), we see that the bosonic condition
$$G_{\ula\ulb\ulc} = \e\du{\ula\ulb\ulc} \uld \nabla_\uld \Phi
\eqno(3.35) $$
relates the two ``field strengths'' $~G_{\ula\ulb\ulc}$~ and
$~\nabla_\ula \Phi$~ in an unexpected way!  Accordingly, the dilaton
superfield $~\Phi$~ is to be a {\it chiral superfield} to comply with the SD
condition, because $~\Tilde{\nabla}_{\Dot{\a}} \Phi = -(1/{\sqrt2})
\Tilde{\chi}_{\Dot{\a}} = 0$.

        Actually, this multiplet plays an important role in the
Wess-Zumino-Novikov-Witten term of the Green-Schwarz
string action, which we
have described in our previous paper [16].  Remarkably, the four $~N=1$~
multiplets of SM, YM, SG and TM allow the SD conditions, and they are
consistent with the Green-Schwarz string.
The self-dual TM was also predicted by a previous work on
the Neveu-Schwarz-Ramond $~N=(2,0)$~ heterotic $~\s\-$model [3].
In particular, it was
noted that the only way the dilaton could be coupled to the string
required that the dilaton must be chiral.  Since the dilaton originates
from the TM, it follows that it must satisfy some type of
chirality condition.  This chirality condition is precisely the
vanishing of $~\Tilde\chi_{\Dot\a}$~ ({\it i.e.} $~\Tilde\nabla_{\Dot\a}
\Phi = 0)$.

        Even though we skip all the details of the consistency check of
the field equations, owing to the supersymmetry of the system it is sufficient
to look into the consistency of our constraints with the SD condition at
dimension $~d\le 1$~\footnotew{It is convenient to assign the dimensions
to our bosonic superfields as
$~\[e\du\ula\ulm\] = \[\Phi] = \[B_{\ulm\uln}] = [A\du\ulm I \] =
0$~ and for fermionic ones as $~\[ \psi\du\ulm{\ul\a i} \]
= \[ \l\du{\ul\a} I \]
= 1/2$.  This is to be universal also for other bosonic and fermionic
fundamental superfields, respectively.  Thus, e.g.,
$~\[T\du{\ula\ulb}{\ul\g k} \] = 1/2,~
\[T\du{\ula\ulb}\ulc \] = 1,~ \[F_{\ul\a i\, \ulb}{}^I\]  = 1/2$~ and
$~\[F_{\ula\ulb}{}^I \] = 1$, {\it etc}.}
{}~{\it without} even checking all the field equations
at $~d\ge 3/2$.  The consistency of the constraints with our SD conditions
under supersymmetry will guarantee the consistency of our SD
conditions.

The curious reader may wonder if the couplings among all the self-dual
multiplets are really consistent to all orders, especially when the
SG multiplet is coupled.  As an intuitive
explanation of how this works, we note the following point:
these multiplets are all re-derived from the corresponding multiplets in
the usual $~D=(1,3)$~ space-time by appropriate replacements of the
{\it dotted} spinors by {\it tilded} spinors, regarding them as
independent quantities.  By doing so, we see that in the
{\it canonical} basis of all the fields, the energy-momentum tensors of all
of them vanish, due to the involvements of at least one of the vanishing
fields that appear in the SD conditions of those multiplets.  The only subtlety
is that the SD condition is to be chosen in the canonical basis, including
the supergravity multiplet itself.  This is because the SD conditions
are {\it not} scale invariant conditions, so that some factors or other
complications may arise.  For example, after a Weyl rescaling,
an unusual factor ~$e^{\xi \Phi}~$ will arise in front of the SD condition on
the field strength $~F_{\ula\ulb}$,
$$F\du{\ula\ulb} I= \half e^{\xi \Phi}
\e\du{\ula\ulb}{\ulc\uld} F\du{\ulc\uld} I ~~,
\eqno(3.36) $$
with some constant $~\xi$.  (See also Appendix C for quantum
corrections.)

\bigskip\bigskip

\noindent {\bf 4. $~N=2$~ Supersymmetry in AW Space-Time}
\medskip\medskip

        Since we have established the $~N=1$~ SDYM as well as the
$~N=1$~ SDSG, the next natural question is about {\it extended}
supersymmetries.    We will show in this section that the answer
is actually in the affirmative, namely we construct the SDYM as well as the
SDSG with $~N=2$~ supersymmetry. However, it turns out to be impossible to
formulate the self-dual $~N=2$~ hypermultiplet, {\it without the
inclusion of additional propagating fields}.

\bigskip\bigskip

\noindent {\bf 4.1 $~N=2$~ Hypermultiplet in AW Space-Time}

        There exists an AW space-time analogue of the $~N=2$~ hypermultiplet
[24], which is straightforward to construct. In the $~N=2~$ superspace, it  is
 defined by the equations [25]
$$D_{\a i}\F_j + D_{\a j}\F_i = \Tilde{D}_{\dt{\a}i}\F_j+ \Tilde{D}_{\dt{\a}j}
\F_i =0 ~~,
\eqno(4.1)$$
in terms of a isoscalar (complex) ~$N=2~$ superfield ~$\F_i\,$, $~{\scst
i~=~1,~2}$. In the absence of central charges, eq.~(4.1) does imply the
equations of motion to be satisfied. Projecting the hypermultiplet into the
{}~$N=1~$ superspace yields two complex scalar ~$N=1~$ superfields ~$\f_i\,$,
$$ \f_i = \F_i\left.\right|~,
\eqno(4.2)$$
which are known to be {\it chiral} ~$(\f_2)~$ and {\it anti-chiral} ~$(\f_1)~$,
 respectively [24,26]. The on-shell hypermultiplet comprises a scalar
isodoublet $\varphi_i(x)~$ and a {\it Dirac} (complex) isoscalar spinor
{}~$\J_{\ul{\a}}=\left( \j_{\a},\Tilde{\k}_{\dt{\a}}\right)$. The $~N=2$~
supersymmetry transformation rules in the AW space-time are
$$\eqalign{&\d\varphi_i = \e\ud{\a}i\j_{\a}
+ \Tilde\e\, \ud{\dt{\a}}i \Tilde{\k}_{\dt{\a}}~, \cr
&\d\j_{\a} = -2i\left( \s^{\ula}\right)_{\a}{}^{\dt{\b}}
\Tilde\e_{\dt{\b}i}\pa_{\ula}\varphi^i~,\cr
&\d\Tilde{\k}_{\dt{\a}} = -2i\left( \st^{\ula}\right)_{\dt{\a}}{}^{\b}
\e_{\b i}\pa_{\ula}\varphi^i~. \cr}
\eqno(4.3)$$
The fields ~$\varphi_2~$ and ~$\j~$ form the on-shell $~N=1$~ chiral multiplet,
while  the fields ~$\varphi_1~$ and ~$\Tilde{\k}~$ enter into the on-shell
$~N=1$~ anti-chiral multiplet.

        As we already know from the ~$N=1~$ considerations of
sect.~3, the SD is associated with the real chiral objects of
definite chirality and the certain choice of the ~$\g\-$matrices. One can use
either of them, but {\it not both} simultaneously in one usual supermultiplet.
Since the ~$N=2$~ hypermultiplet comprises the two ~$N=1~$ chiral multiplets
of {\it different} chirality, one can {\it not} define the self-dual ~$N=2~$
hypermultiplet, even though the ~$N=1~$ self-dual and anti-self-dual
SM do exist.   One can, of course, introduce real chiral and real
anti-chiral ~$N=1$~ scalar superfields to represent self-dual and
anti-self-dual ~$N=1~$ SM, but one can not unify them into one object
with ~$N=2~$
supersymmetry,  since they correspond to the different supersymmetry algebras:
 one for the choice ~$({\bf I})~$ and another for the choice ~$({\bf II})~$ of
 the ~$\g\-$matrices. (See Appendix B).

\bigskip \bigskip

\noindent {\bf 4.2  $~N=2$~ Supersymmetric YM in AW Space-Time}

        The on-shell $~N=2$~ SYM has the field content $~(A\du \ulm
I,\l\du{\a i}I,\Tilde\l\du{\Dot\a i}I,S^I,T^I)$, where the indices
$~{\scst I,~J,~\cdots}$~ are for the
adjoint representations, the indices $~{\scst i,~j,~\cdots~=~ 1,~2}$ are
for the two-dimensional representations of $~Sp(1)$~ group.  The fields $~S^I$~
and $~T^I$~ are real scalars, while  the $~\l^I$~ and $~\Tilde\l^I$~ are
Majorana-Weyl spinors of opposite chiralities, following the same
notation of the $~N=1$~ case.  Our superspace BIds {\it before} imposing
any SD condition are
$$\nabla_{\[A}F\du{B C)} I - T\du{\[A B|}D F\du {D|C)} I \equiv 0~~.
\eqno(4.4) $$
These BIds at the dimensions $~0\le d\le 1$~ are
solved by the constraints
$$\eqalign{&F_{\a i\,\b j}{}^I = 2C_{\a\b} \,\e{\low{i j}}\, T^I ~~, ~~~~
F_{\Dot\a i\,\Dot\b j} {}^I = 2C_{\Dot\a\Dot\b} \,\e{\low {i j}}\, S^I ~~, \cr
& F_{\a i\,\ulb} = - i(\s_\ulb)_{\a\Dot\b} \Tilde\l\udu{\Dot\b} i I ~~,
{}~~~~F_{\Dot\a i\, \ulb} {}^I = - i(\s_\ulb)_{\b\Dot\a} \l\udu \b i I ~~, \cr
&F_{\a i\,\Dot\b j}{}^I = 0 ~~, ~~~~ T\du{\a i\,\Dot\b} {j\,\ulc} = i(\s^\ulc)
_{\a\Dot\b} \,\d\du i j ~~, \cr
& \nabla_{\a i} S^I = -\l\du{\a i} I ~~, ~~~~\Tilde\nabla_{\Dot\a i} S^I
= 0~~, \cr
&\Tilde\nabla_{\Dot\a i} T^I = -\Tilde\l _{\Dot\a i} {}^I ~~, ~~~~\nabla_{\a
i} T^I = 0~~, \cr
&\nabla_{\a i} \Tilde\l\du{\Dot\b j }I = - i \,\e{\low{i j}}\, (\s^\ulc)
_{\a\Dot\b}\nabla_\ulc T^I ~~, \cr
& \Tilde\nabla_{\Dot\a i} \l\du{\b j} I = + i \e{\low{i j}}\, (\s^\ulc)
_{\b\Dot\a} \nabla_\ulc S^I ~~, \cr
&\nabla_{\a i} \l\du{\b j}I = -\fracm 14 \,\e{\low{i j}} \,
(\s^{\ulc\uld} )_{\a\b}
F\du{\ulc\uld} I + f^{I J K} C_{\a\b} \,\e{\low{i j}}\, S^J T^K ~~, \cr
&\Tilde\nabla_{\Dot\a i} \Tilde\l\du{\Dot\b j} I = +\fracm 14 \,\e{\low {i
j}}\, (\Tilde\s^{\ulc\uld}) _{\Dot\a\Dot\b} F\du{\ulc\uld} I + f^{I J K}
C_{\Dot\a\Dot\b} \e{\low {i j}}\, S^J T^K ~~, \cr }
\eqno(4.5)$$
where $~\e_{i j}= - \e_{j i},~\e_{12} = +1$, stands for the
invariant antisymmetric tensor for the $~Sp(1)$~ group.  Needless to say,
the raising and lowering of these $~Sp(1)$~ indices are by $~\e_{i j}$~ and
$\e^{i j}$.

        The BIds of $~3/2\le d\le 2$~ yield the superfield equations:
$$\eqalign{&i(\s^\ula)_{\a\Dot\b} \nabla_\ula \Tilde\l^{\Dot\b i \,I}
- 2f^{I J K} \l\du\a {i J} T^K=0 ~~, \cr
&i(\s^\ula)_{\b\Dot\a} \nabla_\ula \l^{\b i\, I } - 2f^{I J K}
\Tilde\l\du{\Dot\a} {i\,J} S^K = 0~~, \cr
& \nabla_\ulb F^{\ula\ulb\, I} - 2 f^{I J K} (S^J \nabla^\ula T^K + T^J
\nabla^\ula S^K) - 2i f^{I J K} (\l^{i\,J}\s^\ula \Tilde\l\du i K ) = 0~~,
\cr
&\Bo S^I + f^{I J K} (\l^{i\,J} \l\du i K) - 4f^{I K J} f^{J L M} S^K
S^L T^M = 0 ~~, \cr
&\Bo T^I - f^{I J K} (\Tilde\l^{i\, J} \Tilde\l\du i K) + 4f^{I K J}f^{J L
M} S^L T^K T^M = 0 ~~. \cr}
\eqno(4.6) $$
This system has an invariant lagrangian [27] ({\it cf.} Ref.~[28])
$$\eqalign{\Lag_{\rm SYM}^{N=2} = \,&-\fracm 14 (F\du{\ula\ulb}I)^2 - 2i
\Tilde\l^{i\,I}\s^\ulc \nabla_\ulc \l\du i I - 2(\nabla_\ula
T^I)(\nabla^\ula S^I) \cr
&+2f^{I J K} (\l^{i\,I} \l\du i J ) T^K - 2f^{I J K} (\Tilde \l^{i\, I}
\Tilde \l\du i J) S^K \cr
& + 4f^{I J K} f^{K L M} S^I S^L T^J T^M ~~. \cr }
\eqno(4.7) $$
Notice the peculiar kinetic term of $~S^I$~ and $~T^I$, which is
equivalent to two terms with the {\it opposite} signs in the diagonalized
fields.  This indicates the higher-dimensional origin of the system,
such as $~D=(3,3),\,N=1$~ SYM {\it via} simple dimensional reduction.
In the Appendix D, we actually construct such a theory in $~D=(5,5)$.

        We are so far free of any SD condition.  Based on our $~N=1$~
experience, we can easily postulate our SD condition to be
$$ S^I = 0~~.
\eqno(4.8) $$
This condition implies through BIds other superfield equations such as
$$\l\du{\a\,i} I= 0 ~~, ~~~~ F_{\ula\ulb}{}^I = \half\e\du{\ula\ulb}{\ulc\uld}
F_{\ulc\uld}{}^I ~~,
\eqno(4.9) $$
as is desired in Ref.~[27]. Being decomposed into the sub-multiplets with
respect to the $~N=1$~ supersymmetry, the $~N=2$~ SDYM multiplet comprises
the $~N=1$~ SDYM multiplet and the $~N=1$~ on-shell SDSM.

\bigskip\bigskip

\noindent {\bf 4.3 $~N=2$~ Supergravity in AW Space-Time}

An $~N=2$~ SDSG in the AW space-time is also consistently
constructed, based on the {\it non-self-dual} $~N=2$~ SG multiplet first.

        Our on-shell $~N=2$~ SG multiplet has the fields $~(e\du \ula
\ulm, \psi\du\ulm{\a i}, \Tilde\psi\du \ulm{\Dot\a i},  A_\ulm)$, where the
gravitino is in the $~{\bf 2}\-$representation of an
$~SO(2)$~ group, and the graviphoton field $~A\du\ulm {i j}$~ is
the gauge field of the $~SO(2)$.
Our independent BIds {\it before} any SD condition are
$$\eqalign{&\nabla_{\[A}T_{B C)} - T\du{\[ A B|}E T\du{E|C)} D -
R_{\[A B C)}{}^D  \equiv 0~~, \cr
&\nabla_{\[A} F_{B C)} - T\du{\[A B|}D F_{D|C)} \equiv 0 ~~.
\cr }
\eqno(4.10) $$
These BIds at $~0\le d \le 1$~ are solved by the
constraints\footnotew{Here we are using $~SO(2)$~ notation, so that some
constraints look different from the ~$Sp(1)~$ ones.  For example, all the
rasing and lowering of the $~SO(2)$~ are done by the usual Kronecker's
delta, which does {\it not} cause any sign change, and effectively they
do not matter, as opposed to the previous $~Sp(1)$~ case in eq.~(4.5).}
$$\eqalign{&T\du{\a i\,\Dot\b}{j\,\ulc} = i \d\du i j\, (\s^\ulc)_{\a\Dot\b}
{}~~, ~~~~ T_{\a i\,\b j}{}^\ulc = 0 ~~, \cr
&F_{\a i \,\b j} = \e{\low {i j}} C_{\a\b} ~~, ~~~~
F_{\Dot\a}{}^i_{\Dot\b}{}^j =
\e^{i j} C_{\Dot\a \Dot\b} ~~, ~~~~ F_{\a i\Dot\b}{}^j = 0 ~~, \cr
&T\du{\a i\,\ulb} {\Dot\g}{}_k = \fracm i2 \e{\low{i k}}
(\s_\ulc)\du\a{\Dot \g}
(F_{\ulb\ulc} + {\breve F}_{\ulb\ulc}) ~~, ~~~~ T\du{\a i \,\ulb}{\g k} =
0~~, \cr
&T\du{\Dot\a i \,\ulb} {\g k} = \fracm i2 \e{\low i}{}^k (\s_\ulc) \ud
\g{\Dot\a} (F_{\ulb\ulc} - {\breve F}_{\ulb\ulc} ) ~~, ~~~~T\du{\Dot\a
i\, \ulb} {\Dot\g}{}_k =0~~, \cr
&T\du{\ul\a\ul\b}{\ul\g} = 0~~, ~~~~ F_{\ul\a \ulb} = 0~~, ~~~~
T\du{\ula\ulb}\ulc =0~~, \cr
& R_{\a i\,\b j \, \ulc\uld} = \e{\low{i j}} \, C_{\a\b} (F_{\ulc\uld} +
{\breve F}_{\ulc\uld} ) ~~,~~~~
R_{\Dot\a}{}^i{}_{\Dot\b}{}^j{}_{\ulc\uld} = \e^{i j}\, C_{\Dot\a\Dot\b}
(F_{\ulc\uld} - {\breve F}_{\ulc\uld} ) ~~, \cr
&R_{\a i\b j \Dot\g}{}^{k\,\Dot\d}{}_l = - \half C_{\a\b}
(\Tilde\s^{\ule\ulf})\du{\Dot\g}{\Dot\d} \e_{i j} \d\du l k F_{\ule\ulf} ~~,
\cr
&R_{\Dot\a}{}^i{}_{\Dot\b}{}^j{}_{\g k}{}^{\d l} =- \half C_{\Dot\a\Dot\b}
(\s^{\ule\ulf}) \du\g\d \e^{i j} \d\du k l F_{\ule\ulf} ~~, \cr
&R\du{\a i\,\b j\g k}{\d l} = 0~~, ~~~~ R_{\a i \,\Dot\b}{}^j{}
_{\ul\g}{}^{\ul\d} =0~~, \cr }
\eqno(4.11) $$
where the indices $~{\scst i,~j,~\cdots ~=~ 1,~2}$~ are for the
$~{\bf 2}\-$representations of $~SO(2)$,
while {\it underlined} spinorial indices
$~{\scst \ul\a,~\ul\b,~\cdots}$~ are for the pairs of indices
$~{\scst (\a i,~\Dot\a i),~(\b i,~\Dot\b i),~\cdots}$.  The $~{\breve F}
_{\ula\ulb} \equiv (1/2)\e\du{\ula\ulb}{\ulc\uld}
F_{\ulc\uld}$~ is the {\it dual}~ of $~F_{\ula\ulb}$.

        The BIds at ~$d\ge 3/2$~ yield
the superfield equations
$$\eqalign{&i(\s^{\ula\ulb\ulc}) _{\b\Dot\a} T\du{\ulb\ulc}{\b}{}_j =0 ~~,
{}~~~~ i(\s^{\ula\ulb\ulc})_{\a\Dot\b} \Tilde T\du{\ulb\ulc}{\Dot\b j} =
0~~, \cr
&\nabla_\ula F^{\ula\ulb} = 0~~, \cr
&R_{\ula\ulb} + \half \left(F_{\ula\ulc} F\du\ulb\ulc - {\breve F}
_{\ula\ulc} {\breve F} \du\ulb\ulc \right) \cr
&~~~~~ = R_{\ula\ulb} + \left[\, F_{\ula\ulc} F\du \ulb\ulc - \fracm 14
\eta_{\ula\ulb} (F_{\ulc\uld}{})^2 \, \right] = 0~~. \cr }
\eqno(4.12) $$

        Notice the similarity of this system to the $~N=1$~ SG, we saw
in sect.~3, which makes it easier to impose a SD condition.
        Define the superfields $~W_{\a\b}$~ and
$~\Tilde W_{\Dot\a\Dot\b}$~ by
$$W_{\a\b} \equiv \half (\s^{\ulc\uld})_{\a\b} F_{\ulc\uld}~~, ~~~~
\Tilde W_{\Dot\a\Dot\b} \equiv \half (\s^{\ulc\uld})_{\Dot\a\Dot\b}
F_{\ulc\uld}~~,
\eqno(4.13)$$
as the analogues of the $~w_\a$~ and $~\Tilde w_{\Dot\a}$~ in the $~N=1$~
SYM, or, equally, as the analogues of the $~W_{\a\b\g}$ and
$~\Tilde{W}_{\Dot{\a}\Dot{\b}\Dot{\g}}$ in the $~N=1$~ SG.
Then our SD condition is
$$W_{\a\b} = 0~~.
\eqno(4.14) $$
This condition implies the superfield equations
$$\eqalign{&T\du{\ula\ulb}{\g k} = 0~~, ~~~~ \Tilde T\du{\ula\ulb}
{\Dot\g}{}_k =
\half \e\du{\ula\ulb} {\ulc\uld} \Tilde T\du{\ulc\uld} {\Dot\g}{}_k ~~, \cr
&F_{\ula\ulb} = \breve F_{\ula\ulb} ~~, \cr
&R_{\ula\ulb} = 0~~, ~~~~R\du{\ula\ulb} {\ulc\uld} = \half \e
\du{\ula\ulb}{\ule\ulf} R\du{\ule\ulf}{\ulc\uld} ~~, \cr}
\eqno(4.15) $$
{\it via} BIds of $~d\ge 3/2$.  The consistency of our SD condition
(4.14) with all the BIds is also confirmed.
In particular, we can verify that even the $~d=5/2$~ BId is satisfied in a
highly non-trivial way.  First, we notice that under our SD condition
the reduced BId $~\nabla_{\[\ula} T\du{\ulb\ulc\]}{\d l} = 0~$ holds, which
helps us to confirm that
$$\eqalign{&\nabla_{\a i} \left[\, R\du{\ulb\ulc}{\uld\ule} - \half
\e{\low{\ulb\ulc}}{}^{\ulf\ulg} R\du{\ulf\ulg}{\uld\ule} \,\right] \cr
&~~~= \left[ \,i \fracm 12 \s^{\[\uld} \nabla_{\[\ulb} \Tilde
T\ud{\ule\]}{\ulc\]} + i \fracm 12 \s_{\[\ulb}\nabla_{\ulc\]} \Tilde
T^{\uld\ule} \,\right] _{\a i}
-\half \e{\low {\ulb\ulc}}{}^{\ulf\ulg} \left[\,i\half\s^{\[\uld}
\nabla_\ulf \Tilde T\ud{\ule\]} \ulg
+ i\half\s_\ulf \nabla_\ulg \Tilde T ^{\uld \ule} \,\right] _{\a i} \cr
& ~~~= -i \fracm 1 2 (\s_{\ulb\ulc})\du\a \g \nabla^{\[\uld
|}(\s^\ulf \Tilde T\ud{|\ule\]}\ulf )_{\g i} = 0~~, \cr }
\eqno(4.16) $$
by the help of our superfield equations (4.15).
Thus, there is no essential fundamental difference of $~N=2$~ SDSG
from the $~N=1$~ SDSG, and everything is straightforward.

\bigskip\bigskip

\noindent{\bf 5.~~$N=4$~ Self-Dual Supergravity}
\medskip\medskip

        The construction of $~N=4$~ SDSG follows the same pattern
we adopted for the $~N=2$~ SDSG.  We start with the non-self-dual case
first as before, utilizing also the usual $~D=(1,3),\, N=4$~ results.
It has been known that there are {\it three} different superspace versions for
the $~D=(1,3),\,N=4$~ SG, namely the $~SO(4)$~ theory [29], the
$~SU(4)$~ theory [30],
and the $~SU(4)$~ theory related to the heterotic string [31].  Among these it
is easiest to utilize the $~SO(4)$~ theory
consistently truncated down to the $~N=4$~ SDSG theory.

        Our on-shell $~N=4$~ SG with the {\it global}
$~SO(4)$~ symmetry {\it before} the SD condition has the
field content $~(e\du \ula\ulm,\psi\du\ulm{\a i},\Tilde\psi\du\ulm{\Dot\a i},
A\du\ulm{i j},\L_{\a i},\Tilde\L_{\Dot\a i}, A,B)$, where the indices
${\scst i,~j,~\cdots~=~1,~\cdots,~4}$~ are for the $~{\bf
4}\-$representation of the $~SO(4)$~ group.
The central charge gauge fields $~A\du\ulm {i j}$~ are in the
$~{\bf 6}\-$representation.  Since the $~D=(1,3),\,N=4$~ superspace is similar
 to the $~D=(2,2),\,N=4$~ superspace, we can rely on the results of
Ref.~[32].  The essential difference, however, is that the superfields $~W$~
and $~\Bar W$~ in the former [32] are now replaced by the two completely
independent real scalar superfields $~B$~ and $~A$, respectively.
In other words, $~A$~ and $~B$~ are {\it not} related to each other by
complex conjugation in our $~D=(2,2)$ case.

Although the explicit construction to follow concentrates on the $~SO(4)$~
theory, we emphasize that this is purely a matter of choice. In particular,
the $~SU(4)$~ theory [31] that is closely related to the $~D=(1,3),\,N = 4$~
heterotic string can also be made into a self-dual theory.  The key point is
that this theory also possesses two superfield field strengths that permit
the condition in (3.35) to be imposed.  Upon the imposition of this constraint
(3.35) on the $~D=(1,3),\,N = 4$~ supergravity theory, it too can be turned
into a self-dual theory in AW space-time.

        Our independent BIds are
$$\eqalign{&\nabla_{\[ A} T\du{B C)} D - T\du{\[ A B|}E T\du{E|C)} D
-R\du{\[ A B C)} D \equiv 0~~, \cr
& \nabla_{\[ A} F\du{B C)} {i j} - T\du{\[ A B|} D F\du{D|C)} {i j}
\equiv 0 ~~, \cr}
\eqno(5.1) $$
with the superfield strengths $~F\du{A B}{i j}$~
for the the central charge gauge fields.
Utilizing the similarities between the $~D=(2,2),\,N=4$~
SG and the $~D=(1,3),\,N=4$~ SG, we can solve these BIds. In fact, we find that
the appropriate supertorsion, supercurvature, and constituent constraints
are as follows.  First, the supertorsion constraints are
$$\eqalign{&T_{\ul{\a}\ul{\b}}{} ^{\ul{\g}} = \fracm 1 4 A
\L_{(\ul{\a}}\kd {\ul\b )} {\ul{\g}} ~~,~~~~
T_{\ul{\a}\ul{\b}}{} ^{\ul{\dt\g}} ~=~ C_{\a\b}C_{ijkl}
\Tilde\L^{\dt\g l} ~~,~~
T_{\ul{\a}\ul{\b}}{} ^{\ul{c}} ~=~ T_{\ul{\a}\ul{b}}{} ^{\ul{c}}
{}~=~ 0 ~,\cr
&T_{\ul{\a}\ul{\dt\b}}{} ^{\ul{\dt\g}} =
-~ \fracm 1 4 \d_{\ul{\dt\b}}{}^{\ul{\dt\g}} A \L_{\ul{\a}} ~~,~~
T_{\ul{\a}\ul{\dt\b}}{} ^{\ulc} ~=~ 2i \kd {\a} {\g}
\kd {\dt\b} {\dt\g} \kd i k ~~,~~
T_{\ul{a}\ul{\b}}{} ^{\ul{\dt\g}} ~=~ -~ iC_{\a\b}
\Tilde f_{\dt\a}{}^{\dt\g}{}_{jk} ~, \cr
&T_{\ul{a}\ul{\b}\ul{\g}} =
\fracm 14 \lrad{\a\Dot\a} C_{\b\g} \d\du j k ~+~ i\half
C_{\a\b}\L_{\g l}\Tilde\L_{\dt\a}{}^{\[ k}
\d_j{}^{l \]} ~+~i \fracm 1 {16} C_{\a(\b}\L_{\g) l}
\Tilde\L_{\dt\a}{}^{l}\d_j{}^{k} ~~, \cr
&T_{\ul{a} \ul{b}}{} ^{\ul{\g}}  =  - C_{\dt\a \dt\b}
\bigg{[}\S_{\a \b}{}^{\g k} + i \fracm 13 a^{-1} \d_{(\a}{}^{\g}
(D_{\b)\dt\d} B )\Tilde\L^{\dt\d}{}^{k}\bigg{]}  + \fracm 1 4
C_{\a \b}\bigg{[}C^{klmn}\L^{\g}{}_{l}\Tilde f_{\dt\a \dt\b m n} +
ia^{-1} (D^{\g}{}_{(\dt\a}B )\Tilde \L_{\dt\b) }{}^{k}\bigg{]} ~, \cr
&T_{\ul{a}\ul{b}}{} ^{\ul{c}}  =  ~-~i \fracm 1 2 \left[
\d_{\b}{}^{\g}\d_{\dt\a}{}^{\dt\g}\Tilde \L_{\dt\b}{}^{k}\L_{\a k} ~-~
\d_{\a}{}^{\g}\d_{\dt\b}{}^{\dt\g}\Tilde \L_{\dt\a}{}^{k}\L_{\b k}\right] ~.
\cr }
\eqno(5.2) $$
We use the {\it underlined} indices $~{\scst \ul\a,~\ul\b,~\cdots}$~
and $~{\scst \ul{\Dot\a},~\ul{\Dot\b},~\cdots}$~ for $~{\scst \a i,~\b
j,~ \cdots}$~ and $~{\scst \Dot\a i,~\Dot\b j~\cdots}$, while
$~(A{\buildrel\leftrightarrow\over D}_{\a\Dot\a}B)
\equiv A(D_{\a\Dot\a} B) - (D_{\a\Dot\a} A) B~,~~a=a(A,B)\equiv
1-A^2-B^2$.  The new superfields on the r.h.s.~of eq.~(5.2)
above and eq.~(5.4) below are the ones, to which the whole set of the
supertorsions and supercurvatures reduces.
There exist also complementary
constraints, obtained by exchanging
$~A\leftrightarrow B, ~\hbox{\it
dotted}\leftrightarrow \hbox{{\it undotted} indices}$, and $\hbox{\it
tilded}\leftrightarrow \hbox{{\it untilded} superfields}$.  This is
because $~A$~ and $~B$~ superfields are {\it not} related by complex
conjugation in our case.  For example,
$$~T\du{\ula\underline{\Dot\b}}{\underline\g}= - iC_{\Dot\a\Dot\b} f
\dud\a\g{j k}~~.
\eqno(5.3) $$
Their forms are thus parallel and straightforward to derive.

The supercurvatures are:
$$\eqalign{&R_{\ul{\a} \ul{\b} \g\d}  =  0 ~~,~~~~
R_{\ul{\a} \ul{\b}\, \dt\g \dt\d}
{}~=~ 2C_{\a\b} \Tilde f_{\dt\g \dt\d i j} ~,~  \cr
&R_{\ul\a \ul{\dt\b} \g\d}
 =  - \fracm 1 2 C_{\a ( \g} \bigg{[} \L_{\d ) i}
\Tilde \L_{\dt\b}{}^{j}~-~\fracm 1 4 \d\du i j \L_{\d) l}
\Tilde \L_{\dt\b}{}^{l}\bigg{]} ~, \cr
&R_{\ul\a , \b \dt\b, \g\d}  =  i \fracm {7} {16} C_{\a\b} C_{i j k l}
\Tilde \L_{\dt\b}{}^{j} f_{\g\d}{}^{kl} ~+~ i \fracm 1 {32} C_{\b ( \g|}
C_{i j k l} \Tilde \L_{\dt\b}{}^{j} f_{|\d )\a}{}^{kl} \cr
&~~~~~ ~~~~~ ~~~~~ +~ \fracm 3 {16} a^{-1} C_{( \g\vert ( \a}
(D_{\b)  \dt\b} A) \L_{|\d) i}
 ~-~ \fracm 5 {16} C_{\a\b}(D_{( \g\vert \dt\b} A)
\L_{\vert\d ) i}~, \cr
&~R_{\ul\a , \b\dt\b, \dt\g \dt\d}  =  -~ iC_{\a\b}
\Tilde \S_{\dt\b \dt\g \dt\d i} ~-~ i \fracm 1 {16} C_{\dt\b ( \dt\g}
C_{i j k l} \Tilde \L_{\dt\d )}{}^{j} f_{\a\b}{}^{kl} \cr
&~~~~~ ~~~~~ ~~~~~ +~ \fracm 1 {48} a^{-1} C_{\a\b} C_{\dt\b  ( \dt\g\vert}
(D_{\e \vert \dt\d)} A )
\L^{\e}{}_{i} ~+~ \fracm 3 {16} a^{-1} C_{\dt\b ( \dt\g\vert}
(D_{\a\vert \dt\d )} A )
\L_{\b ) i} ~, \cr
&R_{\a\dt\a , \b\dt\b, \g\d}  =  - \fracm 1 2 C_{\dt\a \dt\b}
\bigg{[} V_{\a\b\g\d} ~+~ i \fracm 1 {16} C_{ ( \g\vert ( \a}
\Tilde \L^{\dt\e l} \left\{ D_{\b ) \dt\e} ~+~ \fracm 3 4 a^{-1}
\lrad{\b )\dt\e} \right\} \L_{\vert\d ) l} \cr
&~~~~~ ~~~~~ ~~~~~ +~ C_{\a {(} \g} C_{\d ) \b} \bigg\{ \fracm 1 6 a^{-2} \big(
D_{\e \dt\z}  A \big{)(} D^{\e \dt\z} B  \big)
+ i \fracm 3{16} a^{-1} \lrad{\e\Dot \zeta} \L^{\Dot\zeta i} \L\ud\e i \cr
& ~~~~~ ~~~~~ ~~~~~ ~~~~~ ~~~~~ ~~~~~ ~+~ \fracm {15} {128}
\Tilde \L^{\dt\e l} \Tilde \L_{\dt\e}{}^{m} \L^{\z}{}_{l} \L_{\z m}
\bigg\} \bigg] \cr
& ~~~~~ ~~~~~ ~~~~~ ~+~ \fracm 1 2 C_{\a \b} \bigg{[}f_{\g\d i j}
\Tilde  f_{\dt\a \dt\b}{}^{i j} ~-~ \fracm 1 2 a^{-1}
\big{(}D_{(\g \vert (\dt\a\vert} A \big{)(}
D_{\vert\d ) \vert \dt\b ) } B \big{)} \cr
&~~~~~ ~~~~~ ~~~~~ ~~~~~ ~~~~~ ~~~ ~-~ i \fracm 1 {16} D_{(\g\vert (\dt\a\vert}
\left( \L_{\vert\d ) l} \Tilde \L_{\vert\dt\b )}{}^{l} \right)
{}~-~ \fracm {1} {64} \Tilde \L_{( \dt\a}{}^{l}
\Tilde \L_{\dt\b )}{}^{m}
\L_{\g l} \L_{\d m} \bigg{]} ~. \cr }
\eqno(5.4) $$

The central charge field strengths are:
$$\eqalign{&F_{\ul\a\ul\b}{}^{k l} = 2a^{-1}C_{\a\b} E\du{i j}{k l} ~~, \cr
&E\du{i j}{k l} \equiv\half \left[ \d\du i{\[ k}\d\du j {l\]}
+ AC\du{i j}{k l} \right]
{}~~, ~~~~F\du{\a\Dot\b}{i j} = 0~~,  \cr
&F\du{\a\Dot\a\,\ul\b} {k l} = - i \half a^{-1/2} C_{\a\b}
\Tilde \L\du{\Dot\a} i C_{i j m n} \Tilde E^{m n k l} ~~, \cr
&F\du{\a\Dot\a\,\b\Dot\b} {k l} = \half a^{-1/2} \left[\, C_{\Dot\a\Dot\b}
f\du{\a\b}{m n} E\du{m n}{k l} + C_{\a\b}\Tilde  f_{\Dot\a\Dot\b\, m n}
\Tilde E^{m n k l} \,\right] ~~, \cr
& f\du{\a\b}{k l} = f\du{\b\a} {k l} ~~, \cr}
\eqno(5.5) $$
where $~f\du{\a\b}{i j}$~ and $~\Tilde f\du{\Dot\a\Dot\b}{i j}$~ correspond to
the component YM field strengths as their $~\theta=0\-$sectors.

Finally we need what is called the {\it constituency relations} [32]:
$$\eqalign{&D_{\a i} B  =  (1-A^2 - B^2 ) \L_{\ul\a} ~~~,~~~
\Tilde D_{\Dot\a i} B = 0~~, \cr
&D_{\a i} \L_{\b j}  =  C_{i j k l} f_{\a\b}{}^{kl} ~-~ \fracm 3 4 A
\L_{\a i}\L_{\b j} ~~, ~~~~
{\Tilde D}_{\dt\a}{}^{i} \L_{\b j} ~=~ 2ia^{-1} \d_j{}^{i}(D_{\b
\dt\a} B ) ~+~ \fracm 3 4  B \Tilde\L_{\dt
\a}{}^{i} \L_{\b j} ~,\cr
&{\Tilde D}_{\dt\a}{}^{i}f_{\b\g}{}^{jk}  = \fracm 1 2 B
{\Tilde \L}_{\dt\a}{}^{i}f_{\b\g}{}^{jk} \cr
& ~~~~~ ~~~~~ ~~ ~+~ \fracm 1 2 C^{j k m n}\bigg{[}
\Tilde\L_{\dt\a}{}^{i} \L_{\b m} \L_{\g n}
{}~+~ i\d_m{}^{i}\left\{ D_{( \b\vert \dt\a}
 ~+~i \fracm 1{16} \Tilde\L_{\dt\a}{}^{l} \L_{(\b\vert l}
{}~+~ \fracm 3 4 a^{-1} \lrad{(\b|\Dot\a} \right\}\L_{\vert\g) n}
\bigg{]} ~, \cr
&D_{\a i}f_{\b\g}{}^{jk}  =  \d_i{}^{\[j} \S_{\a\b\g}{}^{k\]}
{}~-~ \fracm 1 2 A \L_{\a i} f_{\b\g}{}^{jk} ~+~ i \fracm 1 3a^{-1}
\d\du {i} {\[j} C_{\a (\b} (D_{\g) \dt\d} B )
\Tilde \L^{\dt\d k \]} ~, \cr
&{\Tilde D}_{\dt\a}{}^{i} \S_{\b\g\d}{}^{j}  =  \fracm 1 4 B
\Tilde \L_{\dt\a}{}^{i}
\S_{\b\g\d}{}^{j} ~-~ \fracm 1 6 \Tilde \L_{\dt\a}{}^{( i} \L_{
( \b |k} f_{|\g\d)}{}^{j)k} ~-~ i \fracm 1 6 C^{i j k l}
(D_{(\b\vert\dt\a} A )
\L_{\vert\g\vert k}\L_{\vert \d) l} \cr
&~~~~~ ~~~~~ ~~~~~ +~ i\bigg{[}D_{(\b\vert\dt\a}  ~-~
i \fracm 3 8 \Tilde \L_{\dt\a}{}^{l} \L_{( \b | l} ~+~ \half
a^{-1} \lrad{(\b|\Dot\a} \bigg{]} f_{\vert \g\d)}{}^{ij}
{}~,\cr
&D_{\a i}\S_{\b\g\d}{}^{j}  =  \d_{i}{}^{j} V_{\a\b\g\d} ~-~ \fracm 1 4 A
\L_{\a i} \S_{\b\g\d}{}^{j} ~-~ \fracm 1 6 a^{-2}
(D_{( \b \vert \dt\g} A) (D_{\vert \g \vert}{}^{\dt\g} B )
C_{\vert \d ) \a} \d\du i j \cr
&~~~~~ ~~~~~ ~~~~~ +~ i \fracm 1 6 \bigg{[}\left\{ D_{(\b\vert \dt\g}
{}~+~ \fracm 34 a^{-1} \lrad{(\b|\Dot\g}
 ~+~i \fracm 1 {16} \Tilde \L_{\dt\g}{}^{l}
\L_{(\b\vert l} \right\} \L_{\vert\g\vert i}\bigg{]}\Tilde \L^{\dt\g j}
C_{\vert\d) \a} \cr
&~~~~~ ~~~~~ ~~~~~ -~ i \fracm 1 8 \bigg{[} \left\{ D_{(\b\vert \dt\g}
 ~+~ \fracm 34 a^{-1} \lrad{(\b\vert \dt\g}   ~+~ i\fracm 1 {16}
\Tilde \L_{\dt\g}{}^{l}
\L_{(\b\vert l} \right\} \L_{\vert\g\vert i}\bigg{]}\Tilde \L^{\dt\g k}
C_{\vert\d) \a}\d_{i}{}^j ~, \cr
&{\Tilde D}_{\dt\a}{}^{i} V_{\b\g\d\e}  =  \fracm 1 {24}
\bigg{[}2i \left\{ D_{(\b\vert\dt\a}  ~-~ \fracm 14
a^{-1}\lrad {(\b\vert\dt\a} \right\}
\S_{\vert \g\d\e )}{}^{i} ~+~ \left(\fracm 3 2 \Tilde\L_{\dt\a}{}^{i}
\L_{(\b| j} ~-~ \fracm {19} {16}\Tilde\L\du{\Dot\a} l \L_{(\b| l}\d\du j
i\right) \S_{\vert \g\d\e )}{}^{j} \cr
&~~~~~ ~~~~~ ~~~~~ ~~~~~ ~~~~~ -~ \left\{ \fracm 9 {16} C_{k l m n}
\Tilde\L_{\dt\a}{}^{l}
f_{(\b\g|}{}^{mn} ~+~ i \fracm 12 a^{-1} \left( D_{(\b\vert\dt\a} A\right)
\L_{\vert\g\vert k} \right\}
f_{\vert\d\e)}{}^{ki} \bigg{]} ~, \cr
&D_{\a i} V_{\b\g\d\e}  =  - \fracm 1 {24} C_{\a (\b|} \bigg[
\left\{ D_{|\g |\dt\e} ~+~ \fracm 1 4a^{-1} \lrad{|\g |\dt\e} \right\}
\left\{ i \fracm 7 {16} C_{i k l m} \Tilde \L^{\dt\e k}f_{|\d\e)}{}^{lm}
{}~-~ a^{-1} (D_{|\d}{}^{\dt\e} A ) \L_{\e) i} \right\} \cr
& ~~~~~ ~~~~~ ~~~~~ ~~~~~ ~~~~~ -~ \fracm {15} {16} \left\{
\fracm 3 {16} C_{i k l m}
\Tilde \L^{\dt\e k}f_{|\g\d|}{}^{lm} ~+~ ia^{-1} (D_{|\g}{}^{\dt\e} A
\big{)} \L_{\d| i}\right\} \L_{|\e) n} \Tilde \L_{\dt\e}{}^n \bigg] ~. \cr}
\eqno(5.6) $$

We now concentrate on the question of SD condition.  The slight
difference of the present system from the $~N=1$~ or $~N=2$~ counterpart is
that the scalar superfields $~A$~ and $~B$~ are the most fundamental
superfields.  For this reason, the natural SD condition is to put one of
these scalars to be zero.  The convenient choice is
$$B=0 ~~.
\eqno(5.7) $$
This generates other superfield equations, such as
$$\li{&\L_{\a i} = 0~~,
&(5.8{\rm a}) \cr
&f\du{\a\b}{i j} = 0~~,
&(5.8{\rm b}) \cr
&\S\du{\a\b}{\g k}= 0  ~~,
&(5.8{\rm c}) \cr
&V_{\a\b\g\d} = 0~~.
&(5.8{\rm d}) \cr } $$
through the BIds of $~d\ge 3/2$.
Eqs.~(5.7) and (5.8) are the complete set of field equations in the
system.  Eq.~(5.8b) is nothing else than the SD condition on the
central charge field
strength, (5.8c) is the $~N=4~$ analogue of the $~N=1$~ condition
$~W_{\a\b\g} = 0$~ in eq.~(3.30).  Eq.~(5.8d) implies the SD condition
 $~R\du{\ula\ulb}{\ulc\uld} = (1/2) \e\du{\ula\ulb} {\ule\ulf}
R\du{\ule\ulf}{\ulc\uld}$~ for the Riemann tensor and
the Ricci-flatness $~R_{\ula\ulb} =
0$~ because of the vanishing torsion $~T\du{\ula\ulb}\ulc = 0$~
under eq.~(5.8a).
The consistency of the SD condition (5.7) with all the Bianchi identities can
be easily confirmed by the inspection of the above constraints,
especially of the constituency relations (5.6).

\bigskip\bigskip

\noindent{\bf 6.An Apparent ~$N=4$ ``No-Go'' Barrier for SDYM}
\medskip\medskip

Before discussing the SD condition in the ~$N=4~$ SYM case, we need to
formulate the {\it non-self-dual} version of this theory in $~D=(2,2)$.

Our ~$N=4$~ SYM field content is ~$\left( A_{\ula}{}^I\, , \l_{\a i}{}^I\, ,
\Tilde{\l}_{\dt{\a}i}{}^I\, , S^{\hat{i}I}\, , T^{\hat{i}I}\right)$, where the
indices $~{\scst I,~J,~\cdots}~$ are for the adjoint representation of a gauge
group, the
indices $~{\scst i,~j,~\cdots ~=~1,~2,~3,~4}~$ are used for the ~$SO(4)~$
representations, while the indices ~${\scst \hat{i},~\hat{j},~\cdots ~
=~1,~2,~3}~$ are reserved for the 3-dimensional vector representations of the
{}~$SU(2)~$ factors in the ~$SO(4)\cong SU(2)\otimes SU(2)$. The
{}~$S^{\hat{i}}~$
 and ~$T^{\hat{i}}~$ can equivalently be represented by
the {\it self-dual}$\,:~S_{i j}=(1/2) \e_{ijkl}S_{kl}$, and
{\it anti-self-dual}$\,:~T_{i j}=-(1/2)\e_{ijkl}T_{kl}$~
tensors of ~$SO(4)$, respectively. The fields ~$S~$ and ~$T~$
are {\it real}
scalars, while the ~$\l$'s and ~$\Tilde{\l}$'s are MW spinors,
essentially in the same notation used for the ~$N=2~$ SYM case,
considered in sect.~4.

Similarly to the familiar ~$D=(1,3),\,N=4~$ SYM theory of
Ref.~[21], we impose the ~$N=4~$ superspace constraints on the
{}~$N=4~$ SYM field strengths as
$$ F_{\a i\b j}{}^I=2C_{\a\b}U_{i
j}{}^I~,~~F_{\dt{\a}i\dt{\b}j}{}^I=2C_{\dt{\a}
\dt{\b}}W_{i j}{}^I~,$$
$$F_{\a i\dt{\b}j}{}^I=0~,\eqno(6.1)$$
where the real scalar ~$N=4$~ superfields ~$U$~ and ~$W$~ have been introduced,
{}~$U^{i j}=-U^{j i}~$ and ~$W^{i j}=-W^{j i}$.

Our goal is to obtain the irreducible component spectrum containing one YM
field, so we impose the additional constraint beyond eq.~(6.1), in order to get
rid of redundant components, while maintaining the theory to be
{\it non-trivial}, namely
$$ W_{i j}={\fracm 12}\e_{ijkl}U_{kl}\equiv \breve{U}_{ij}~.
\eqno(6.2)$$
In fact, this additional constraint also goes along the lines of the ~$N=4$~
superspace formulation of the $~D=(1,3),\,N=4~$ SYM.  ({\it cf.}
Ref.~[21]).

As the consequences of the ~$N=4~$ superspace BIds, we have two equations
for the ~$U~$ superfield to satisfy:
$$\eqalign{
\nabla_{\a i}U_{j k} + \nabla_{\a j}U_{i k} & =0 ~,\cr
\Tilde{\nabla}_{\dt{\a} i}\breve{U}_{jk} + \Tilde{\nabla}_{\dt{\a} j}
\breve{U}_{ik} & =0~.}
\eqno(6.3)$$
It is important to notice that eq.~(6.3) gives {\it all} the information about
the ~$U~$ superfield which follows from the BIds and the constraints (6.1) and
(6.2). In particular, eq.~(6.3) does imply the equations of motion for the
{}~$N=4~$ SYM theory to be satisfied.

It is now an easy exercise to check that we are left with the components
$$U_{i j}~,~~\nabla\du{\a}i U_{i j}~,~~\Tilde{\nabla}_{\dt{\a}}{}^i
\breve{U}_{ij}~, ~~F_{\ula\ulb}~,
\eqno(6.4)$$
as the {\it only independent} ones. The self-dual and anti-self-dual parts of
{}~$U$~ represent the ~$S$~ and ~$T$~ above:
$$\eqalign{
U^+_{i j}\equiv {\fracm 12}\left( U_{i j} + \breve{U}_{ij}\right) & =
S_{i j}~,\cr
U^-_{i j}\equiv {\fracm 12}\left( U_{i j} - \breve{U}_{ij}\right) & = T_{i
j}~,}
\eqno(6.5)$$
the covariant spinor derivatives of ~$U$~ in eq.~(6.4) are identified with the
{}~$\l$'s and ~$\Tilde{\l}$'s,
$$F_{\a i\ulb}^I = -i\left( \s_{\ulb}\right)_{\a\Dot\b}\Tilde{\l}^{\dt{\b}}
{}_i{}^I~,~~F_{\dt{\a}i\ulb}{}^I=-i\left(\s_{\ulb}\right)_{\b\dt{\a}}\l^{\b}
{}_i{}^I~,
\eqno(6.6)$$
whereas ~$F_{\ula\ulb}~$ stands for the (non-self-dual) YM field strength:
$$\eqalign{&\nabla\du{\a}i\l_{\b j}{}^I=-{\fracm 14}\d\du j i
\left(\s^{\ulc\uld}\right)_{\a
\b}F_{\ulc\uld}{}^I + {\fracm
12}f^{I J K}C_{\a\b}\e^{ipkl}U_{kl}{}^J U_{p j}{}^K~,\cr
&\Tilde{\nabla}\du{\dt{\a}} i\Tilde{\l}_{\dt{\b}j}{}^I={\fracm 14}\d\du
j i \left(
\st^{\ulc\uld}\right)_{\dt{\a}\dt{\b}}F_{\ulc\uld}{}^I + {\fracm 12}f^{I J K}
C_{\dt{\a}\dt{\b}}\e^{ipkl}U_{kl}{}^J U_{p j}{}^K~.\cr }
\eqno(6.7) $$
Any higher number of covariant derivatives acting on ~$U~$ can be reduced,
using the results above.

In the {\it component} formulation, the $~N=4$~ supersymmetry transformation
rules in the AW space-time take the form:
$$\eqalign{&\d A_{\ula}{}^I=-i\Tilde{\e}\,\st_{\ula}\l^I
- i\e\s_{\ula}\Tilde{\l}^I~, \cr
&\d\l^I=-\fracm14\s^{\ula\ulb}\e F_{\ula\ulb}{}^I
- {\fracm 12}\a_{\hat{i}}\s^{\ula}
\Tilde{\e}\, \nabla_{\ula} S_{\hat{i}}{}^I
-{\fracm 12}\b_{\hat{i}}\s^{\ula}
\Tilde{\e}\, \nabla_{\ula} T_{\hat{i}}^I +
{\fracm i4}\e^{\hat{i}\hat{j}\hat{k}}f^{IJK}\a_{\hat{i}}\e S_{\hat{j}}{}^J
S_{\hat{k}}{}^K \cr
&~~~~~ ~~~~~ -{\fracm i4}\e^{\hat{i}\hat{j}\hat{k}}f^{IJK}\b_{\hat{i}}\e
T_{\hat{j}}{}^J
T_{\hat{k}}{}^K - {\fracm 12}f^{I J K}\a_{\hat{j}}\b_{\hat{k}}\e S_{\hat{j}}
{}^J T_{\hat{k}}{}^K~, \cr
&\d\Tilde{\l}^I=-\fracm14\st^{\ula\ulb}\Tilde{\e}F_{\ula\ulb}
- {\fracm 12}\a_{\hat{i}}
\st^{\ula}\e\nabla_{\ula} S_{\hat{i}}{}^I
+{\fracm 12}\b_{\hat{i}}\st^{\ula}\e
\nabla_{\ula} T_{\hat{i}}{}^I + {\fracm i4}
\e^{\hat{i}\hat{j}\hat{k}}f^{IJK}\a_{\hat{i}}\Tilde{\e} S_{\hat{j}}{}^J
S_{\hat{k}}{}^K \cr
& ~~~~~ ~~~~~ -{\fracm i4}\e^{\hat{i}\hat{j}\hat{k}}
f^{I J K}\b_{\hat{i}}\Tilde{\e}\,
T_{\hat{j}}{}^J T_{\hat{k}}{}^K + {\fracm 12}f^{I J K}\a_{\hat{j}}\b_{\hat{k}}
\Tilde{\e} S_{\hat{j}}{}^J T_{\hat{k}}{}^K~, \cr
&\d S_{\hat{i}I}=i\e\a_{\hat{i}}\l^I + i\Tilde{\e}\a_{\hat{i}}\Tilde{\l}^I~,
{}~~~~\d T_{\hat{i}I}=i\e\b_{\hat{i}}\l^I - i\Tilde{\e}\b_{\hat{i}}\,
\Tilde{\l}^I ~, \cr  }
\eqno(6.8) $$
where we have introduced the ~$4\times 4~$ $~SO(4)~$ gamma matrices
{}~$\a~$ and ~$\b$,
which satisfy an algebra [33]
$$\left\{ \a^{\hat{i}},\a^{\hat{j}}\right\}=2\d^{\hat{i}\hat{j}}~,~~
\[\a^{\hat{i}},\a^{\hat{j}}\]=2i\e^{\hat{i}\hat{j}\hat{k}}\a^{\hat{k}}~,$$
$$\left\{ \b^{\hat{i}},\b^{\hat{j}}\right\}=2\d^{\hat{i}\hat{j}}~,~~
\[\b^{\hat{i}},\b^{\hat{j}}\]=2i\e^{\hat{i}\hat{j}\hat{k}}\b^{\hat{k}}~,$$
$$\[ \a^{\hat{i}},\b^{\hat{j}}\] =0~.
\eqno(6.9)$$
All the  ~$\a^{\hat{i}}_{ij}$~ and ~$\b^{\hat{i}}_{ij}~$ matrices are
supposed to be antisymmetric on their ~$SO(4)~$ indices $~{\scst i}~$
and $~{\scst j}$, where ~${\scst i,~j~=~1,~2,~3,~4}$~ and $~{\scst
\hat{i},~\hat{j}~=~1,~2,~3\,}$.  The ~$SO(4)$~ indices as well as $~{\scst
\a,~\dt{\a}}~$
are implicit in eq.~(6.8).  The ~$N=4$~ supersymmetry algebra reads in
this case as
$$\[\d_1,\d_2\]=\d_t(\x^{\ula})+\d_g(\L_g)~,
\eqno(6.10)$$
where the space-time translation parameter $\x$ and the gauge parameter $\L_g$
are given by
$$\eqalign{&\x^{\ula}=i\left[ \left( \Tilde{\e}_1\st^{\ula}\e_2\right)-\left(
\Tilde{\e}_2\st^{\ula}\e_1\right)\right]~,\cr
&\L_g{}^I  = -i\left[ \left( \Tilde{\e}_1\a^{\hat{i}}\Tilde{\e}_2\right) +
\left(\e_1\a^{\hat{i}}\e_2\right)\right]S_{\hat{i}}{}^I
- i\left[ \left(\Tilde{\e}_1\b^{\hat{i}}\Tilde{\e}_2\right)
- \left(\e_1
\b^{\hat{i}}\e_2\right)\right] T_{\hat{i}}{}^I~. \cr}
\eqno(6.11)  $$
An invariant lagrangian and the equations of motion could also be written down
here, which will eventually turn out to be unnecessary in what
follows in this section.\footnotew{However, we
will give it in a different context in eq.~(7.11).}

We are now prepared enough to discuss the SD condition for this ~$N=4~$ theory.
When trying to implement it, we encounter that the ~$N=4~$ SYM theory in
$~D=(2,2)$~ does {\it not} follow the pattern developed for other
supersymmetric
theories in the previous sections. In superspace, the basic reason for this is
that the relevant superfield ~$U~$ is {\it not} chiral. Instead, both self-dual
and anti-self-dual constituents of the YM field strength are contained in the
one superfield, and there is no way to separate them consistently with the
{}~$N=4~$ supersymmetry. Any additional constraint, like the chirality or the
SD of ~$U$, makes the theory {\it trivial}, as can be
easily checked by the use
of eq.~(6.3).  The result is always negative, in the trial of
constructing the ~$N=4$~ SDYM.

To make it even more clear, let us consider the ~$D=(2,2),\,N=4~$
SYM theory by decomposing it into the ~$N=2~$
``sub-multiplets'' it contains.  The ~$N=4~$ non-self-dual SYM multiplet
consists of the ~$N=2$~ non-self-dual SYM multiplet {\it and} the
{}~$N=2~$ non-self-dual hypermultiplet. But we already know from subsect.~4.1
that the self-dual hypermultiplet does {\it not} exist. In
summary, the ~$N=4~$ SDYM does {\it not} exist because there is {\it no} real
chiral (or self-dual) ~$N=2~$ hypermultiplet. The latter does {\it not}
exist, simply because one needs both chiral {\it and} anti-chiral scalar
{}~$N=1~$ superfields to introduce the ~$N=2~$ hypermultiplet which is no
longer
 represented by the chiral ~$N=2$~ superfield.

        We finally give another reasoning for a non-existence of $~N=4$~
supersymmetric SDYM, based on the transformation rules (6.8).
Let us for simplicity think
of the Abelian case in eq.~(6.8), and try to put $\l^I=0$.  The obvious
obstruction is that the r.h.s.~of $~\d\l^I$~ is to vanish, requiring
that the $\a$ and $~\b\-$matrix terms are to vanish.  Since these two
sorts of matrices are algebraically {\it independent} of each other, unless
we kill {\it both} $~S^{\hat i I}$~ and $~T^{\hat i I}\-$fields, but we
can {\it not} satisfy such
a requirement, and thus the supersymmetry itself is truncated from $~N=4$.

The failure to construct the maximally extended ~$D=(2,2),\,N=4$~ SDYM
theory gives evidence that the ~$N\-$extended SDSGs for ~$N>4~$ should
not exist either in the same sense, namely without the inclusion of
additional propagating fields. The basic reason for this feature seems
to be the absence of
appropriate {\it chiral} ~$N\-$extended superfield strengths in their
superspace on-shell formulations. The chirality of the relevant superfields
turns out to be essential to separate the self-dual and anti-self-dual parts
of the components.
\vglue.2in

\bigskip\bigskip

\def\hati{{\hat i}} \def\hatj{{\hat j}} \def\hatk{{\hat k}}

\noindent {\bf 7.~~To Bypass No-Go Barrier}
\medskip\medskip

        The SD condition, when considered as an equation of motion,
presents a long unsolved problem.  In particular, we would like to have
an action whose variation yields
  the SD condition as an equation of motion.  As a more stringent
requirement, we might demand that such an action should contain {\it no}
additional degrees of freedom.  Under these restrictions there is {\it
no}
known action that is satisfactory, as we have seen in the previous
sections.  However, dropping
the latter condition allows one to simply use a Lagrange multiplier.
Variation of the gauge field yields an equation of propagation for the
Lagrange multiplier.  The type of action suggested by Parkes [34] and
supersymmetrized by Siegel [35] is exactly of this nature.  This can
clearly be seen even at the level of superfields.  In this section we
give the superfield formulation for $~N=1$~ and $~N=2$~ SDYM and SDSG,
and also $~N=4$~ SDYM in component formulation to bypass the no-go
barrier.

Let us start with the $~N=1$~ SDYM theory, and consider the following action
$$I{\,}_{\rm SDYM} = \int d^4 x d^2 \theta \, \L^{\a\,I} W\du\a I ~~,
\eqno(7.1) $$
where $~W_\a$~ is the usual field strength for a SYM multiplet.  In this
action $~\L_\a$~ is a chiral $~(\Tilde \nabla_{\Dot\a} \L_\b = 0)$~
lagrange multiplier.  In order to prove that this is precisely the $~N=1$~
Parkes-Siegel (PS) action, it is convenient to define the components
contained in $~\L_\a$:
$$\eqalign {&\L_\a {}^I | = \r \du\a I ~~, ~~~~ \nabla_\a\L_\b {}^I | =
(\s^{\ula\ulb})_{\a\b} G_{\ula\ulb} {}^I + C_{\a\b} \varphi^I ~~, \cr
& \nabla^\g \nabla_\g \L_\b  {}^I | = \psi _\b {}^I ~~, \cr }
\eqno(7.2a) $$
and the components of $~W_\a$~ have their usual form:
$$\eqalign{ & W_\a {}^I | = \l_\a {}^I ~~, ~~~~ \nabla_\a W_\b {}^I | =
(\s^{\ula\ulb})_{\a\b} F_{\ula\ulb} {}^I (A) + C_{\a\b} D {}^I ~~, \cr
& \nabla^\g \nabla_\g W_\a {}^I | = i (\s^\ula)_{\a\Dot\b} \nabla_\ula
\Tilde\l ^{\Dot\b\, I} ~~. \cr }
\eqno(7.2b) $$
The superfield action above then reduces to
$$\eqalign{I{\,}_{\rm SDYM} = \int d^4 x \Big[ & \half G^{\ula\ulb\, I}
(F_{\ula\ulb} {}^I (A) - \half \e\du{\ula\ulb}{\ulc\uld}
F_{\ulc\uld} {}^I (A) )
+i\r ^{\a\, I} (\s^\ula)_{\a\Dot\b} \nabla_\ula\Tilde\l^{\Dot \b\, I}
+ \varphi^I D {}^I + \psi^{\a\,I} \l_\a {}^I
\Big] ~~, \cr }
\eqno(7.3) $$
We thus clearly see the first two terms correspond to the
``propagating'' fields in the PS SDYM action.  The latter two terms only
yield trivial equations for auxiliary fields.\footnotew{It is interesting
to note that $~\L_\a$~ corresponds to a TM.  Usually in a
WZ gauge $~\r_\a=0$~ and $~\varphi$~ and $~\psi_\a$~ correspond to
physical fields.}

The existence of the self-dual $~N = 1$~ TM (3.31) - (3.35) as a solution to a
set of superspace BIds implies that the philosophy of the
PS can be applied to that theory also.  An $~N=1$~ off-shell PS formulation
is given by
$$I_{\rm SDTM}^{N=1} = \int d^4 x d^2 \theta \Tilde \L^{\Dot\a} \Tilde
D_{\Dot\a} \Phi ~~,
\eqno(7.4) $$
where $~\Tilde D_{\Dot\g} \Tilde\L^{\Dot \a} = 0 $, and $~D^2\Phi =
\Tilde D^2 \Phi=0$.  It is a simple matter to show this action imposes
eq.~(3.35) as an equation of motion.

        It is also obvious how to generalize the PS formulation
to the case of SDSG.  Namely we just make the replacements $~\L^\a
\rightarrow \L^{\a\b\g}$~ and $~W_\a \rightarrow W_{\a\b\g}$, where
$~\L^{\a\b\g}$~ is a Lagrange multiplier and $~W^{\a\b\g}$~ is
the superfield conformal supergravity field strength.  So the obvious action
for $~N=1$~ PS SDSG becomes
$$I{\,}^{N=1}_{\rm SDSG} = \int d^4 x d^2 \theta \, \varphi ^3
\L^{\a\b\g} W_{\a\b\g} ~~,
\eqno(7.5) $$
where $~\varphi^3$~ is the chiral density multiplet.  The fact that
SG fields appear only (besides the density factor) through $~W_{\a\b\g}$,
the conformal field strength, implies that this action possesses {\it
space-time superconformal symmetry!}  This means that using a component tensor
calculus approach, only superconformal
$~Q\-$variations are required to prove the superinvariance of this action.
Since
only superconformal components enter eq.~(7.5), the usual $~S~$ and $~P$~
auxiliary fields are {\it absent} from this action.  Superconformal symmetry
also tells us a lot more about eq.~(7.5).  For example, the component
fields $~\l_\a,~F_{\ula\ulb}$~ and $~D$~ are replaced respectively by
$~\psi\du{\ula\ulb}\g , ~W_{\ula\ulb\ulc\uld},~ \partial_{\[ \ula}
A_{\ulb \]} ~$ and $~ \phi\du{\ula\ulb} \g$.  These latter quantities
correspond to the gravitino field strength, the anti-self-dual part of
the Weyl tensor, the curl of the SG axial vector auxiliary field, and
finally the $~S\-$supersymmetry gauge field strength.  (Of course, we
are in a second order formulation so the spin-connection
$~\omega_{\ula\ulb\ulc}$~ and $~S\-$supersymmetry gauge field
$~\phi\du\ula \g$~ are expressed in terms of $~e\du\ula\ulm$~ and
$~\psi\du \ula \g$.)  The component content of the Lagrange multiplier
$~\L_{\a\b\g}$~ is also perfectly clear.  The corresponding components
to $~\r _\a ,~G_{\ula\ulb},~ \varphi$~ and $~\psi_\a$~ are given by
$~\r _{\a\b\g},~\L_{\ula\ulb\ulc\uld},~t_{\ula\ulb}$~ and
{}~$\s_{\a\b\g}$.  The equations of motion for these respectively yield
the vanishing gravitino field strength, the vanishing
self-dual Weyl tensor, the vanishing self-dual auxiliary axial
vector field strength and the vanishing $~S\-$supersymmetry gauge
field strength tensor.

        Once the $~N=1$~ superfield formulation of PS-type constructions
is clear, we can immediately generalize it to higher $~N$.  For example,
the $~N=2$~ SDYM PS action is just
$$I{\,}_{\rm SDYM} ^{N=2} = \int d^4 x d^4 \theta \, \L^I S^I ~~,
\eqno(7.6) $$
where $~\L^I$~ is a chiral scalar Lagrange multiplier superfield and
$~S^I$~ is the $~N=2$~ SYM vector multiplet field strength.  Similarly for
$~N=2$~ SG, we have
$$I{\,}_{\rm SDSG}^{N=2} = \int d^4 x d^4 \theta \, \e^{-1} \L^{\ula\ulb}
W_{\ula\ulb} ~~,
\eqno(7.7) $$
with $~\e^{-1}$~ denoting the $~N=2$~ chiral density multiplet,
$~W_{\ula\ulb} $~ is the $~N=2$~ conformal SG field strength tensor
[36] and $~\L^{\ula\ulb}$~ is an $~N=2$~ chiral Lagrange
multiplier.  Finally this construction works for the $~N=4$~ {\it
conformal} SG theory [37] as well:
$$I{\,}_{\rm SDSG}^{N=4} = \int d^4x d^8 \theta \, \e^{-1} \L W ~~ .
\eqno(7.8) $$

We now give the PS formulation of the $~N=4$~ supersymmetric SDYM
system in components.  This is extremely interesting, because it is {\it
not} known how to write the $~N=4~$ analog of eq.~(7.1) in superspace
consistently.  Nevertheless, by proceeding with a component
formulation, we are able to write down such a theory.
Due to the complication in the {\it off-shell}
formulation for $~N=4$, we give the {\it on-shell} results in component fields.
The field content is
$~(A\du\ula I,G\du{\ula\ulb} I, \r^I,\Tilde\l^I, S\du {\hat i}I, T\du
{\hat i}I)$,
where $~{\scst \hat i,~\hat j,~\cdots~=~1,~2,~3}$~ are the indices
for the $~\a$~ and $~\b\-$matrices exactly as in sect.~6.  We first
give the invariant action $~I\,{}^{N=4}_{\rm SDYM} = \int d^4x\,
\Lag\,{}^{N=4}_{\rm SDYM}$, where
$$ \eqalign{\Lag{\,}^{N=4}_{\rm SDYM}= \,& - \fracm 12 G^{\ula\ulb\,I}
(F\du{\ula\ulb}I- \half\e\du{\ula\ulb}{\ulc\uld} F\du{\ulc\uld} I) + \fracm 12
(\nabla_\ula S\du{\hat i} I)^2 - \fracm 12 (\nabla_\ula T\du{\hat i}I)^2
+ 2i (\r^I \s^\ula D_\ula \Tilde\l^I)  \cr
& - i f^{I J K} \left[ (\Tilde \l^I \a_{\hat i} \Tilde\l^J)
S\du{\hat i} K + (\Tilde\l^I \b_{\hat i} \Tilde\l^J) T\du{\hat i} K
\right] ~~. \cr }
\eqno(7.9) $$
Because of the peculiar coupling
between $~G$~ and $~F$, the {\it self-dual} part of $~G\du{\ula\ulb}I$
is a ``gauge'' degree of freedom.  This $~G\du{\ula\ulb}I$~ plays a similar
role to the {\it anti-self-dual} part of $~F\du{\ula\ulb}I$~ in
the {\it non-self-dual} SYM.  The supertranslation rule is
$$\eqalign{&\d A\du \ula I = - i(\e \s_\ula \Tilde\l^I) ~~, \cr
&\d G_{\ula\ulb} {}^I = i (\Tilde\e\, \Tilde\s_{\[\ula} \nabla_{\ulb\]}
\r^I) -  i \fracm 12 f^{I J K} \left[ (\e \a_\hati\s_{\ula\ulb}\r^J)
A\du\hati K - (\e \b_\hati \s_{\ula\ulb} \r^J) B\du \hati K \right] ~~, \cr
&\d \r^I = - \fracm 14 \s^{\ula\ulb} \e \, G\du{\ula\ulb} I
- \half \a_{\hat i} \s^\ula \Tilde\e\, \nabla_\ula S\du\hati I
- \half \b_{\hat i} \s^\ula \Tilde \e \, \nabla_\ula T_{\hat i}{}^I \cr
& ~~~~~ ~~~ + i \fracm 14 \e^{\hat i\hat j\hat k} \a_\hati \e\, f^{I J K}
S\du{\hat j} J S\du \hatk K -i \fracm 14 \e^{\hat i\hat j\hat k}
\b_{\hat i} \e\, f^{I J K} T\du{\hat j} J T\du{\hat k} K
- \half f^{I J K} \a_{\hat j} \b_{\hat k} \e\, S\du{\hat j} J T\du{\hat
k} K ~~, \cr
&\d \Tilde \l^I = - \fracm 14 \Tilde\s^{\ula\ulb}\, \Tilde\e\,
F\du{\ula\ulb} I -
\fracm 12 \a_{\hat i} \Tilde\s^\ula \e\, \nabla_\ula S\du{\hat i}I + \half
\b_{\hat i} \Tilde\s^\ula \e\, \nabla_\ula T\du{\hat i}I ~~, \cr
&\d S\du{\hat i} I = i (\e\a_{\hat i} \r^I) + i (\Tilde\e \a_{\hat i}
\Tilde\l^I) ~~, ~~~~
\d T\du{\hat i} I = i(\e \b_{\hat i} \r^I) - i (\Tilde\e \b_{\hat i}
\Tilde\l^I) ~~. \cr }
\eqno(7.10) $$

        To compare the $~N=4$~ SDYM (7.9) and (7.10)
with the usual $~N=4$~ {\it non-self-dual} SYM,
we first give the lagrangian of the latter:
$$\eqalign{\Lag{\,}^{N=4}_{\rm SYM} = &\, - \fracm 14
(F\du{\ula\ulb} I )^2 +
\half (\nabla_\ula S\du\hati I)^2 - \half (\nabla_\ula T\du\hati I)^2 + 2i
(\l^I \s^\ula\nabla_\ula \Tilde\l^I) \cr
& - i f^{I J K} \left[ (\l^I \a_\hati \l^J) + (\Tilde\l^I \a_\hati \Tilde\l^J)
\right] S \du \hati K
+ i f^{I J K} \left[ (\l^I \b_\hati \l^J) - (\Tilde\l^I \b_\hati \Tilde\l^J)
\right] T\du \hati K \cr
& - \fracm 14 (f^{I J K} S\du\hatj J S\du \hatk K)^2
- \fracm 14 (f^{I J K} T\du\hatj J T\du\hatk K)^2
+ \half (f^{I J K} S\du\hatj J T\du\hatk K)^2 ~~, \cr }
\eqno(7.11) $$
which is invariant under supersymmetry (6.8).
The negative sign for the $~T\-$kinetic term in eq.~(7.11) is
peculiar to our $~D=(2,2)$.

The most important property of
the lagrangian (7.9) is the {\it absence} of quartic terms in scalar fields.
Furthermore, the Yukawa interactions of the $~\r\r S$~ or $~\r\r T\-$type
are also absent there, while $~\l\l S$~ or $~ \l\l T\-$terms are present
in eq.~(7.11).  The transformation law for $~G\du{\ula\ulb}I$~ in eq.~(7.10)
is very similar to that of the {\it anti-self-dual} part of $~F\du{\ula\ulb}I$~
derivable from eq.~(6.8) except for the appearance of the bilinear terms.
The $~A\du \ula I$~
transformation law in eq.~(7.10) lacks the $~\Tilde\e\, \Tilde\s_\ula
\r\-$term, which is expected from eq.~(6.8).  Another
important difference is the absence of the bilinear bosonic terms for
$~\d\r$~ in eq.~(7.10).  The absence of some Yukawa terms and
the quartic terms in eq.~(7.9) are closely related to these properties.  Since
$~\r\du i I$~ is a superpartner of multiplier $~G\du{\ula\ulb} I$,
there is natural {\it asymmetry} between $~\r$~ and $~\Tilde\l$~ in
eq.~(7.10).

The {\it on-shell} closure of supersymmetry in eq.~(7.10) is easily confirmed,
especially by the use of the $~A_\ula$~ and $~G~\-$field equations.
Even though such a
check is parallel to the {\it non-self-dual} SYM case, there appear to be
important differences, which result in the absent terms as well as the new
ones in the transformation rule (7.10).

We point out the advantages as well as the drawbacks of the PS formulation
for supersymmetric SDYM.  The no-go barrier has been bypassed by
the introduction of the additional propagating fields $~G\du{\ula\ulb}I$~
and $~\r\du iI$.  This has some
advantage, especially when we are interested in the lower-dimensional
integrable systems with such strong supersymmetry as $~N=4$.
However,
the price to be paid is the inclusion of additional modes even for
spin-one helicities out of $~G\du{\ula\ulb}I$~ increasing the field
content in lower-dimensions after dimensional reductions.  This may be
a {\it disadvantage}, when the tight-field content under the
SD condition plays an important role for integrability in lower-dimensions.
Actually a similar situation is known in the context of $~D=10$~
Type IIA and Type IIB superstrings, both of which give the
{\it same} zero-mass spectrum in lower-dimensions after the simple
dimensional reduction.  The formulation in Ref.~[35] seems similar to Type IIB
superstring construction, due to its {\it chiral} features.

We should also point out that there is another more radical way which
seems to hold the promise of overcoming the barrier.  Although we will
not discuss this other method in detail, it also holds the promise of
providing very powerful new insights into the long unsolved problem
of off-shell $~D=4,\,N=4$~ SYM theory.  Let us return to the first
assumptions (6.1) in deriving the no-go barrier.  It is well known that
for the $~N=4$~ case (unlike any lower SYM theory) that eq.~(6.1) actually
implies the YM equation of motion for the gauge field in the multiplet.
{}From this viewpoint, therefore, it is now suprising that the SD condition is
{\it inconsistent} with the starting point in eq.~(6.1).  We may ask if it is
possible to somehow change this starting point.  The answer turns out to
be in the affirmative.  Instead we may write
$$ \eqalign{&F_{\a i\b j}{}^I=2C_{\a\b}U_{i j}{}^I + f_{(\a \b)~ (i j)}{}^I
{}~~,~~~F_{\Dot{\a}i\Dot{\b}j}{}^I=2C_{\Dot{\a}
\Dot{\b}}W_{i j}{}^I + {\Tilde f}_{(\Dot{\a} \Dot{\b})~ (i j)}{}^I ~~,
\cr
&F_{\a i\Dot{\b}j}{}^I=0~~. \cr }
\eqno(7.12)$$
where ~$f_{(\a \b)~ (i j)}{}^I~$ and $~{\Tilde f}_{(\Dot{\a} \Dot{\b})~
(i j)}{}^I ~$ are independent auxiliary superfields that have
not been seen previously.  When these superfields are present, the usual
YM equations of motion are no longer imposed on the gauge
field in the multiplet.  A rather suggestive way to impose an asymmetry
between ~$U_{i j}{}^I$~ and ~$W_{i j}{}^I~$ is to impose the vanishing of
only one of the auxiliary superfields (say ~${\Tilde f}_{(\Dot{\a} \Dot{\b})~
(i j)}{}^I = 0$).  This might achieve the goal of making the $~N=4$~ case more
similar to the lower ~$N~$ cases.  Note for example that for lower ~$N$~
values in the PS formulation, the Lagrange multiplier always comes from a
Lagrange multiplier superfield that is different from the gauge
supermultiplet upon which the SD condition is imposed.  Note that this
is contrary to the component level $~N=4$~ theory presented above.

In closing, we note that it has been suggested by Siegel [35,38] that
the space-time supersymmetric $~N=4$~ versions of the theories above
are closely linked to the $~N=2$~ and/or $~N=4$~
world-sheet supersymmetric string theories.  In particular, in Ref.~[35]
even an $~N=8$~ SDSG version was suggested in a way to be related to
strings.  Should this prove to be
the case, we will be in the marvelous position of being able to find an
$~N=8$~ {\it superconformal multiplet} for the first time!
Even in the case of the $~N=4$~ theory, study of the open strings should
yield new insight into the structure of $~N=4$~ SYM theory.  This
promises to be an area that begs further exploration.

\bigskip\bigskip

\noindent{\bf 8.~~Supersymmetric SDYM Zero Modes}
\medskip\medskip

The Euclidean formulation of a quantum field theory is well-known to be not
just a matter of convenience but an appropriate framework to define the Green's
function generating functional of the theory. Another advantage of the
Euclidean formulation is its relation with topology, which manifests itself in
the existence of the {\it instantons} -- classical solutions of the field
equations with  non-trivial topology.

When trying to connect the AW space-time formulation of the quantum field
theory to the Euclidean one by a {\it double} Wick rotation, we immediately
encounter the obstruction since, strictly speaking, this continuation can only
be performed with the {\it complex} fields. The reason is that the reality
conditions are sensitive to a choice of the space-time signature: there are
both Majorana and MW spinors in ~$2 + 2~$ dimensions, but there are
no MW spinors in ~$1 + 3~$ dimensions and there are no Majorana spinors
at all in the Euclidean
space-time. Restricting ourselves to the case of the ~$N\-$extended
supersymmetric SDYM theories for simplicity, we choose to consider the \nt
supersymmetric SDYM model first. That theory can be defined in the AW
space-time in terms of the complex chiral spinors, and they can be rotated to
the Euclidean space-time easily.\footnotew{The relevance of the \nt
supersymmetric  YM theory in this context has been noticed in Ref.~[39].}

Given the Euclidean version of the \nt ~SDYM theory, we can take advantage of
the existing results about the YM instantons [1,40] and their quantum effects
in supersymmetric gauge theories [41]. The well-known Euclidean BPSTH-instanton
solution for the YM vector field,
$$A_{\ula}{}^I(x) = \fracm 2 g\,\h^I_{\ula\ulb}\,{(x-x_0)^{\ulb}\over (x-x_0)^2
+ \r^2}~,
\eqno(8.1)$$
and the associated SDYM field strength,
$$F_{\ula\ulb}{}^I(x)=-\fracm 4 g\,\h^I_{\ula\ulb}\,{\r^2\over \left[ (x-x_0)^2
+ \r^2\right]^2 }~,
\eqno(8.2) $$
are parametrized by the center-of-instanton coordinate ~$(x_0)_{\ula}~$ and the
instanton size (the scale of the instanton) ~$\r$. In eqs.~(8.1) and (8.2) the
{}~$\h^I_{\ula\ulb}~$ represents the constants known as the 't Hooft symbols
[40],\footnotew{In fact, one amusing point to note is that the $~\a$~
and $~\b$~ matrices of Ref.~[33] (see sect.~7) are exactly the same as the
$~\eta$~ and
$~\Bar\eta$-symbols of Ref.~[40].}~whereas the ~$g~$ is the gauge
coupling constant. We restrict ourselves to the
case of the ~$SU(2)~$ gauge group as usual, the extension to a general gauge
group ~$G~$ could, in principle, be done in a Chevalley basis for ~$G$.

The self-dual solution (8.1) can be interpreted as the vector zero mode, which
is accompanied in the \nt  supersymmetric SDYM case by the Dirac zero
modes (see subsect.~4.2 for our notation):

$$\eqalign{
\left(\Tilde{\l}_{\rm s s}\right)_{\dt{\a}}{}^{iJ}(x) & = g\left(\st_{\ula\ulb}
\right)_{\dt{\a}}{}^{\dt{\b}}F^{\ula\ulb\,J}(x)\,\Tilde{\a}_{\dt\b}{}^i~,\cr
\left(\Tilde{\l}_{\rm s c}\right)_{\dt{\a}}{}^{iJ}(x) & = g\left(\st_{\ula\ulb}
\right)_{\dt{\a}}{}^{\dt{\b}}F^{\ula\ulb\,J}(x)\,
i\left(\st_{\ulc}\right)_{\dt{\b}}
{}^{\g}\left(x - x_0\right)^{\ulc}\b_{\g}{}^i ~.}
\eqno(8.3)$$

They do satisfy the Dirac equation, provided the YM field strength ~$F~$ is
{\it self-dual}. The appearance of the ~$\Tilde{\l}_{\rm s s}~$ is due to
supersymmetry,
while the existence of another zero mode ~$\Tilde{\l}_{\rm s c}~$ is due to
superconformal symmetry of the theory [41]. In eq.~(8.3) the ~$\Tilde{\a}~$ and
{}~$\b~$ represent anti-chiral and chiral Grassmannian constants respectively,
which parametrize the solutions of the supersymmetric equations of motion. As
was noted in Ref.~[41], the ~$(x_0,\Tilde{\a})~$ form the chiral superspace
coordinates, while the collective coordinates ~$(\r,\b)~$ are united into a
superfield parameter. Without adding new propagating fields, some kind of
supersymmetry among the solutions survives by allowing the instanton size
{}~$\r$, as well as the ~$\b$, to transform [41]. The unusual (and incomplete)
supersymmetry of Ref.~[41] can be recast into the usual \nt  supersymmetry if
the theory is extended by the inclusion of a complex scalar, i.e. within the
\nt supersymmetric YM theory constrained by the supersymmetric SD
condition similar to that used in subsect.~4.2  but now in the Euclidean space.
The last observation has first been made by Zumino [39], and we are not going
to repeat here his presentation.

However, this is not the end of the story. Given the  \nt  SDYM spin-1 and
spin-1/2 zero modes, there should be in addition, the spin-0 or {\it scalar}
zero modes, which are supposed to satisfy the equation of motion presented in
the last line of eq.~(4.6) to be transformed to the Euclidean space. To this
end, we are going to find the \nt  SDYM scalar zero modes explicitly.

The scalar equation of motion in the \nt  SDYM theory with the ~$SU(2)~$ gauge
group takes the form
$$\eqalign{\pa^{\ula}\pa_{\ula}T^I + g\e^{IJK} & \left(
\pa^{\ula}A_{\ula}{}^J\right)T^K +
2g\e^{IJK}A^{\ula\,J}\pa_{\ula}T^K  \cr
&+g^2\e^{ILM}\e^{M N P}\left(A_{\ula}{}^L A^{\ula\,N}\right)T^P
= \e^{IJK} \left( \Tilde{\l}^{iJ}\Tilde{\l}_i^K\right)~, \cr }
\eqno(8.4)$$
where ~$\e^{IJK}~$ are the ~$SU(2)~$ structure constants; ~${\scst
I,~J,~K~=~1,~2,~3}$. Using
the explicit solutions for the fields ~$A_{\ula}{}^I(x)~$ and
{}~$\Tilde{\l}_{\dt{\a}}{}^{iJ}(x)~$ given above in eqs.~(8.1), (8.2)
and (8.3), and the identities for the ~$\h\-$symbols:
$$\eqalign{&\h^I_{\ula\ulb} =
\half\e_{\ula\ulb\ulc\uld}\h^I_{\ulc\uld}~,\cr
&\h^I_{\ula\ulb}\h^J_{\ula\ulc} = \d^{IJ}\eta_{\ulb\ulc} +\e^{IJK}
\h^K_{\ulb\ulc}~, \cr
&\e^{IJK}\h^J_{\ula\ulb}\h^K_{\ulc\uld}=\eta_{\ula\ulc}\h^I_{\ulb\uld} -
\eta_{\ula\uld}\h^I_{\ulb\ulc} - \eta_{\ulb\ulc}\h^I_{\ula\uld}
+ \eta_{\ulb\uld} \h^I_{\ula\ulc}~, \cr}
\eqno(8.5) $$
and for the ~$\s\-$matrices:
$$\eqalign{&\st_{\ula\ulb} = \half
\e_{\ula\ulb\ulc\uld}\st_{\ulc\uld}~,\cr
&\st_{\ula\ulb}\st_{\ulc\uld}=- \e_{\ula\ulb\ulc\uld} +
\half \left(\eta_{\ula\uld}\st_{\ulb\ulc} + \eta_{\ulb\uld}
\st_{\ulc\ula} + \eta_{\ula\ulc}\st_{\uld\ulb} + \eta_{\ulb\ulc}
\st_{\ula\uld}\right)~, \cr
&\s^{\ulc}\st^{\ula\ulb}= \half \left(\eta^{\ulc\ula}\s^{\ulb} -
\eta^{\ulc\ulb}\s^{\ula} + \e^{\ula\ulb\ulc\uld}\s^{\uld}\right)~, \cr}
\eqno(8.6) $$
allows us to considerably simplify eq.~(8.4) to the form
$$\pa^{\ula}\pa_{\ula}T^I(x) - {8(x-x_0)^2\over \left[ (x-x_0)^2 +
\r^2\right]^2}
T^I(x) + {4\r^4\over \left[ (x-x_0)^2 + \r^2\right]^4}q^I(x) = 0~,
\eqno(8.7)$$
where we have introduced the notation
$$q^I(x) = \left\{ \eqalign{ q^I_{\rm s s}(x) & =
16\h^I_{\ula\ulb}\left( \Tilde{\a}^i\st^{\ula\ulb}\,\Tilde{\a}_i\right)
\equiv q^I ~, \cr
q^I_{\rm s c}(x) & = -16\h^I_{\ula\ulb}\left( \b^i\s_{\ulc}\st^{\ula\ulb}\,
\Tilde{\s}_{\uld}\,\b_i\right)(x-x_0)^{\ulc}(x-x_0)^{\uld}  \cr
{} & \equiv p^I_{(\ulc\uld )}(x-x_0)^{\ulc}(x-x_0)^{\uld}~.} \right.
\eqno(8.8)$$
In eq.~(8.8) the ~$q^I~$ and ~$p^I_{\ulc\uld}~$ are the covariant
nilpotent constants.  In particular, the ~$p_{\ula\ulb}~$ is traceless,
$$p^I_{\ula\ula}=0~,
\eqno(8.9)$$
because of the identity
$$\s^{\ulc}\,\st^{\ula\ulb}\,\st_{\ulc}=0~.
\eqno(8.10)$$

We now substitute
$$T^I_{\rm s s}(x) = t_{\rm s s}(y)q^I~,~~~y=(x-x_0)^2~,
\eqno(8.11)$$
in the case of the {\it supersymmetric} scalar zero mode ~$T_{\rm s s}(x)$,
which
results in the ordinary differential equation\footnotew{In our notation here,
prime always means a differentiation, e.g. ~$t^{\prime}(y) \equiv d t/d y
{}~,~~t^{\prime\prime}(y) \equiv d^2 t/d y^2~,~~$etc.}
$$y t^{\prime\prime}_{\rm s s} + 2t^{\prime}_{\rm s s} - {2y\over \left(y +
\r^2
\right)^2}t_{\rm s s} + {\r^4\over \left( y + \r^2\right)^4}=0~.
\eqno(8.12)$$
After the substitution
$$t_{\rm s s}(y) = {\r^4f_{\rm s s}(y)\over \left(y + \r^2\right)^2}
\eqno(8.13)$$
eq.~(8.12) takes the form
$$y(y + \r^2)^2f^{\prime\prime}_{\rm s s} + 2(y + \r^2)(\r^2 -
y)f^{\prime}_{\rm
 s s} -4\r^2 f_{\rm s s} + 1 =0~.
\eqno(8.14)$$
This equation has a {\it constant} particular solution,
$$f_{\rm s s} = {1\over 4\r^2}~,
\eqno(8.15)$$
while the two independent solutions of the associated homogeneous equation
{}~$(q=0)~$ take the form
$$\eqalign{&f_1(y) = {1\over 3\r^6}y^3 + {4\over 3\r^4}y^2
+ {2\over \r^2}y + 1~,\cr
& f_2(y) = {\r^2\over y} + 1 ~. \cr}
\eqno(8.16) $$
Both solutions in eq.~(8.16) seem to be irrelevant to the instantons, since the
scalar field ~$T^I(x)~$ defined by eqs.~(8.11) and (8.13) either goes to
infinity
when ~$x\rightarrow \infty~$ in the case of the ~$f_1(y)~$ to be added to the
solution (8.15), or has a singularity at ~$y=0~$ in the case of the added
{}~$f_2(y)$.

Therefore, we conclude that the correct scalar zero mode corresponding by
supersymmetry to ~$A^I_{\ula}(x)~$ and ~$\Tilde{\l}_{\rm s s}^{iI}(x)$, takes
the form
$$T^I_{\rm s s}(x) = {4\r^2\over \left[ (x-x_0)^2 + \r^2\right]^2}\,
\h^I_{\ula\ulb}\left( \Tilde{\a}^i\st^{\ula\ulb}\,\Tilde{\a}_i\right)=\left(
\Tilde{\a}_i\,\Tilde{\l}_{\rm s s}^{iI}\right)~.
\eqno(8.17)$$
This solution is to be expected since in the second equality of eq.~(8.17) we
just get the \nt  supersymmetry transformation law of the \nt  SDYM theory.

Let us turn now to the {\it superconformal} zero mode ~$T_{\rm s c}(x)~$ and
substitute
$$T^I_{\rm s c}(x) = {f_{\rm s c}(z)\over \r^2 z(z-1)^3}p^I_{\ula\ulb}
(x-x_0)^{\ula}
(x-x_0)^{\ulb}~,
\eqno(8.18)$$
where we have introduced the new argument, $z$,
$$ z = \r^{-2}y + 1 \equiv {(x-x_0)^2\over \r^2} + 1 ~.\eqno(8.19)$$

The substitution (8.18) reduces the partial differential equation (8.7)
to a second-order ordinary differential equation, because of the use of
eq.~(8.9). It leads to the equation of the {\it Fuchsian
type} [42] and, ultimately, to the {\it hypergeometric equation}
of Gauss.\footnotew{We appreciate discussions with G.~Gilbert, who
pointed out this fact.}~~The equation for the ~$f_{\rm s c}(z)$~
takes the form
$$ f_{\rm s c}^{\prime\prime} - 2\left( {1\over z-1} + {1\over z}\right)
f_{\rm s c}^{\prime} + {2\over z(z-1)}f_{\rm s c} +{(z-1)^2\over z^3}
=0~.
\eqno(8.20)$$

The general solution of eq.~(8.20) reads
$$f_{\rm s c}(z) = - F(z)\int^z {z_1^2(z_1 - 1)^2\over F^2(z_1)}d z_1 \,
\int^{z_1} {F(z_2)
\over z_2^5} \,d z_2~~
\eqno(8.21)$$
in terms of the solution ~$F(z)$~ to the associated homogeneous
equation, which is just the standard {\it hypergeometric function}
{}~$F(\a,\b,\g;z)~$ [43] with the parameters ~$\a$, $\b~$ and ~$\g~$ to be
determined from the
coefficients of eq.~(8.20) according to the rule [43]
$$ \a + \b =-5~,~~~\a\b=2~,~~~\g=-2~.
\eqno(8.22)$$
Since the ~$\g~$ is a negative integer, the solution to the homogeneous
hypergeometric equation is in fact
[43]
$$F(z)={F(\a,\b,-2;z)\over \G(-2)}= \fracm16\,\a(\a +1)(\a + 2)\b(\b
+ 1)(\b + 2) z^3 F(\a + 3,\b + 3,4;z)~,
\eqno(8.23)$$
where
$$\a = {-5 -\sqrt{17}\over 2}~,~~~\b = {-5 +\sqrt{17}\over 2}~.
\eqno(8.24)$$

The behaviour of the solution ~$T_{\rm s c}(x)~$ at
the infinity ~$x\rightarrow \infty~$ is given by
$$ T^I_{\rm s c}(x)\rightarrow {\r^4\over
2(x^2)^3}p^I_{\ula\ulb}x^{\ula}x^{\ulb}~.
\eqno(8.25)$$

Near the instanton center, where ~$x\rightarrow x_0$~, we have
$$ T^I_{\rm s c}(x)\rightarrow\; -{(x-x_0)^2\over 4\r^4}p^I_{\ula\ulb}
(x-x_0)^{\ula}(x-x_0)^{\ulb}~.
\eqno(8.26)$$

Eq.~(8.25) guarantees the finiteness of the Euclidean action for the scalar
superconformal zero modes, whereas the vanishing of the scalar solution at
{}~$x_0=0~$ in eq.~(8.26) clearly matches the similar property of the
corresponding Dirac zero mode in the second line of eq.~(8.3).\footnotew{The
vanishing of the solutions at ~$x_0=0~$ is due to the fact that we are in fact
talking about the {\it orbital} part of the superconformal transformations
in question. The spin part of them is zero for the scalars. The superconformal
symmetry among the solutions is trivially restored at
{}~$x\rightarrow\infty$.}~~The exact solution in eq.~(8.21) is
supposed to be interpolating between the
two asymptotic solutions in eqs.~(8.25) and (8.26). It has been known for
some time that the ``standard'' scalar equation of motion in the instanton
background, ~$D^{\ula}D_{\ula}\F (x)=0$, does {\it not} have well-behaved
asymptotically flat Euclidean solutions, and eq.~(8.23) is in accordance with
that fact. It is the {\it Yukawa coupling} in the supersymmetric gauge theory
that is responsible for the drastic improvement of the behaviour of the scalar
solution both at the infinity and at the center of the instanton, as indicated
by eqs.~(8.25) and (8.26). Contrary to the supersymmetric scalar zero mode in
eq.~(8.17), the superconformal scalar zero mode {\it can not} be represented as
the superconformal transformation of the superconformal Dirac zero mode
because, though the \nt  SDYM equations of motion are superconformally
invariant, their instanton solutions are {\it not}.\footnotew{The
superconformal invariance may be recovered in the quantum theory after
integrating over ~$\r~$ and ~$\b~$ [41].}

There seem to be no obstructions in a continuation of the Euclidean solutions
to the AW space-time.\footnotew{Of course, the relation with topology becomes
much more obscure, if any, in the AW space-time compared to the Euclidean
space, and there is no well-defined value for the scalar action in ~$2 + 2~$
dimensions any more.}~~In many respects, there are analogues between various
objects in ~$D=(4,0)~$ and ~$D=(2,2)~$ dimensions due to a similarity of the
``Lorentz'' group structure in both cases:
$$SO(4)\cong SU(2)\otimes SU(2)~,$$
$$SO(2,2)\cong SU(1,1)\otimes SU(1,1)\cong Sp(2)\otimes Sp(2)~.
\eqno(8.27)$$
In particular, the 't Hooft's ~$\h\-$symbols in both cases are nothing but the
generators of the self-dual subgroups ~$SU(2)~$ or ~$Sp(2)$, respectively,
$$J^I_{\rm SD}= \frac{1}{4}\h^I_{\ula\ulb}M^{\ula\ulb}~,~~~J^I_{\rm ASD}=
\frac{1}{4}\Bar{\h}^I_{\ula\ulb}M^{\ula\ulb}~,
\eqno(8.28)$$
where we have introduced the ``Lorentz'' generators ~$M^{\ula\ulb}~$ and the
generators of the self-dual ~$(J^I_{\rm SD})~$ and anti-self-dual ~$(J^I_{\rm
ASD})~$ subgroups.

Finally, it is worthwhile to take a look at the $~N=4$~ supersymmetric SDYM
theory
which exists in the PS formulation. The Euclidean version of that theory
could easily be read off from the results of sect.~7. First, let
us look at the equations of motion which follow from the lagrangian
(7.9) for the scalars ~$S_\hati{}^I(x)~$ and ~$T_\hati{}^I(x)$, and the spinors
{}~$\r_j{}^I(x)~$ and ~$\Tilde{\l}_j{}^I(x)$. The search for their solutions
associated with the YM instanton leads to the same equations as were considered
above in this section, after taking into account rather obvious redefinitions
of the covariant nilpotent constants ~$q^I~$ and ~$p^I_{\ula\ulb}$. The
appearance of the additional internal symmetry indices does not play any
essential role here, and the dynamical factors for the scalar and spinor
instanton solutions uncovered in this section remain intact. The only novel
feature is due to the presence of the additional propagating field
$G_{\ula\ulb}(x)$ whose equation of motion resembles the YM case:
$$D^{\ula}G^{-}_{\ula\ulb} \equiv \half D^{\ula}\left( G_{\ula\ulb} -
\frac{1}{2}\e_{\ula\ulb\ulc\uld}G^{\ulc\uld}\right)=0~.
\eqno(8.29)$$
Eq.~(8.29) simply states that the solution for the ~$G^{-}_{\ula\ulb}~$
should be represented by the anti-self-dual field:
$$\left(G^-\right)^I_{\ula\ulb}=const\,\cdot \Bar{\h}^I_{\ula\ulb}\,{\r^2\over
\left[ (x-x_0)^2 + \r^2\right]^2}~,
\eqno(8.30)$$
where the symbols ~$\Bar{\h}^I_{\ula\ulb}~$ have also been introduced by
't Hooft [40].

\bigskip\bigskip

\noindent{\bf 9.~~Concluding Remarks}
\medskip\medskip

The relevance of a $~D=4$~ space-time with the ~$(2,2)~$ signature
(the AW space-time) appears to be based on the assumption that
the underlying ``master'' theory of all exactly solvable (integrable) models
generated by the SDYM theory in $~D=(2,2)$~ might be just the $~N=2~$
 heterotic string. The search for the underlying theory of integrable systems
is conceptually justified by noticing that it could explain the origin of
hidden symmetries in a variety of physically interesting examples of such
systems. Space-time supersymmetry by itself is a good motivation to
introduce the supersymmetric SDYM and SDSG theories, which have a good chance
to be the generating theories for all {\it supersymmetric} integrable models
in lower dimensions.  Just another motivation comes from the anticipated
space-time supersymmetry in the $~N=2$~ superstring theory [38].
The explicit superspace constructions given above prove
the consistency between SD and (extended) supersymmetry in $~D=(2,2)$.
The indefinite ~$D=(2,2)~$ space-time signature and the existence of
MW spinors in
this space-time were crucial in all of the constructions. The real difference
between the superspace formulations of various supersymmetric gauge theories
and supergravities with extended supersymmetry in ~$D=(1,3)~$ and
$~D=(2,2)~$ stems
from the {\it reality conditions} imposed on the spinor derivatives, the
superspace torsion and curvature components with spinor indices. The maximally
extended $~N=4~$ SYM and the $~N>4~$ SG appear to be incompatible with the SD
condition without the inclusion of new propagating fields.
The reason for this is the apparent importance of the MW property
for the relevant superfield strengths, which is only possible for $~N\leq 1~$
in the case of SM, $~N\leq 2~$ in the case of SYM and
$~N\leq 4~$ in the case of SG.\footnotew{The $~N=2$~ {\it twisted} SYM theory
in the {\it Euclidean} space-time is relevant for the
{}~$D=4$~ Donaldson theory [44] describing topology of low-dimensional
manifolds,
as advocated by Witten [16]. However, there seem to be no reason to link our
construction of the ~$N=2~$ SDYM theory directly with the Witten's ~$N=2~$
SYM theory, since they are very different.  Nevertheless, there might be a
possibility to exploit the {\it topological} information provided by the
{}~$N=2~$ SDYM theory to get some insights into the Witten's topological
$~N=2$~ SYM theory.}

Both SDYM and SDG equations of motion are  non-linear partial
differential equations which allow an infinite number of conservation laws.
The supersymmetry does not violate the basic reason of integrability of the
SDYM equations, namely their interpretation as the {\it zero curvature
conditions} in the Hamiltonian formulation, which makes it possible to apply
the inverse scattering method for their integration [45]. This is simply
because the SD condition on the YM field strength is {\it not} modified in
(extended) supersymmetry, but just accompanied by the associated equations on
the other components of the self-dual supermultiplet under consideration. This
 is particularly amusing in extended SDSGs, where both scalars and vectors are
added to the graviton field, and the SDG condition {\it implies} the
SDYM condition. As was argued by Mason [46] (see also Ref.~[47]), the
dimensional reduction of the SDYM equations with two null vectors and the
gauge group $~SL(\infty )~$ describes the SD Einstein equations in the
so-called Plebanski [48] form,\footnotew{The Plebanski equation is nothing but
the constancy condition on the metric determinant, which
ensures the Ricci-flat condition in the case of K\"ahler geometry. It is the
theorem in $~D=4$~ [23] that a Ricci-flat 4-manifold is a hyper-K\"ahler
one when
the curvature is self-dual.} re-discovered by Penrose [49] in the context of
general relativity, by Park [50] in the context of ~$W\,$-algebras, and by
Ooguri and Vafa [3] in the context of
strings. Therefore, from the viewpoint of integrability, the SDG
equations may not be considered as independent.

Another understanding of integrability of SDYM and SDG equations
in $~D=4$~ is based on their equivalence to the {\it two-dimensional}
non-linear sigma-models with a pure Wess-Zumino term in their actions
[50].
This observation allows to connect four-dimensional self-dual theories with
two-dimensional conformal field theories. In particular [50], the twistor
construction of the $~D=4$~ gravitational instantons given by Penrose
[49], can be identified with the inverse scattering
approach to the relevant $~D=2$~ non-linear sigma-model. It has been
known for a while [51], that the moduli space of $~D=4$~ YM instantons and
that of $~D=2$~ instantons valued in the loop group, are equivalent.
The important issue in this respect is the relevant gauge symmetry. As
was shown in Ref.~[50], in the case of SDG the symplectic diffeomorphisms of a
 $~D=2$~ surface act as a gauge group, while in the case of SDYM one has a
specific infinite-dimensional gauge group too, which generalizes the Kac-Moody
 algebra as well as the universal $~W_{\infty}\-$algebra. This conclusion
agrees with the statement made by Mason [46], that the transformation group
$~SL(\infty)~$ among the solutions of the SD equation is a loop group of
area-preserving diffeomorphisms of a $~D=2$~ surface.
It indicates an existence of some general relation between the SD and
the $~W_{\infty}~$-symmetries which can be interpreted as the special
area-preserving diffeomorphisms [52]. Our construction of the $~N=4$~ SDSG
gives the first evidence for the existence of a $~N=4~$ extended ~$W_{\infty
}\-$algebra.

The obvious application of our construction is the {\it reduction} of the
supersymmetric self-dual systems from ~$D=4$~ to $~D=2$, the work of
which is in progress now.  The reduction means imposing an appropriate
ansatz for
the gauge connection after introducing two Killing symmetries. As for the SDYM
is concerned, the use of the compact Lie group and Euclidean signature yields
the $~D=2$ ~Toda field theory [53], while the use of non-compact
$~SL(2)~$ group
and the $~(2,2)~$ signature gives rise to the KdV equation [54] (see also
Ref.~[55] as for the further generalizations of this construction). Therefore,
 one now has a natural
way to connect the $~D=2$~ super-Toda field equations with the $~D=(2,2)$~
supersymmetric SDYM, and supersymmetrize the KdV equation as well as the
well-known KdV flows [56].  It
has recently been conjectured [55], that SDG in the AW space-time
provides a universal integrable system of equations, which yields {\it all}
the KdV-flows by truncation, so that it is just enough to supersymmetrize
the reduction of the $~D=4$~ SDSG in order to get the super-KdV flows.

In Appendix C we argue that the $~N=2$~ superstring is
finite to all orders in string loops.  Though there is now some evidence
[35,38] that the $~N=2$~ as
well as $~N=4$~ superstrings do possess space-time supersymmetry,
a precise connection between the ~$N=2$~ superstrings and the space-time
supersymmetry is yet to be elaborated.  It is well-known that
the Green-Schwarz superstrings, when coupled to SYM and/or SG backgrounds, do
require them to satisfy their equations of motion [57]. Technically, it
originates from requiring the Siegel's invariance of the coupled
Green-Schwarz action,
which is crucial for the consistency of the theory. Taking the supersymmetric
space-time background on-shell is just enough for the consistency of the
theory, but it is the non-trivial and important observation that the SD
condition in addition does {\it not} violate that consistency [17].

        In Appendix D, we indicate an interesting possibility that a
SG plus SYM theory in $~D=(5,5)$~ may be even a more fundamental theory,
that generates the $~D=(2,2)$~ SDSG and SDYM, {\it via} simple
dimensional reductions.  If this is indeed the case, the SDYM and SDSG in
$~D=(2,2)$~ themselves are some effective theories of the more fundamental
theory.  Since the $~D=10$~ is supposed to be the highest dimension,
where SG and SYM theories are possible notwithstanding particular
signatures [8], it is {\it not unlikely} that the $~D=(5,5)$~ SG + SYM theory
is singled out in a significant way, as in the case of $~N=1$~
superstring model.

\bigskip\bigskip

\noindent {\bf Acknowledgements}

We are grateful for R.~Brooks, D.~Depireux, T.~H{\" u}bsch, G.~Gilbert,
A.~Schwarz, W.~Siegel, C.~Vafa and E.~Witten for stimulating discussions.

\bigskip\bigskip

\vfill\eject

\noindent{\bf Appendix A: General Notation and Conventions}
\medskip\medskip

For the flat space-time with four dimensions and signature $~(2,2)$, the
metric form in {\it real} coordinates ~$y^{\ula} = \left(
y^1,y^2,y^3,y^4\right)~$ reads
$$d s^2=\h^{(R)}_{\ula\ulb} d y^{\ula} d y^{\ulb} \equiv (d y^1)^2 + (d y^2)^2
 - (d y^3)^2 - (d y^4)^2~~.
\eqno(A.1) $$
The very form of the flat ~$(2,2)~$ metric suggests to use {\it complex}
coordinates
$$x^1 = \fracm 1 {\sqrt 2}\left( y^1 + i y^2\right)~,~~~x^2 = \fracm 1{\sqrt2}
\left( y^3 + i y^4\right)
\eqno(A.2) $$
and their conjugates, ~$x^{\ula} = \left( x^1,x^2,\bar{x}^1,
\bar{x}^2\right)$,
instead of the real ones. The complex notation is natural in the ~$(2,2)~$
space-time, because of the classical isomorphism [9] ~$SO(2,2;{\bf R})\cong
SU(1,1;{\bf C})\otimes SU(1,1;{\bf C})$.
In other words, there is an obvious complex structure
in this space-time which gives the simplest example of a $~D=4$~
hermitian manifold.
In the complex notation, the metric takes the form
$$d s^2 = \h_{\ula\ulb}dx^{\ula}dx^{\ulb}\equiv 2\left( d x^1d\bar{x}^1 -
d x^2d\bar{x}^2\right)~,
\eqno(A.3) $$
where we have introduced the notation
$$x^{a}=(x^1,x^2)~~, ~~~~x^{\ab} = \bar{x}^{a}=(\bar{x}^1,\bar{x}^2)~~,~~~~~~
({\scst \ula = (a,\ab);~~a = 1,2;~\ab = \bar{1},\bar{2}})~~.
\eqno(A.4) $$
The metric tensor ~$\h_{\ula\ulb}~$ in this notation is
$$ \h_{\ula\ulb} = \left( \begin{array}{cc} 0 & \h_{a\bb} \\
\h_{\ab b} & 0 \end{array} \right)~~,~~~~
\h_{a\bb} = \h_{\bb a} \equiv \left( \begin{array}{cc} 1 & 0 \\ 0 & -1
\end{array} \right) ~~.
\eqno(A.5) $$
The metric form can now be rewritten as
$$d s^2 = 2\h_{a\bb}dx^{a}dx^{\bb}~.
\eqno(A.6) $$
The inner product of ~$(2,2)~$ vectors ~$V~$ and $~U$~ reads: $V^{\ula}
U_{\ula}\equiv V^a U_a + V^{\ab}U_{\ab}=V^a\h_{a\bb}U^{\bb} + V^{\ab}\h_{\ab b}
U^b$.

For the $~D=4$~ vectorial indices we use the {\it underlined} letters
{}~${\scst\ula~\equiv ~(a,\ab),~\ulb~\equiv ~(b,\bb),~\ldots\,}$,
which are equivalent to the
pairs of two-dimensional complex coordinate indices ~${\scst a,~b,~\ldots
{}~=~1,~2}~$ and their conjugates $~{\scst ~\ab ,~\bb ,~\ldots ~=~ \bar{1},
{}~\bar{2}}$. As for the spinorial indices,
we use the {\it undotted} and {\it dotted} letters $~{\scst\a ,~\b ,~\ldots
{}~=~ 1,~2}~$ and $~{\scst \dt{\a} ,~\dt{\b} ,~\ldots ~=~ \dt{1} ,~\dt{2}}~$
for
Weyl spinors of each
chirality. We also use the {\it underlined} spinorial indices ~${\scst
\ul{\a}~\equiv~
(\a,~\dt{\a}),~\ul{\b}~\equiv~ (\b,~\dt{\b}),~\ldots\,}$, denoting the pairs of
{\it dotted} and {\it undotted} indices. The chiral spinor dotted and undotted
indices can be raised and lowered by the use of the antisymmetric charge
conjugation matrix ~$C_{\a\b},~C^{\a\b}~$ and ~$C_{\dt{\a}\dt{\b}},
{}~C^{\dt{\a}\dt{\b}}$.  (See Appendix B for details).

As a rule for $~N=1$~ supersymmetry, we use capital
Greek letters to represent any Majorana spinor
field in the four-component
notation for spinors, while {\it the same} lower case Greek letter is used to
identify the chiral (Weyl) constituents of the same spinor, e.g.
$$ \J_{\ul{\a}} =\left( \begin{array}{c} \j_{\a} \\ \Tilde{\j}_{\dt{\a}}
\end{array}\right)~.
\eqno(A.7) $$
The same rule is sometimes used for the diagonal products of gamma matrices,
e.g. the spinor generators ~$\S^{\ula\ulb}~$ (see the Appendix B) are
decomposable as
$$\S^{\ula\ulb} = \left(\begin{array}{cc} \s^{\ula\ulb} & 0 \\ 0 &
\st^{\ula\ulb} \end{array}\right)~~.
\eqno(A.8) $$

A real ~$(2,2)\-$dimensional vector ~$V_{\ula} =\left( \cv_1,\cv_2,\cv_3,\cv_4
\right)~$ can be represented by a complex pair
{}~$V_{a}=\left(V_1,V_2\right)~$ in
the complex notation, where ~$V_1=\cv_1 + i\cv_2~$ and ~$V_2=\cv_3 + i\cv_4$.
Then we have ~$V_{\ab}=\left( V_{\bar{1}},V_{\bar{2}}\right)=\left( \bar{V}_1,
\bar{V}_2\right)$, respectively.

The ~$\s\-$matrices of eq.~(B.16) below can be used to convert a vector index
into a pair of spinor indices,
$$ V^{\a\dt{\a}}=V_{\ula}\left(\s^{\ula}\right)^{\a\dt{\a}}~,~~V_{\ula}=
\fracm 1 2\left( \st^{\ula}\right)_{\dt{\a}\a}V^{\a\dt{\a}}~.
\eqno(A.9)$$

\newpage

\noindent{\bf Appendix B: Dirac Matrices}
\medskip\medskip

The {\it Klein-Gordon} equation for a scalar ~$\F~$ in the real coordinates,
$$\left( \bo - m^2\right)\F (y)\equiv \left( \h^{\ula\ulb}\pa_{\ula}\pa_{\ulb}
- m^2\right)\F (y) = 0~,
\eqno(B.1) $$
is rewritten in the complex coordinates as
$$\left( \bo - m^2 \right) \F (x) \equiv \left( 2\h^{a\bb}\pa_{a}\pa_{\bb} -
m^2 \right) \F (x) = 0~.
\eqno(B.2) $$

The {\it Dirac equation} results from the factorization procedure of the
Klein-Gordon kinetic operator in eq.~(B.1), and in the real notation it takes
the form
$$\left( i\G^{\ula} \pa_{\ula} + m \right) \J (y) = 0~,
\eqno(B.3) $$
where the Dirac gamma matrices ~$\G^{\ula}~$ satisfy the Clifford algebra
$$ \left\{ \G^{\ula},\G^{\ulb}\right\} = 2\h_{(R)}^{\ula\ulb}~.
\eqno(B.4) $$
Eq.~(B.2) can also be factorized, and it leads to the Dirac equation in the
form
$$ \left(i\g^{\ula}\pa_{\ula} + m\right)\J (x) \equiv \left(i\g^{a}\pa_{a} +
i\g^{\ab}\pa_{\ab} + m \right) \J (x) = 0~,
\eqno(B.5) $$
where we have introduced the corresponding Dirac matrices ~$\g^{\ula}=\left(
\g^{a},\g^{\bb}\right)$. The latter satisfy an algebra
$$\li{&\g^{a}\g^{\bb} + \g^{\bb}\g^{a} = 2\h^{a\bb}~,
&(B.6a) \cr
&\g^{a}\g^{b} + \g^{b}\g^{a} = 0 ~,
&(B.6b) \cr
&\g^{\ab}\g^{\bb} + \g^{\bb}\g^{\ab} = 0 ~.
&(B.6c) \cr } $$
In particular, all of those ~$\g^{\ula}\-$matrices are nilpotent: ~$\left(
\g^{\ula}\right)^2 = 0~$ (no sum).

The representation theory of the Clifford algebra (B.4) can be developed along
the lines of the familiar ~$D=(1,3)$~ case. There exists only one
non-trivial ~$4\times 4~$ matrix representation of eq.~(B.4), and its
(equivalent) explicit forms can easily be constructed by the use of ~$2\times
2~$ Pauli matrices,
$$ \t_1 =\left( \begin{array}{cc} 0 & 1 \\ 1 & 0
\end{array}\right)~~,~~~\t_2 =
\left( \begin{array}{cc} 0 & -i \\ i & 0 \end{array}\right)~~,~~~~\t_3 = \left(
\begin{array}{cc} 1 & 0 \\ 0 & -1\end{array}\right)~~.
\eqno(B.7) $$

The {\it Majorana} representation of the ~$\G\-$matrices takes the form
$$\eqalign{
\G^1 =\left(\begin{array}{cc} 0 & -i\t_3 \\ i\t_3 & 0\end{array}\right)~, &
{}~~\G^2 =\left(\begin{array}{cc} 0 & -i\t_1 \\ i\t_1 &
0\end{array}\right)~,\cr
\G^3 =\left(\begin{array}{cc} 0 & \t_2 \\ -\t_2 & 0\end{array}\right)~, &
{}~~\G^4 =\left(\begin{array}{cc} iI_2 & 0 \\ 0 & -iI_2\end{array}\right)~,}
\eqno(B.8) $$
where ~$0~$ and ~$I_2~$ are the ~$2\times 2~$ zero and identity matrices,
respectively. In this representation  all of
the components of its ~$\G\-$matrices are {\it pure imaginary}, i.e. all the
entries of the ~$i\G^{\ula}~$ in the Dirac equation are purely {\it real}.

Another explicit representation, which is particularly useful in supersymmetry,
has the {\it diagonal} ~$\G_5\-$matrix.  It provides a preferred
basis for introducing the two-component notation for spinors in four space-time
dimensions, and generically has the form
$$\G^{\ula} = \left(\begin{array}{cc} 0 & \s^{\ula} \\ \st^{\ula} & 0
\end{array}\right)
\eqno(B.9) $$
with some ~$2\times 2\-$dimensional entries ~$\s^{\ula}~$ and ~$\st^{\ula}$.
The appropriate explicit representation is given by
$$\eqalign{
\G^1 = \left(\begin{array}{cc} 0 & -i\t_1 \\ +i\t_1 & 0\end{array}\right)~, &
{}~~\G^2 =\left(\begin{array}{cc} 0 & -i\t_2 \\ +i\t_2 &
0\end{array}\right)~,\cr
\G^3 = \left(\begin{array}{cc} 0 & +\t_3 \\ -\t_3 & 0 \end{array}\right)~, &
{}~~\G^4 = \left(\begin{array}{cc} 0 & +iI_2 \\ +iI_2 & 0 \end{array}\right)~,}
\eqno(B.10) $$
since in this representation we have
$$\G_5\equiv \G^1\G^2\G^3\G^4 = \left(\begin{array}{cc} I_2 & 0 \\ 0 & -I_2
\end{array}\right)~.
\eqno(B.11) $$

Given an explicit representation of the ~$\G\-$matrices, it is easy to
construct an explicit representation of the ~$\g\-$matrices (B.6) by forming
the linear combinations
$$V_{\pm}\equiv \fracm 1{\sqrt2}\left( \G^1 \pm i\G^2\right),~~W_{\pm} =
\fracm 1{\sqrt2}\left( \G^3 \pm i\G^4\right)~.
\eqno(B.12) $$
It follows that
$$V_{\pm}^2 = W_{\pm}^2 = 0~~,~~~~\left\{ V_+,V_-\right\}
=-\left\{ W_+,W_-\right\} = 2~~,~~~~V W + W V =0~~,
\eqno(B.13) $$
the latter being true for any assignment of the subscripts ~$\pm$. Therefore,
each choice,
$$({\bf I}):~\g^{a}=\left( V_+,W_+\right),~\g^{\bb}=\left( V_-,W_-\right),~~
({\bf II}):~\g^{a}=(V_+,W_-),~\g^{\bb}=(V_-,-W_+)~,
\eqno(B.14)$$
yields the corresponding representation for the ~$\g\-$matrices. The two
possibilities are  just related to either the SD
{}~$({\bf I})~$ or the ASD ~$({\bf II})$. For simplicity,
we restrict ourselves to the SD here and choose ~$({\bf I})$~
for the rest of the Appendix.

 Explicitly, we find
$$\eqalign{
{}~~~\g^1 = \left( \begin{array}{cccc} 0 & 0 & 0 & 0 \\ 0 & 0 & -i\sqrt{2} & 0
\\
0 & 0 & 0 & 0 \\ +i\sqrt{2} & 0 & 0 & 0 \end{array} \right),
& ~~~\g^2 = \left(
\begin{array}{cccc} 0 & 0 & 0 & 0 \\ 0 & 0 & 0 & -\sqrt{2} \\ -\sqrt{2} & 0
& 0 & 0 \\ 0 & 0 & 0 & 0 \end{array}\right), \cr
{}~~~\g^{\bar{1}} = \left( \begin{array}{cccc} 0 & 0 & 0 & -i\sqrt{2} \\
0 & 0 & 0 & 0 \\ 0 & +i\sqrt{2} & 0 & 0 \\ 0 & 0 & 0 & 0\end{array}\right), &
{}~~~\g^{\bar{2}}=\left(\begin{array}{cccc} 0 & 0 & +\sqrt{2} & 0 \\ 0 & 0 & 0
&
 0 \\ 0 & 0 & 0 & 0 \\ 0 & +\sqrt{2} & 0 & 0\end{array}\right).}
\eqno(B.15) $$
This representation has the structure (B.9) which allows to introduce the
{}~$2\times 2~$ ~$\s\-$matrices as
$$\g^{a} = \left(\begin{array}{cc} 0 & \s^{a} \\ \st^{a} & 0 \end{array}
\right),~~\g^{\ab} = \left(\begin{array}{cc} 0 & \s^{\ab} \\ \st^{\ab} & 0
\end{array} \right).
\eqno(B.16) $$
The ~$\s-$matrices satisfy an algebra
$$\li{ &\s^{a}\st^{\bb} + \s^{\bb}\st^{a} = 2\h^{a\bb}~,
&(B.17a) \cr
&\s^{a}\st^{b} + \s^{b}\st^{a} = 0~,
&(B.17b) \cr
&\s^{\ab}\st^{\bb} + \s^{\bb}\st^{\ab} = 0~,
&(B.17c) \cr } $$
which follows from eqs.~(B.6) and (B.16).

We find it convenient to define a basis in the space of ~$2\times 2~$
matrices by introducing the following set
$$\eqalign{
P_+ = \left( \begin{array}{cc} 1 & 0 \\ 0 & 0 \end{array}\right)~, &
{}~~~P_- = \left( \begin{array}{cc} 0 & 0 \\ 0 & 1 \end{array}\right)~,\cr
\t_+ = \left( \begin{array}{cc} 0 & 1 \\ 0 & 0 \end{array}\right)~, &
{}~~~\t_- = \left( \begin{array}{cc} 0 & 0 \\ 1 & 0 \end{array}\right)~.}
\eqno(B.18)$$
Comparing eqs.~(B.15), (B.16) and (B.18), we find
$$\eqalign{
\s^{a} = \left(-i\sqrt{2}\t_{-},-\sqrt{2}P_{-}\right), &
{}~~\st^{a} = \left( +i\sqrt{2}\t_{-},-\sqrt{2}P_{+}\right), \cr
\s^{\ab} = \left( -i\sqrt{2}\t_{+},+\sqrt{2}P_{+}\right), & ~~
\st^{\ab} = \left(+i\sqrt{2}\t_{+},+\sqrt{2}P_{-}\right)~.}
\eqno(B.19)$$

Linearly independent covariant products of ~$\g\-$matrices form a basis in the
space of all ~$4\times 4~$ matrices, and our choice of the basis is

\vfill\eject

$$ I_4~; $$
$$ \g^{a}~,~ \g^{\ab}~; $$
$$ \g_3~,~~\S^{a\bb}~,~~\g_{\3b}~;
\eqno(B.20)$$
$$ \g^{a}\g_{\3b}~,~~\g_{3}\g^{\ab}~; $$
$$ \g_5~,$$
where ~$\g_5\equiv \G_5$, while ~$\g_3$~ and ~$\g_{\3b}$~ represent the {\it
covariant} products
$$\g^{a}\g^{b}=2i\e^{ab}\g_3~,~~\g^{\ab}\g^{\bb}=2i\e^{\ab\bb}\g_{\3b}
{}~.
\eqno(B.21) $$
In eq.~(B.21) the Levi-Civita symbols have been introduced, ~$\e^{ab}=
-\e^{ba}
 =-\e_{ab}=\e_{ba},~\e_{12}=-1;~~\e^{\ab\bb}=-\e^{\bb\ab}=-\e_{\ab\bb}=
\e_{\bb\ab},~\e_{\bar{1}\bar{2}}=1\,$. The ~$\S^{a\bb}~$ in eq.~(B.20) are
similar to their ~$(3+1)\-$dimensional counterparts and take the form
$$\S^{a\bb} = -\fracm 14\left( \g^{a}\g^{\bb} - \g^{\bb}\g^{a} \right)~.
\eqno(B.22)$$
Those four matrices are covariantly reducible as
$$\S^{a\bb} = \fracm 12\h^{a\bb}\S + \widehat{\S}^{a\bb}~,
\eqno(B.23)$$
where we have introduced the irreducible pieces of ~$\S^{a\bb}~$ as
$$ \S \equiv \h_{a\bb}\S^{a\bb}=\g_{\3b 3}-\g_{3\3b}~~,~~~~
\h_{a\bb}\widehat{\S}^{a\bb} =0~.
\eqno(B.24) $$

The commutators involving the ~$\g_3,~\g_{\3b}~$ and ~$\S^{a\bb}~$ take the
form
$$\[ \g_3 , \g_{\3b}\] = -\h_{a\bb}\S^{a\bb}=-\S ~,~~
\left[ \g_3 , \S^{a\bb}\right] =\h^{a\bb}\g_3~,~~\left[ \g_{\3b},\S^{a\bb}
\right] = -\h^{a\bb}\g_{\3b}~,$$
$$\left[ \S^{a\bb},\S^{c\db}\right] = \h^{a\db}\S^{c\bb} -
\h^{c\bb}\S^{a\db}~.
\eqno(B.25)$$
This set of matrices  forms the Lie algebra which is in fact isomorphic to
{}~$so(2,2)$, as it should.  The ~$\left\{\S^{a\bb}\right\}~$ form the
subalgebra
which is isomorphic to ~$u(2)$. Clearly, the  ~$\left\{ \g_3,\S^{a\bb},\g_{\3b}
\right\}~$ are nothing but the ~$SO(2,2)~$ generators in the spinor
representation, while ~$\left\{ \S^{a\bb}\right\}~$ represent the generators
of the ~$U(2)~$ subgroup in the same representation. Eq.~(B.23) reflects the
relevant Lie algebra decomposition ~$u(2)=su(2)\oplus u(1)\,$.

In the explicit representation (B.20), the ~$\g_3$, ~$\S~$ and
{}~$\g_{\3b}~$ take the form
$$\g_3 = \left( \begin{array}{cc} \t_{-} & 0 \\ 0 & 0 \end{array} \right),
{}~~\S =\left( \begin{array}{cc} \t_3 & 0 \\ 0 & 0\end{array}\right)~,
{}~~\g_{\3b} = \left( \begin{array}{cc} \t_{+} & 0 \\ 0 & 0 \end{array}
\right)~,
\eqno(B.26)$$
and therefore, they separately form an ~$su(2)~$ subalgebra.

The ~$\g_3~$ and ~$\g_{\3b}~$ are nilpotent and hermitian-conjugated
to each other, ~$\g_3^{\dg} = \g_{\3b}$. All matrices in eq.~(B.26) obviously
commute with the ~$\g_5$~ and satisfy the funny property:
$$ \g_3 = \g_5 \g_3~,~~\g_{\3b} = \g_5 \g_{\3b}~,~~\S =\g_5\S ~.
\eqno(B.27) $$
The ~$\Hat{\S}^{a\bb}~$ is also commuting with the ~$\g_5\,$, but has
the property
$$ \g_5\widehat{\S}^{a\bb}=-\Hat{\S}^{a\bb}~.
\eqno(B.28)$$
In addition, we find the relations:
$$ \g^{a}\g_3 = \g_3\g^{a} = \g_{\3b}\g^{\ab} = \g^{\ab}\g_{\3b} = 0~,$$
$$\[ \g^{a},\g_{\3b}\] = -i\e^{ab}\h_{a\cb}\g^{\cb}~,~~\[
\g^{\ab},\g_3\] = -i\e^{\ab\bb}\h_{c\bb}\g^{c}~.\eqno(B.29) $$
and some useful identities:
$$ \h_{a\bb}\g^{a}\g^{\bb} = 2 - 2\S ~,~~\S\g^{a} = -i\e^{ac}\h_{c\bb}\g_3
\g^{\bb}~.
\eqno(B.30) $$

The {\it chiral} projection operators can be rewritten as
$$\fracm 12\left( 1 + \g_5 \right) = \left( \g_3 + \g_{\3b}\right)^2 =
\g_3\g_{\3b} + \g_{\3b}\g_3 = 2\g_3\g_{\3b} + \S ~,
\eqno(B.31a) $$
$$\fracm 1 2\left( 1 -\g_5 \right) = 1 - \g_{\3b}\g_3 - \g_3\g_{\3b} =
1 - \S - 2\g_3 \g_{\3b}~.
\eqno(B.31b) $$
It is interesting that there are more {\it hermitian projectors} for spinors
in $~D=(2,2)$.
Defining
$$\g_{3\3b}\equiv \g_3 \g_{\3b}~,~~\g_{\3b 3}\equiv \g_{\3b}\g_3~,
\eqno(B.32)$$
we find
$$\g^2_{3\3b}=\g^{\dg}_{3\3b}=\g_{3\3b}~,~~\g^2_{\3b 3} = \g^{\dg}_{\3b 3}
=\g_{\3b 3}~~, $$
$$\left( \g_3\g_{\3b}\right)^2=\g_3\g_{\3b}~,~~\left(\g_{\3b}\g_3\right)^2 =
\g_{\3b}\g_3~.
\eqno(B.33)$$

In the explicit representation (B.15) or (B.19), we have
$$\g_{\3b 3} =\g_{\3b}\g_3 = \left( \begin{array}{cc} P_{+} & 0 \\ 0 & 0
\end{array}\right),~~\g_{3\3b}=\g_3\g_{\3b} = \left( \begin{array}{cc} P_{-}
& 0 \\ 0 & 0 \end{array}\right)~.
\eqno(B.34) $$
Clearly, they are commuting with the ~$\g_5$. In addition, we find
$$\eqalign{
\[ \g^{a},\g_{3\3b}\] = -i\e^{ac}\h_{c\bb}\g_3\g^{\bb}~~, &
{}~~~~\[ \g^{a},\g_{\3b 3}\] = -i\e^{ac}\h_{c\bb}\g^{\bb}\g_3~,\cr
\[ \g^{\ab},\g_{3\3b}\] = -i\e^{\ab\cb}\h_{b\cb}\g^{b}\g_{\3b}~~, &
{}~~~~\[ \g^{\ab},\g_{\3b 3}\] = -i\e^{\ab\cb}\h_{b\cb}\g_{\3b}\g^{b}~.}
\eqno(B.35) $$
\vglue.2in

\newpage

\noindent{\bf Appendix C: Finiteness of $~N=2$~ Superstrings}
\medskip\medskip

Within the standard background field approach [58] to the quantum
four-dimensional SYM theory, it is not difficult to extend the {\it
non-renormalization theorem} (NRT) [59] from the flat to the super-instanton
background [41]. In other words, none of the possible counter-terms survives,
when the background fields are restricted to be self-dual. In terms of the
super-instanton solution, any correction is supposed to be invariant under
the shifts of the ~$\Bar{\theta}\-$coordinate (NRT!). but there is no way to
cancel it because of the absence of an anti-chiral instanton collective
coordinate (the instanton superfield is chiral!) [41]. In terms of the
relevant superfield strengths, there is simply no way to write down any
counter-term for the supersymmetric SDYM systems, since the super-SD is always
dictated by the conditions that chiral {\it or} anti-chiral superfields are
to vanish. The physical significance of vanishing quantum corrections is the
{\it integrability} of the initial classical system combined with
supersymmetry.

Unlike the SYM theories, the SGs are non-renormalizable for general
backgrounds. Nevertheless, as was shown in Ref.~[60], the SD condition
eliminates all the counter-terms to all orders, except the {\it topological}
 ones, the latter being quadratic in the curvature. The SDSG finiteness follows
as a consequence of both supersymmetry and fermionic chiral invariance
[60].
The main assumption used in Ref.~[60] is just the validity of supersymmetry or
the existence of the supersymmetrical regularization.

In parallel with the supersymmetric SDYM instanton solutions, the SDSG
instantons could also be developed. The solutions to eq.~(3.28) in the
Euclidean space are known as the {\it gravitational instantons}. They are
Ricci-flat and may be used to describe tunneling between different
gravitational (or string) vacua, as well as  ``topological'' fluctuations of
the gravitational field (the Hawking's foam) [61]. According to Ref.~[61], all
asymptotically locally flat non-singular solutions of the Einstein equations
are self-dual (or anti-self-dual). Therefore, it seems to be reasonable to
take on the only known compact solution -- the so-called {\it Eguchi-Hanson}
(EH) instanton [62] -- as a basis. In the AW space-time it is described by
the metric
$$\eqalign{d s^2 = \, & {dr^2\over  1 - 1/r^4} - {\frac 14}r^2
\left[  d\vq^2 +\sinh^2 \vq\,d\f^2\right] \cr
& +{\frac 14}r^2\left( 1 - 1/r^4\right) \left[ d\j^2 + 2\cosh\vq\,d\j
d\f + \cosh^2\vq\,d\f^2\right]~, \cr}
\eqno(C.1)$$
where we have introduced the parametrization
$$\eqalign{
x^1 = & r\cosh\left( {\frac 12}\vq\right) \exp\left[ {\frac 12}i(\j + \f)
\right]~,\cr
x^2 = & r\sinh\left( {\frac 12}\vq\right) \exp\left[ {\frac 12}i(\j -\f)\right]
{}~.}\eqno(C.2)$$
The AW space-time complex coordinates ~$(x^1,x^2)~$ are defined in the Appendix
A, see eqs.~(A.1)--(A.4).

The EH-instanton solution allows a supersymmetrization within the
$~N\-$extended SDSG ~$(N\leq 4)~$ to a full ~$N\-$extended {\it chiral}
superfield.
The chirality implies the UV-finiteness through the NRT.  Again, the
{\it extended} SG is needed to move freely between the AW and Euclidean
space-times in quantum theory. The typical example is the possible {\it
three-loop} correction to the $~N=1$~ quantum SG [63]
$$I_{N=1}^{(3)} = \int d^4 x\, d^2 \theta \, d^2 \Tilde \theta\,
W_{\a\b\g} W^{\a\b\g} \Tilde W_{\Dot\a\Dot\b\Dot\g} \Tilde
W^{\Dot\a\Dot\b\Dot\g} ~~, \eqno(C.3) $$
which comprises the supersymmetrization of the Bel-Robinson tensor squared
[63],
$$T^2_{\ula\ulb\ulc\uld} =\left[ R^{\ule}{}_{\ulc}{}^{\ulf}{}_{\ula}R_{\ule
\uld\ulf\ulb} + \breve{R}^{\ule}{}_{\ulc}{}^{\ulf}{}_{\ula}\breve{R}_{\ule\uld
\ulf\ulb}\right]^2~,
\eqno(C.4)$$
where the $~\breve R_{\ula\ulb\ulc\uld}$~ is the dual of
$~R_{\ula\ulb\ulc\uld}$.
Obviously, if we restrict the SG background to be {\it self-dual} by setting
$~W_{\a\b\g} = 0$, this integral is to vanish, since it contains {\it both}
chiral and anti-chiral superfields $~W_{\a\b\g}$~ and $~\Tilde W_{\Dot\a\Dot\b
\Dot\g}$ multiplicatively.  This property persists to all orders in the quantum
perturbations [60].   In the usual
$~D=(1,3)$~ space-time, we could {\it not} impose $~W_{\a\b\g} = 0$,
while keeping $~\Tilde W_{\Dot\a\Dot\b\Dot\g}\neq 0$, which is now
{\it possible} in our $~D=(2,2)$ case, since $~\Tilde W_{\Dot\a\Dot\b\Dot\g}$~
is {\it no longer} the complex conjugate of $~W_{\a\b\g}$.

The above considerations can equally be applied to the (heterotic) $~N=2$~
superstring, whose backgrounds are just supersymmetric SDYM and SDSG, to prove
its finiteness. If all the counter-terms are to be {\it
local} at a given string-loop order, as they are expected to be from the
experience with the $~N=1$~ superstring, then {\it any} quantum corrections
at {\it any} genus of the world-sheet are to possess the same features
as the {\it tree} level counter-terms. Combined with the arguments above, it
means the total finiteness of the $~N=2$~ superstring theory.

Another finiteness argument comes from the anticipated  equivalence between the
$~N=2~$ and $~N=4$~ superstrings [38]. In the $~N=4$~ formulation the
$~SO(2,2)~$ (or ``Lorentz'') covariance is manifest, which makes it possible to
clearly distinguish between the Neveu-Schwarz and Ramond sectors of the theory
 and prove a presence of space-time supersymmetry in it. For the $~N=4$~
superstring, the $\s\-$model field equations are implied {\it
classically}, rather
than by quantum $\b\-$function calculations. In other words, the $~N=4$~
superstring action remains {\it unrenormalized} against quantum
corrections because
of the $~N=4$~ supersymmetry on the world-sheet [64]. The $~N=4$~
supersymmetry is associated with the hyper-K\"ahler geometry [64] which is
nothing but the self-dual geometry for the {\it four-dimensional} target
[23].
 Hence, it is the $~N=4$~ world-sheet supersymmetry (which is quite obscure in
 the $~N=2$~ formulation) that is responsible for the $~N=2$~ superstring
finiteness. The string-loop corrections may even vanish for the $~N=2$~
superstring, which would then constitute a {\it topological field theory}. The
last viewpoint has recently been advocated by Siegel [35].

        It is amazing that the SD condition associated with the {\it
chiral} superfields plays such an important role for the $~N=2$~ superstrings.
 In the $~N=1$~ superstring theory, there is no such a strong condition as
the SD condition on the backgrounds, so that the $~N=1$~ superstring
finiteness at the string-loop level is rather obscure.

\newpage

\noindent {\bf Appendix D: $D=(5,5),\,N=1$~ SG $\,+\,$ SYM and Dimensional
Reduction}
\medskip\medskip

        In this appendix we give a $~D=10$~ theory in the
space-time signature $(5,5)$~ with $~N=1$~ local supersymmetry, which
produces our $~D=(2,2)$~ SYM and SG with various values $~N~$ of
supersymmetry, after dimensional reduction and compactifications.  Due
to the no-go barrier described in sect.~6, only some of these resultant
models can be consistently truncated to generate the SDSG and
supersymmetric SDYM
theories in $~D=(2,2)$.  We give one example of a $~D=(2,2),\,N=1$~ system
after such dimensional reduction/compactification of our $~D=(5,5),\,N=1$~
initial theory.

        The interesting case related to the SDSG and SD SYM is the
$~D=(2,2),\,N=1$~ theory,\footnotew{We mention that the $~D=(2,2),\,N=4$~
theory with {\it global} $~SU(4)$~ is easily obtained
by what is called the {\it simple dimensional reduction}, described in
Ref.~[65].
Since this theory does {\it not} yield the SD theory due to the global
symmetry $~SU(4)$~ as mentioned in sect.~5, we do not give the details of
the derivation or the results.} based on a technique of {\it dimensional
reduction}, similar to what we call dimensional reduction (DR)
{\` a} la Witten [66].  This technique has been developed in component
formulation, based on the {\it Calabi-Yau} (CY) compactification [67]
in the usual $~D=(1,9),\,N=1$~ heterotic superstring theory.
The {\it non-compactness} [68] of the ``extra'' $~E=(3,3)$~ dimensional CY
space with the indefinite signature $~(3,3)$~ poses {\it no} problem,
as we will discuss later.

        We first establish our initial $~D=(5,5),\,N=1$~ SG + SYM
system.  To this end, we have to clarify the spinorial properties of
the space-time.  There are good similarities as well as
important differences among the $~D=(5,5)$, $~D=(1,9)$~ and
$~D=(2,2)$~ cases, which we can utilize to
simplify the calculation.  First of all, there exist {\it Majorana}, {\it
Weyl} as well as {\it Majorana-Weyl} spinors in the $~D=(5,5)$~
space-time, as can be easily noticed by the help of Ref.~[8].
The main difference of the $~D=(5,5)$~ case from
the $~D=(1,9)~$ case is about the {\it anti-chiral} Weyl spinors,
namely the {\it chirality-conjugates} are {\it not} complex conjugates of
the {\it undotted} Weyl spinors similarly to the $~D=(2,2)$~ case.
This is due to the reason very similar to the $~D=(2,2)$~ case, i.e.~the
matrix $~\G^{11} \equiv \G_1 \G_2 \cdots \G_{10} $~ [8] gives
$~(\G^{11})^2 = 1$, so that $~\G^{11}$~ can be {\it purely real}.
Therefore, the chiral projectors $~P_\pm \equiv (1/2) (1\pm\G^{11} )$~
are {\it not} complex conjugate to each other, just like the $~D=(2,2)$~
case.  It is to noted also that the {\it bars} in our $~D=(5,5)$~
case denote just the spinorial scalar products, exactly like the $~D=(1,9)$~
case considered in Ref.~[69].

After this arrangement of notation, we can construct our initial
$~D=(5,5),\,N=1$~ SG + SYM theory with the field content $~(e\du a
m,\psi\du m \a, B_{m n}, \chi_\a, \Phi; A\du m I, \l^{\a I})$, where
instead of the {\it dots} for spinorial indices, the {\it
superscripts} and  {\it subscripts} of spinorial indices denote respectively
the {\it chiral} and {\it anti-chiral} spinors in
$~D=10$.  Therefore, $~\psi\du m \a$~ and $~\l^{\a I}$~
are all chiral Majorana-Weyl spinors, while $~\chi_\a$~ is anti-chiral
Majorana-Weyl spinor.  This system is described by the BIds:
$$\eqalign{&\nabla_{\[ A} T\du{B C)} D - T\du{\[ A B|} E T\du{E|C)} D -
R\du{\[A B C)} D \equiv 0 ~~, \cr
&\fracm 16 \nabla_{\[ A} G_{B C D)} - \fracm 14 T\du{\[ A B|} E G_{E|C
D)} \equiv \fracm 14 F\du{\[ A B} I F\du {C D)} I \equiv X^4_{A B C D} ~~,
\cr
&\nabla_{\[ A} F\du{B C)} I - T\du{\[ A B |} D F\du{D| C)} I \equiv 0 ~~.
\cr }
\eqno(D.1) $$
These BIds are solved by the following set of constraints:
$$\eqalign{&T\du{\a\b}{c} = i(\G^c)_{\a\b}~~,
{}~~~~G_{\a\b\g}=G_{\a bc} = 0~~, ~~~~
G_{\a\b c} = i\fracm 12(\G_c)_{\a\b} ~~, ~~~~T\du{\a b}{c}= 0~~, \cr
&T\du{\a b}{\g} = i\fracm 1{96} (\G_b \G^{efg})\du\a\g
(\Bar\l^I\G_{efg} \l^I) ~~, \cr
&T\du{\a\b}{\g} = -{\sqrt 2}\[\d\du{(\a}{\g}\chi_{\b)} +
(\G^e)_{\a\b}(\G_e \chi)^{\g}\]~~, \cr
&T_{a b}{}^c = -2 G_{a b}{}^c ~~,
{}~~~~~\nabla_\a \Phi = -\fracm 1 {\sqrt 2}\chi_{\a} ~~, \cr
&F\du{\a b} I =i \fracm 1{\sqrt 2} (\G_b \l^I) _\a ~~, \cr
&\nabla_{\a} \chi{\low\b} = - i\fracm 1{\sqrt 2} (\G^c)_{\a\b}
\nabla _c \Phi +
i\fracm 1{12\sqrt 2}(\G^{cde})_{\a\b} \left( G_{c d e}-i\fracm 14
\Bar\chi\G_{cde}\chi +i \fracm 18 \Bar\l^I \G_{cde} \l^I\right) ~~, \cr
&\nabla_\a\l^{\b I} = -\fracm 1{2{\sqrt 2}} (\G^{cd} )\du\a\b
F\du{cd} I - \fracm 1{2{\sqrt 2}} (\G^{cd})\du \a\b (\Bar\chi\G_{cd}
\l^I) - {\sqrt 2} \chi_\a \l^{\b I}
-\fracm 1{\sqrt 2} \d\du\a\b (\Bar\chi \l^I) ~~, \cr
&R_{\a\b c d} =-2i(\G^e)_{\a\b} G_{c d e}
- \fracm1{48} (\G\du{cd}{efg})_{\a\b} (\Bar\l^I\G_{efg} \l^I)
-\fracm18 (\G^e)_{\a\b} (\Bar\l^I\G_{cde}\l^I) ~~, \cr
&\nabla_\a G_{b c d}= -i \fracm 14(\Gamma_{[b}T_{cd\, ]})_\a
+i\fracm 1{\sqrt 2} (\G_{\[ b|}\l^I)_\a  F\du{|cd\]} I ~~, \cr
&R_{\a b c d} = -i(\Gamma_{\[c}T_{d \]b})_\a
+i\fracm 1{\sqrt 2} (\G_{\[ b|}\l^I)_\a  F\du{|cd\]} I ~~. \cr }
\eqno(D.2) $$
Here the $~G_{m n p}$~ contains the Chern-Simons term of the YM:
$$G_{m n p} \equiv \fracm 12 \partial_{\[ m} B_{n p\]} + X_{m n p}^3 ~~,
\eqno(D.3) $$
where $~X_{A B C}^3$~ satisfies the relation
$$\fracm 16 \nabla_{\[ A} X_{B C D)} ^3 -\fracm14 T\du{\[ A B|}E X_{E|
C D)}^3 - X_{ABCD} ^4 \equiv 0 ~~.
\eqno(D.4) $$
This set of constraints is a $~D=(5,5)$~ analogue of the constraints
developed for the $~D=(1,9)$~case in Ref.~[70] in order to simplify
the $~\beta\-$function calculations in Green-Schwarz $~\s\-$models,
owing to the vanishing exponential
functions of the dilaton field $\Phi$, as well as the simplification of
superfield equations, whose explicit forms we skip here.

        We next look into the issue of compactification to the
$~D=(2,2),\,N=1$~ theory, based on the DR {\` a} la Witten on a CY
manifold [66].  Due to the similarity
between $~D=(1,9),\,N=1$~ and $~D=(5,5),\,N=1$~ theories, we can utilize
the main results in the compactification of $~D=(1,9),\,N=1$~
heterotic string, applying them to the special {\it non-compact}
manifold [71] with the signature ~$(3,3)$.  However, as has been
well-known, the {\it Kaluza-Klein} (KK)-type compactification on a CY manifold,
keeping all the curved background, is very hard in practice.  Fortunately the
DR {\` a} la Witten provides us a convenient way [66] to handle this problem
relying on a simple principle in a component field formulation.
The main principle is that we keep only the
$~SU(3)~$ singlet quantities among all the zero mass fields upon the
compactification, where $~SU(3)~$ is the holonomy group of the CY manifold.
By this prescription, the component field computation is drastically
simplified, because all the fields we maintain are invariant
under the ~$SU(3)$~ group.  Our task is to apply this method to our external
non-compact space-time with the signature $~(3,3)$.

Our next question is the validity of this prescription in
superspace.  Fortunately, a superspace version of this prescription
has also been developed in Ref.~[72] for the purely SG sector.
Due to another similarity between
the $~D=(2,2),\,N=1$~ and $~D=(1,3),\,N=1$~ SG systems, we can borrow the
results in Ref.~[72] for our purposes rather easily.

Keeping these points in mind, we establish our
notation first.  The indices we use are
$~{\scst a,~b,~ \cdots}$~ and $~{\scst \a,~\b,~ \cdots}$~ for respectively
vector and spinor indices in $~D=(5,5),\,N=1$~ superspace, while
$~{\scst \ul a,~\ul b,~ \cdots}$~ and
$~{\scst \a,~\b,~ \cdots}$~  (or $~{\scst \Dot\a,~\Dot\b,~ \cdots}$) for
$~D=(2,2),\,N=1$~ vector and
spinor (or antispinor) indices, respectively.  (The $~_{\a,~\b,~ \cdots}$~
indices are used both in $~D=(5,5)$~ and ~$D = (2,2)$, as long as they are not
confusing from the context.)  For $~E=(3,3)$~ we use
$~{\scst \hat a,~\hat b,~ \cdots}~$ as their vectorial indices, while we use no
 indices for spinors, because they are not
needed in our calculation.  For example, our $~\G$-matrices in $~D =
(5,5)$~ are decomposed as
$$(\G^a)_{\a\b} ~~\rightarrow  ~~\cases{~~(\s^{\ul a})_{\a\Dot\b}
{}~~,\cr ~~\G^{\hat a} ~~~~~. \cr}
\eqno(D.5) $$
Accordingly, we have $~\G_{11} = \G_7 \G_5$, where $~\G_7$~
characterizes the chiralities in $~E=(3,3)$.

        We can now study the detailed significance of the $~E=(3,3)$~ space,
which is
a {\it non-compact} analog of the $~E=(0,6)$~ CY space.  There are
similarities as well as differences between our {\it ``non-compact} CY'' (NCCY)
and the usual CY space [67].  One of the differences is that the holonomy
group of our NCCY is no longer $~SU(3)$, but it is a subgroup of
$~SL(2,C)$, which can be understood by noticing the isomorphism
$~SO(3,3)\approx SL(4,R)$.  If we exclude the torsion
and the dilaton condensate: $~G_{a b c} = 0,~\Phi = 0$~ in the purely
bosonic background, the gravitino transformation
rule obtained from (D.2) implies that the Ricci-flatness solves its
Killing spinor condition.\footnotew{Due to the indefinite signature,
there may be other solutions, but this Ricci-flat solution gives the
Killing spinor condition.}~~Accordingly, the covariant
constant spinor enables us to show the covariant constancy of an almost
complex structure, as in the usual CY case [67].  Therefore the extra
space-time can be Ricci-flat and K{\"a}hler.
Relevantly, there must be a Killing spinor, which is a singlet under a
little group of $~SO(3,3)\approx SL(4,R)$.
A typical example is $~SL(4,R) \rightarrow SO(2) \otimes SO(3) \otimes
SO(3)$~ under
which the original spinor $~{\bf 4}$~ of $~SL(4,R)$~ goes to $~{\bf 4}
\rightarrow {\bf (2,1)} + {\bf (1,2)}$.  Since we need at least a singlet,
we have to go
further down.  A simplest choice is $~SL(4,R) \rightarrow SO(2) \otimes
SO(2) \otimes SO(3)$, whereupon $~{\bf 4} \rightarrow {\bf 1} + {\bf 1}
+ {\bf 2}$, regarding $~SO(2) \otimes SO(3)$~ as the holonomy group of
the NCCY.  Eventually, the original supersymmetry
$~\e^\a$~ yields the $~N=1$~ supersymmetry in $~D=(2,2)$.
Now the only remaining Killing spinor equation is from the gaugino:
$$~\d \l^{\a I} = \fracm 1 {2\sqrt2} ( \G^{\hat c \hat d}) \du \b \a \e^\b
F_{\hat c\hat d} = 0 ~~,
\eqno(D.6) $$
implying the vanishing first Chern class.  Thus we see that our
NCCY shares the similar properties: the
Ricci-flatness, K{\" a}hlerness, and vanishing first Chern class, with
the usual CY manifold [67].\footnotew{We acknowledge T.~H{\" u}bsch for
explaining this point.}

This can be also understood, considering the spinors in $~E=(3,3)$.
The $~D=(5,5)$~ chirality
operator $~\G_{11}$~ is a product of those for $~E=(3,3)$~ and
$~D=(2,2)$: ~$\G_{11} = \G_7 \G_5$.  This implies that the Majorana-Weyl
spinor with the positive chirality in $~D=(5,5)$~ has either $~(+,+)$~
or $~(-,-)$~ chiralities in $~E=(3,3)$~ and $~D=(2,2)$, respectively.
We can also impose the Majorana condition additionally on the Weyl spinor
in $~E=(3,3)$, and the resulting $~D=(2,2)$~ theory will have the $~N=2$~
supersymmetry.  For these extended supersymmetries, we can keep the
holonomy group $~SO(3,3)$~ for $~E=(3,3)$, while to get the $~N=1$~
supersymmetry, we need to reduce the holonomy group into $~SO(2)
\otimes SU(2)$, as above.

We now review the DR {\` a} la Witten for the usual CY.
Its principle was that
we truncate all the quantities which are {\it non-invariant} under
the holonomy group
$~SU(3)$~ of the CY extra space.  Under this principle, any quantity with
indices $~{\scst\hat a,~\hat b,~ \cdots}$~ it put to zero, except for
the antisymmetric tensor, which can be proportional to the $~SU(3)$~
invariant tensor [67] $~\e_{\hat a\hat b}$.  As for spinors, their
indices are singlets under the $~SU(3)$~ group.  The NCCY analog of this
prescription is to truncate all the quantities, which carry
non-singlet indices with respect to the holonomy group $~SU(2)$, as we
described.  Hence all the fields carrying $~E=(3,3)$~ vectorial
indices are truncated, and all
the $~D=(5,5)$~ spinorial indices are reduced to be singlets under the
$~SU(2)$~ holonomy group of our NCCY.\footnotew{Eventually what we will get
after our DR is formally independent of the details of the holonomy group for
our NCCY.  This is due to the fact that we are reducing one of the two
${\bf 2}$'s in $~{\bf 4} \rightarrow {\bf 2}+{\bf 2}$ under $~SL(4,R)
\rightarrow SO(2) \otimes SL(2,C)$~ into singlets, reducing $~N=2$~ into
$~N=1$~ in $~D=(2,2)$, as described above.}

     We next show how our DR works for our constraints (D.2).  As an
example, we consider the case of $~\nabla_\a\chi{\low \b}$~ in (D.2):
$$\nabla_{\a}\chi{\low\b}
{}~~\rightarrow~~\cases{~\nabla_{\a} \Tilde
\chi_{\Dot\b} = - \fracm i{\sqrt 2} (\s^{\ul c} )_{\a\Dot\b}\nabla_{\ul c} \Phi
\cr
{}~~~~~ ~~~~~ ~~~~ +{i\over{12\sqrt 2}} (\s^{\ul{cde}})_{\a\Dot\b}(G_{\ul{cde}}
-i\half         \chi \s_{\ul{cde}}\Tilde\chi + i \fracm 14 \l^I
\s_{\ulc\uld\ule} \l^I ) ~~,   \cr
{}~~~ \cr
{}~\nabla_{\a}\chi{\low\b} = 0 ~~. \cr }
\eqno(D.7) $$
Needless to say, we can make use of the Fierz identities, such as
$~(1/6) (\s^{\ula\ulb\ulc})_{\a\Dot\b} (\s_{\ula\ulb\ulc})_{\g\Dot\d}$
$=(\s^\ula)_{\a\Dot\b} (\s_\ula) _{\g\Dot\d} = - 2C_{\a\g}
C_{\Dot\b\Dot\d}$~ in $~D=(2,2)$.  (See Ref.~[72] for other details.)
As a result of this
rule, we get the following $~D=(2,2),\,N=1$~ constraints:
$$\eqalign{&T\du{\a\Dot\b} \ulc = i(\s^\ulc)_{\a\Dot\b} ~~, ~~~~
\nabla_\a \Phi = -\fracm 1{\sqrt 2} \chi_\a ~~, \cr
&G_{\a\Dot\b \ulc} = i\half (\s_\ulc) _{\a\Dot\b} ~~,~~~~ G_{\a\b\ulc} =
0~~, ~~~~ G_{\ul\a\ulb\ulc} = 0~~, ~~~~ G_{\ul\a\ul\b\ul\g}=0~~, \cr
&T\du{\a\b} {\Dot\g} = 0~~, ~~~~T\du{\a\Dot\b} \g = {\sqrt 2} \d\du\a\g
\Tilde\chi_{\Dot \b} ~~, ~~~~ T\du{\a\b} \g = -{\sqrt 2}\d\du {(\a}\g
\chi{\low{\b)}} ~~, \cr
& \nabla_\a\Tilde\chi_{\Dot\b} = -i\fracm 1{\sqrt 2} (\s^\ulc) _{\a\Dot\b}
\nabla_\ulc \Phi + i \fracm 1{12{\sqrt2}}(\s^{\ulc\uld\ule})_{\a\Dot\b}
G_{\ulc\uld\ule} - \fracm 1 {4{\sqrt 2}} (\s^\ulc)_{\a\Dot\b} (2\chi
\s_\ulc\Tilde\chi-\l^J\s_\ulc \Tilde\l^J) ~~, \cr
& T\du{\a\ulb}\g =-i \fracm 1 8 (\s_\ulb\Tilde\s^\ulc)\du\a\g
(\l^I\s_\ulc \Tilde\l^I)  ~~, ~~~~T\du{\ul\a\ulb}\ulc = 0~~, ~~~~
T\du{\ula\ulb} \ulc = - 2 G\du{\ula\ulb} \ulc ~~, \cr
& R_{\a\Dot\b\ulc\uld} = -2i (\s^\ule)_{\a\Dot\b} G_{\ulc\uld\ule} ~~, \cr
& F\du{\a\ulb} I =i \fracm 1{\sqrt 2} (\s_\ulb \Tilde\l^I) _\a ~~, \cr
&\nabla_\a\l^{\b\, I} =- \fracm 1{2\sqrt2} (\s^{\ulc\uld})\du\a\b
F\du{\ulc\uld} I  - \fracm 1{2{\sqrt 2}} (\s^{\ulc\uld})\du \a\b
(\chi\s_{\ulc\uld} \l^I)
-{\sqrt 2} \chi_\a\l^{\b\, I} -\fracm 1{\sqrt 2} \d\du\a\b
(\chi\l{}^I+\Tilde\chi \Tilde\l^I)  ~~, \cr
& \nabla_\a\Tilde\l^{\Dot\b\, I} = -{\sqrt 2} \chi_\a\Tilde\l ^{\Dot\b\, I}
{}~~. \cr}
\eqno(D.8) $$
As is easily seen, this contains a SG multiplet
$~(e\du\ula\ulm,\,\psi\du\ulm\a,\,\Tilde\psi\du\ulm{\Dot \a})$, a TM
$~(B_{\ulm\uln},$ $\chi_\a,$ $\Tilde\chi_{\Dot\a},$ $\Phi)$~ and a YM
multiplet $~(A\du\ulm I,\,\l\du\a
I,\,\Tilde\l\du{\Dot\a} I )$.
By putting the YM and TM to zero, we can re-obtain the purely SG
 constraints in eqs.~(3.25) and (3.27), after appropriate field rescalings.
At this stage, the truncation into the SDSG or the supersymmetric
SDYM is straightforward.

What we have performed so far, have been the practically simplified DR
rules to get the theories in $~D=(2,2)$~ (the AW space-time), which do {\it
not} contain any effect of KK modes in $~E=(3,3)$.  Our DR
rule fortunately bypasses the problem associated with the {\it
non-compactness} of the extra space, because we do {\it not} keep the
KK {\it massive} modes.  In the usual field theory, the unitarity of the
initial theory in higher-dimensions is rigorously required, and, therefore,
the indefinite signature $~(5,5)$~ is {\it not} acceptable for such purposes.
  In our case, however, this does {\it not} concern us, since the
resultant theory of $~D=(2,2)$~ itself has the indefinite metric.

We have considered only the $~E=(3,3)\-$type CY extra space, but we
can also  perform the usual KK compactification now on the {\it compact}
$~E=(0,6)$~ CY internal manifold, which yields the {\it usual}
$~D=(1,3),\,N=1$~ theory!  However, in such a case the resulting theory
will perhaps have some problems with unitarity, caused by the indefinite
metric in the original $~(5,5)$~ theory itself.  We also note that the
result for such a DR is {\it formally} the same as the above case.
This is also why we could utilize the result of the DR on the
usual CY [72].

Even though we have considered only the case of $~D=(5,5)$ as the initial
higher-dimensional theory, we could formulate other supersymmetric
theories equally as well, such as a $~D=(3,3)$~theory in six-dimensions, which
is similar to a $~D=(1,5),\,N=2$~ theory constructed in Ref.~[73].
Actually, we could perform similar DR as above, to get the $~D=(2,2),\,N=1$~
SDSG and supersymmetric SDYM {\it after} subsequent truncation, the
details of which we skip in this paper.

        We finally mention the possibility that
such a $~D=(5,5)$~ theory with the indefinite metric is related to some
superstring theory, that we may have overlooked in the past.  Since
even the $~N=2$~ superstring has the non-compact target space such as
$~D=(2,2)$~ with {\it two} time directions, it is not totally senseless to
introduce more time coordinates in a target space-time for some
superstring theories.

\newpage

\refs
\normalsize
\baselineskip 7pt

\items{1} A.A. Belavin, A.M. Polyakov, A.S. Schwartz and Y.S.~Tyupkin,
\pl{59}{75}{85};  \\
R.S. Ward, \pl{61}{77}{81}; \\
M.F. Atiyah and R.S. Ward, \cmp{55}{77}{117}; \\
E.F. Corrigan, D.B. Fairlie, R.C.~Yates and P.~Goddard,
\cmp{58}{78}{223}; \\
E.~Witten, \prl{38}{77}{121}. \\

\items{2} M.F.~Atiyah, unpublished; \\
R.S.~Ward, Phil.~Trans.~Roy.~Lond.~{\bf A315} (1985) 451 ;\\
N.J.~Hitchin, Proc.~Lond.~Math.~Soc.~{\bf 55} (1987) 59 . \\

\items{3} H.~Ooguri and C.~Vafa, \mpl{5}{90}{1389};
\np{361}{91}{469}; \ibid{367}{91}{83};\\
H.~Nishino and S.J.~Gates, ``{\it $N=(2,0)$~ Superstring as the
Underlying Theory of Self-Dual Yang-Mills Theory''}, Maryland preprint,
UMDEPP 92-137, to appear in Mod.~Phys.~Lett.

\items{4} M.~Ademollo, L.~Brink, A.~D'Adda, R.~D'Auria, E.~Napolitano,
S.~Sciuto, E.~Del Giudice, P.~Di Vecchia, S.~Ferrara, F.~Gliozzi, R.~Musto
and R.~Pettorino, \pl{62}{76}{105}; \\
M.~Ademollo, L.~Brink, A.~D'Adda, R. D'Auria, E. Napolitano, S. Sciuto, E. Del
Giudice, P. Di Vecchia, S. Ferrara, F. Gliozzi, R. Musto, R. Pettorino and
J.H. Schwarz, \np{111}{76}{77}; \\
L. Brink and J.H. Schwarz, \np{121}{77}{283}.

\items{5} E.S. Fradkin and A.A. Tseytlin, \pl{106}{81}{63}; \\
S.D. Mathur and S. Mukhi, \np{302}{88}{130}.

\items{6} I.V. Volovich, \pl{123}{83}{329}; \\
C.R. Gilson, I. Martin, A. Restuccia and J.C. Taylor,
\cmp{107}{86}{377};
J.~Lukierski and W.~Zakrzewski, \pl{189}{87}{99}.

\items{7} S.V.~Ketov, S.J.~Gates, Jr.~and H.~Nishino, {\it Majorana-Weyl
Spinors and Self-Dual Gauge Fields in $~2+2~$ Dimensions}, Maryland
preprint, UMDEPP-92-163 (February 1992).

\items{8} T. Kugo and P.D. Townsend, \np{221}{83}{357}.

\items{9} R. Gilmore, {\it Lie Groups, Lie Algebras, and Some of Their
Applications}, Wiley-Interscience, 1974, p. 52.

\items{10} Y.A. Gol'fand and E.P. Lichtman, Pis'ma ZhETF {\bf 13} (1971)
452;\\
J. Wess and J. Bagger, {\it Supersymmetry and Supergravity}, Princeton
University Press, 1983.

\items{11} M. Sakamoto, \pl{151}{85}{115};\\
C.M. Hull and E. Witten, \pl{160}{85}{398};\\
R. Brooks, F. Muhammad and S.J. Gates Jr., \np{268}{86}{599}.

\items{12} C.~Lee, K.~Lee and E.J.~Weinberg, \pl{243}{90}{105}.

\items{13} J. Wess and B. Zumino, \np{70}{74}{39}; \pl{49}{74}{52}.

\items{14} C.~Vafa and N.P.~Warner, \pl{218}{89}{51};\\
W.~Lerche, C.~Vafa and N.P.~Warner, \np{314}{89}{427}.

\items{15} J. Wess and B. Zumino, \np{78}{74}{1}; \\
S. Ferrara and B. Zumino, \np{79}{74}{413}; \\
A. Salam and J. Strathdee, \pl{51}{74}{353}.

\items{16} E. Witten, \cmp{117}{88}{353}.

\items{17} H. Nishino, S.J. Gates, Jr. and S.V. Ketov, {\it Supersymmetric
Self-Dual Yang-Mills and Supergravity as Background of Green-Schwarz
Superstring}, Univ. of Maryland preprint, UMDEPP 92--171 (February 1992).

\items{18} Y. Brihaye, D.B. Fairlie, J. Nuyts and R.G. Yates, Jour. Math.
Phys. {\bf 19} (1978) 2528; \\
K. Pohlmeyer, Comm. Math. Phys. {\bf 72} (1980) 37.

\items{19} J. Wess and B. Zumino, \pl{66}{77}{361}.

\items{20} S.J. Gates Jr., M.T. Grisaru, M. Rocek and W. Siegel, {\it
Superspace}, Benjamin/Cummings, Reading, MA, 1983.

\items{21} M.F. Sohnius, \np{136}{78}{461}.

\items{22} E. K\"ahler, Abh. Math. Sem. Univ. Hamburg. {\bf 9} (1933) 173.

\items{23} M.F. Atiyah, H.J. Hitchin and I.M. Singer, Proc. Roy. Soc. Lond.
{\bf A262} (1978) 425.

\items{24} P. Fayet, \np{113}{76}{135}.

\items{25} M.F. Sohnius, \np{138}{78}{109}.

\items{26} A.S. Gal'perin, E.A. Ivanov and V.I. Ogievetskii, Sov. J. Nucl.
Phys. {\bf 35} (1982) 458.

\items{27} S.J. Gates, Jr., H. Nishino and S.V. Ketov, {\it Extended
Supersymmetry and Self-Duality in $2 + 2$ Dimensions}, Univ. of Maryland
preprint, UMDEPP 92--187 (March 1992).

\items{28} R. Grimm, M. Sohnius and J. Wess, \np{133}{78}{275}.

\items{29} A. Das, Phys. Rev. {\bf D15} (1977) 2805; \\
E. Cremmer and J. Scherk, \np{127}{77}{259}.

\items{30} E. Cremmer, J. Scherk and S. Ferrara, \pl{74}{78}{61}.

\items{31} S.J. Gates, Jr. and J. Durachta, \mpl{21}{89}{2007}.

\items{32} S.J. Gates, Jr., \np{218}{83}{409}.

\items{33} J. Scherk, {\it Extended Supersymmetry and Extended Supergravity
Theory}, Carg\'{e}se Lecture Notes, LPTENS preprint 78/21 (September 1978).

\items{34} A.~Parkes, {\it ``A Cubic Action for Self-Dual Yang-Mills''},
Z{\" u}rich preprint, ETH-TH/92-14 (March, 1992).

\items{35} W.~Siegel, {\it ``The $~N=2\,(4)$~ String is Self-Dual $~N=4$~
Yang-Mills''}, Stony Brook preprint, ITP-SB-92-24 (May, 1992).

\items{36} E.~Bergshoeff, M.~de Roo and J.W.~van Holten, {\it ``Extended
Conformal Supergravity and Its Applications}, in {\it Superspace and
Supergravity}, Proceedings of Nuffield Workshop, Cambridge, July 1980),
eds.~S.W.~Hawking and M.~Ro{\v c}ek, Cambridge University Press,
Cambridge (1981).

\items{37} E.~Bergshoeff, M.~de Roo and B.~de Wit, \np{182}{81}{173}.

\items{38} W.~Siegel, {\it ``The $~N=4$~ string is the Same as the $~N=2$~
String''}, Stony Brook preprint, ITP-SB-92-13 (April, 1992).

\items{39} B.~Zumino, \pl{69}{77}{369}.

\items{40} G.~'t Hooft, \pr{14}{76}{3432}.

\items{41} A.I.~Vainstein, V.I.~Zakharov, V.A.~Novikov and M.A.~Shifman,
Sov.~Phys.~Usp.~{\bf 25} (1982) 195;
V.A.~Novikov, M.A.~Shifman, A.I.~Vainstein and V.I.~Zakharov,
\np{229}{83}{381}; 394; 407; \pl{139}{84}{389}.

\items{42} E.D.~Rainville, {\it ``Intermediate Course in Differential
Equations''}, John Wiley and Sons, London, 1943.

\items{43} I.S.~Gradshteyn and I.M.~Ryzhik, {\it ``Table of Integrals
Series and Products''}, Academic Press, New York, 1980.

\items{44} S. Donaldson, J. Differ. Geom. {\bf 18} (1983) 269; {\it ibid}
{\bf 26} (1987) 397.

\items{45} A.A. Belavin and V.E. Zakharov, \pl{73}{78}{53} ;\\
S.P. Novikov, S.V. Manakov, L.P. Pitaevskii and V.E. Zakharov,
{\it Theory of Solitons}, Consultans Bureau, N.Y., 1984; \\
C. Rebbi and G. Soliani, {\it Solitons and Particles}, World Scientific,
Singapore, 1984 ;\\
L.D. Faddeev and L.A. Takhtajan, {\it Hamiltonian Methods in the Theory of
Solitons}, Springer-Verlag, Heidelberg, 1987; \\
A. Das, {\it Integrable Models}, World Scientific, Singapore, 1989.

\items{46} L.J. Mason, Twist. Newslett. {\bf 30} (1990) 14.

\items{47} N. Hitchin, {\it Yang-Mills Equations for Diffeomorphism Groups},
Talk delivered at the Workshop of the 1990 Yukawa Intern. Seminar ``Common
Trends in Mathematics and Quantum Field Theories''.

\items{48} J.F. Plebanski, Jour. Math. Phys. {\bf 16} (1975) 2395; \\
C.P. Boyer and J.F. Plebanski, Jour. Math. Phys. {\bf 26} (1985) 229.

\items{49} R. Penrose, Gen. Rel. Grav. {\bf 7} (1976) 31.

\items{50} Q.-H. Park, \pl{238}{90}{287}; \ibid{257B}{91}{105}.

\items{51} M.F. Atiyah, Comm. Math. Phys. {\bf 93} (1984) 437.

\items{52} E. Sezgin, {\it Area-Preserving Diffeomorphisms,
$W_{\infty}$~ Algebras and $~W_{\infty}$~ Gravity},
Texas A \& M preprint, CTP-TAMU-13/92 (February 1991).

\items{53} A.N. Leznov and M.V. Saveliev, \cmp{74}{80}{111}.

\items{54} L. Mason and G. Sparling, \pl{137}{89}{29}.

\items{55} I. Bakas and D.A. Depireux, \mpl{6}{91}{399}; 1561; 2351.

\items{56} J.-B. Zuber, {\it KdV and W - Flows}, Saclay preprint
SPhT/91-052 (January 1991).

\items{57} E. Witten, \np{268}{86}{79}; \\
A. Sen, \np{268}{86}{287}; \\
C.M. Hull, \pl{178}{86}{357}; \\
M.T. Grisaru, A. van de Ven and D. Zanon, \pl{173}{86}{423};
\np{277}{86}{388}.

\items{58} L.F.~Abbott, M.T.~Grisaru and H.J.~Schnitzer,
\pr{16}{77}{3002};

\items{59} M.T.~Grisaru, W.~Siegel and M.~Ro{\v c}ek, \np{159}{79}{429}.

\items{60} C.M.~Christensen, S.~Deser, M.J.~Duff and M.T.~Grisaru,
\pl{84}{79}{411}; R.E.~Kallosh, JETP Lett.~{\bf 29} (1979) 172;
\np{165}{80}{119}.

\items{61} S.~Hawking, in {\it ``Recent Developments in Gravitation''},
Cargese, 1978; \np{114}{78}{349}.

\items{62} T.~Eguchi and A.~Hanson, \ap{120}{79}{82};\\
E.~Calabi, Ann.~Sci.~Ec.~Norm.~Sup.~{\bf 12} (1979) 269.

\items{63} S. Deser, J.H. Kay and K.S. Stelle, Phys. Rev. Lett. {\bf 38}
(1977) 527; \\
S. Deser and J.H. Kay, \pl{76}{78}{400}; \\
S. Ferrara and P. van Nieuwenhuizen, \pl{78}{78}{573}; \\
R. Kallosh, Pis'ma ZhETF {\bf 29} (1979) 493; \\
P. Howe and U. Lindstr\"om, {\it Counter-Terms for Extended Supergravity}, in
``Superspace and Supergravity'', Cambridge Proceedings 1980, ed.~by
S.W.~Hawking and M.~Rocek, p.~413; \\
P. Howe and U. Lindstr\"om, \np{181}{81}{487}.

\items{64} S.V.~Ketov, {\it ``Non-Linear Sigma-Models in Supersymmetry,
Supergravity and Strings''}, Nauka, Novosibirsk, 1992.

\items{65} J. Scherk and J.H. Schwarz, \np{153}{79}{61}.

\items{66} E. Witten, \pl{155}{85}{151}.

\items{67} P.~Candelas, G.~Horowitz, A.~Strominger and E.~Witten,
\np{253}{85}{46}.

\items{68} I. Ya. Aref'eva and I.V. Volovich, \pl{164}{85}{287}; \\
A.D. Popov, \pl{259}{91}{256}.

\items{69} S.J. Gates, Jr. and H. Nishino, \pl{173}{86}{46} and 51;
\np{282}{87}{1}; \np{291}{87}{205}; \pl{266}{91}{14}.

\items{70} M.T. Grisaru, H. Nishino and D. Zanon, \pl{206}{88}{625};
\np{314}{89}{363}.

\items{71} I. Ya. Aref'eva and I.V. Volovich, \np{70}{87}{297};
Fisika {\bf 18} (1986) 150; Theor. Math. Phys. {\bf 70} (1987) 297.

\items{72} H. Nishino, \np{338}{90}{386}.

\items{73} H. Nishino and E. Sezgin, \pl{144}{84}{187}; \np{278}{86}{353}.

\end{document}